\documentclass[a4paper,11pt]{article}
\pdfoutput=1 

\usepackage[usenames,dvipsnames,svgnames,table]{xcolor} 
\usepackage{jcapmod}
\usepackage{verbatim}

\usepackage[T1]{fontenc} 
\usepackage[latin9]{inputenc}

\definecolor{lightgray}{gray}{0.9}
\definecolor{Amber}{rgb}{1.0, 0.75, 0.0}
\definecolor{blizzardblue}{rgb}{0.67, 0.9, 0.93}
\definecolor{burningsand}{RGB}{220, 148, 129}
\definecolor{burgundy}{rgb}{0.5, 0.0, 0.13}

\usepackage{float}
\usepackage{graphicx}
\usepackage{hyperref}
\usepackage{amsmath}
\usepackage{amssymb}
\usepackage{booktabs}
\usepackage{microtype}
\usepackage[shortlabels]{enumitem}
\usepackage{cancel}
\usepackage{amsbsy}
\usepackage{color}
\usepackage[capitalise]{cleveref}
\usepackage{breakurl}
\usepackage[normalem]{ulem}
\usepackage{mathtools}
\usepackage{csquotes}
\usepackage{bigints}
\usepackage{lmodern}
\usepackage{bm}
\usepackage{dsfont}
\usepackage{bbold}
\usepackage{empheq}
\usepackage{caption}
\usepackage{subcaption}
\usepackage{makecell}
\usepackage{multirow}
\usepackage{siunitx}

\makeatletter
\g@addto@macro\bfseries{\boldmath}
\makeatother

\pdfpageheight\paperheight
\pdfpagewidth\paperwidth

\DeclareSIUnit\parsec{pc}

\usepackage[textwidth=0.71\paperwidth, textheight=0.8\paperheight]{geometry}

\renewcommand*{\vec}[1]{\bm{#1}}

\newcommand{\dd}{\mathrm{d}}

\newcommand*{\E}[1]{\texorpdfstring{\ensuremath{E_{#1}}}{E#1}}

\newcommand{\xobs}{\bm{\tilde{x}_{0}}}
\newcommand{\xobslss}{\boldsymbol{x_0}}
\newcommand{\xvec}{\bm{x}}
\newcommand{\kvec}{\boldsymbol{k}}
\newcommand{\kpvec}{\boldsymbol{k'}}
\newcommand{\hatkvec}{\boldsymbol{\hat{k}}}
\newcommand{\nvec}{\boldsymbol{n}}
\newcommand{\Tvec}{\boldsymbol{T}}
\newcommand{\alm}{a_{\ell m}}
\newcommand{\hatp}{\hat{p}}

\newcommand{\calR}{\mathcal{R}}
\newcommand{\calP}{\mathcal{P}}
\newcommand{\calPR}{\calP^{\calR}}

\newcommand{\LA}{L_{A}}
\newcommand{\LB}{L_{B}}
\newcommand{\LLSS}{L_{\mathrm{LSS}}}
\newcommand{\Lcircle}{L_{\mathrm{circle}}}
\newcommand{\Lmod}{L_\text{mod}}
\newcommand{\DKL}{D_{\textsc{kl}}}

\newcommand{\catboost}{\texttt{CatBoost}}
\newcommand{\shap}{\texttt{SHAP}}

\newcommand{\lcdm}{\(\Lambda\)CDM}

\newcommand{\para}[1]{\par\vspace{2mm}\noindent\textbf{#1}\,---\,}

\usepackage[table]{xcolor}

\newcommand{\TP}[1]{\ifmmode\text{\textcolor{olive}{\textbf{ #1}}}\else\textcolor{olive}{\textbf{[TP: #1]}}\fi}

\allowdisplaybreaks

\makeatletter
\DeclareRobustCommand{\rcite}[1]{%
  \rcite@aux#1,\@nil{#1}%
}
\def\rcite@aux#1,#2\@nil#3{%
  \if\relax#2\relax
    Ref.~\cite{#3}%
  \else
    Refs.~\cite{#3}%
  \fi
}
\makeatother

\subheader{IFT-UAM/CSIC-26-12}

\title{Cosmic topology. Part IIc. Detectability with non-standard primordial power spectrum}

\author[a,b]{Joline Noltmann,}
\author[a,c,d]{Andrius Tamosiunas,}
\author[d,e]{Deyan P. Mihaylov,}
\author[a,d,f]{Yashar Akrami,}
\author[a]{Javier Carr\'on Duque,}
\author[g]{Thiago S. Pereira,}
\author[d]{Glenn D. Starkman,}
\author[a]{George Alestas,}
\author[h,i,j]{Stefano Anselmi,}
\author[d]{Craig J. Copi,}
\author[d,k]{Fernando Cornet-Gomez,}
\author[f]{Andrew H. Jaffe,}
\author[l]{Arthur Kosowsky,}
\author[a,m]{Mikel Martin Barandiaran,}
\author[d]{Anna Negro,}
\author[d]{Amirhossein Samandar}

\collaboration{(COMPACT Collaboration)}
\collaborationImg{\includegraphics[width = 0.1\textwidth]{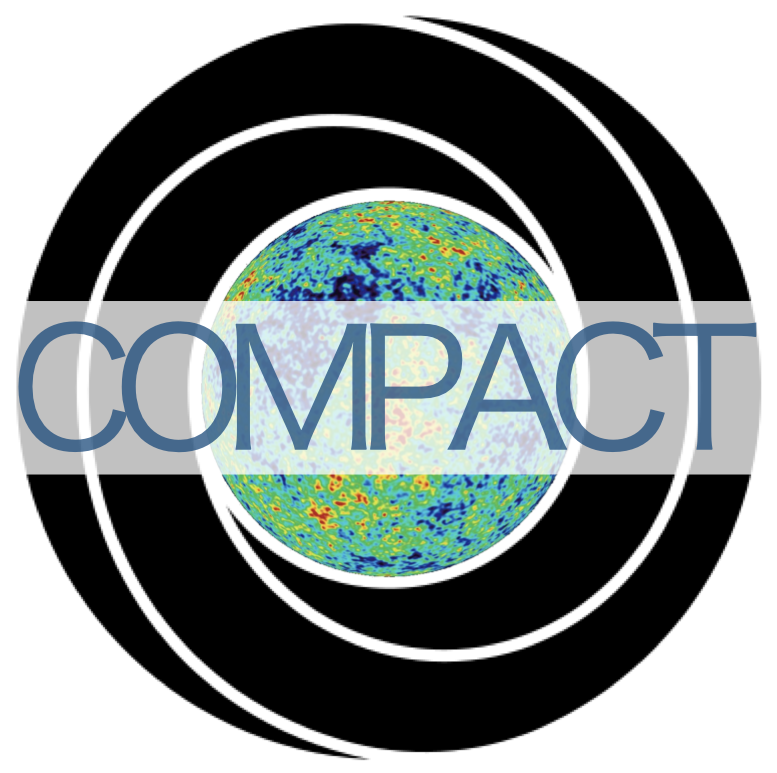}}

\affiliation[a]{Instituto de F\'isica Te\'orica (IFT) UAM-CSIC, C/ Nicol\'as Cabrera 13-15, Campus de Cantoblanco UAM, 28049 Madrid, Spain}
\affiliation[b]{Institute for Theoretical Particle Physics and Cosmology, RWTH Aachen University, Templergraben 55, 52062 Aachen, Germany}
\affiliation[c]{Institute of Theoretical Astrophysics, University of Oslo, P.O. Box 1029 Blindern, N-0315 Oslo, Norway}
\affiliation[d]{CERCA/ISO, Department of Physics, Case Western Reserve University, 10900 Euclid Avenue, Cleveland, OH 44106, USA}
\affiliation[e]{Department of Astronomy, Faculty of Physics, Sofia University ``St~Kliment~Ohridski'', 5 James Bourchier Blvd, 1164 Sofia, Bulgaria}
\affiliation[f]{Astrophysics Group \& Imperial Centre for Inference and Cosmology, Department of Physics, Imperial College London, Blackett Laboratory, Prince Consort Road, London SW7 2AZ, United Kingdom} 
\affiliation[g]{Departamento de F\'{i}sica, Universidade Estadual de Londrina, Rod. Celso Garcia Cid, Km 380, 86057-970, Londrina, Paran\'{a}, Brazil}
\affiliation[h]{INFN, Sezione di Padova, via Marzolo 8, I-35131 Padova, Italy}
\affiliation[i]{Dipartimento di Fisica e Astronomia ``G. Galilei'', Universit\`a degli Studi di Padova, via Marzolo 8, I-35131 Padova, Italy}
\affiliation[j]{Laboratoire Univers et Th\'eories, Observatoire de Paris, Universit\'e PSL, Universit\'e Paris Cit\'e, CNRS, F-92190 Meudon, France}
\affiliation[k]{Departamento de F\'isica, Universidad de C\'ordoba, Campus Universitario de Rabanales, Ctra. N-IV Km. 396, E-14071 C\'ordoba, Spain}
\affiliation[l]{Department of Physics and Astronomy, University of Pittsburgh, Pittsburgh, PA 15260, USA}
\affiliation[m]{Departamento de F\'isica Te\'orica, Universidad Aut\'onoma de Madrid, 28049 Madrid, Spain}

\emailAdd{joline.noltmann@rwth-aachen.de}
\emailAdd{andrius.tamosiunas@astro.uio.no}
\emailAdd{deyan.mihaylov@case.edu}
\emailAdd{yashar.akrami@csic.es}
\emailAdd{javier.carron@csic.es}
\emailAdd{tspereira@uel.br}
\emailAdd{glenn.starkman@case.edu}
\emailAdd{g.alestas@csic.es}
\emailAdd{stefano.anselmi@pd.infn.it}
\emailAdd{craig.copi@case.edu}
\emailAdd{fernando.cornetgomez@case.edu}
\emailAdd{a.jaffe@imperial.ac.uk}
\emailAdd{kosowsky@pitt.edu}
\emailAdd{mikel.martin@uam.es}
\emailAdd{anna.negro@case.edu}
\emailAdd{amirhossein.samandar@case.edu}

\abstract{
    Non-trivial spatial topology of the Universe can imprint potentially observable signatures on the cosmic microwave background (CMB).
    In this study, we investigate how deviations from the standard nearly-scale-free primordial power spectrum impact observables for the fully compact, orientable Euclidean topologies (\E{1}--\E{6}).
    We examine how such deviations modify the detectability of the underlying topology, depending on whether they are an intrinsic consequence of non-trivial topology or independent of it.
    We compute CMB temperature correlation matrices across a range of topologies, fundamental domain sizes, and observer locations for both standard and modified primordial power spectra.
    The impact of these modifications on the detectability of topology is quantified using the Kullback-Leibler divergence, providing an estimate of the distinguishability of non-trivial and simply-connected topologies based solely on CMB temperature observations. 
    In addition, we employ the \catboost\  machine learning algorithm to classify harmonic-space realizations of CMB temperature maps and thereby assess the observational prospects for topology detection. 
    Signatures of non-trivial topology are encoded in the off-diagonal structure of the CMB temperature correlation matrices and are most prominent on the largest angular scales. 
    Deviations from the simple power-law primordial spectrum at these scales can substantially alter the detectability of topology, either enhancing its characteristic CMB imprints or suppressing them below observational sensitivity. 
    Our results demonstrate that uncertainties in the primordial power spectrum must be carefully accounted for in robust searches for cosmic topology using the CMB.
}


\begin{document}
\maketitle
\flushbottom

\section{Introduction}
\label{secn:intro}

The global topology of the Universe remains a fundamental unresolved question in modern cosmology: while general relativity governs the local geometry of spatial hypersurfaces, it does not fix the global connectivity, leaving open the possibility of multiply-connected space. Cosmic microwave background (CMB) anisotropies provide a particularly sensitive probe of such non-trivial topology. Their statistical properties are observationally accessible while being precisely predicted within the standard \(\Lambda\) cold dark matter (\lcdm) framework.

In this framework, the large-scale Universe is modeled by the class of homogeneous and isotropic Friedmann--Lema\^{i}tre--Robertson--Walker (FLRW) spacetime metrics. Current observations \cite{Planck18,ACT:2023kun} favor a local geometry that is very close to the spatially flat metric, for which the line element in spherical coordinates is
\begin{equation}\label{eq:flrw-metric}
    \dd s^2 = -c^2 \dd t^2 + a(t)^2\left[\dd r^2+r^2(\dd\theta^2 + \sin^2\theta \, \dd\phi^2)\right]\,,
\end{equation}
where \(a(t)\) is the scale factor.

Departures from this background description can largely be treated with linear perturbation theory. In brief, this consists in linearizing Einstein's field equations around the background FLRW solution, yielding a set of partial differential equations describing the evolution of independent scalar, vector, and tensor types of perturbations. These form the foundation for calculating cosmological observables, allowing, in particular, a precise characterization of the statistics of CMB anisotropies. 

Scalar perturbations provide the dominant contribution to the observed CMB temperature anisotropies. Because the relevant physics is linear, the statistical properties of the observed temperature fluctuations are linearly related to those of the primordial  curvature perturbation specified during inflation, with recombination (at \(z \approx 1100\)) providing the main surface from which the photons free-stream, and with all subsequent evolution and projection effects encoded in transfer functions.

The observed CMB temperature fluctuations $\Delta T$ on the celestial sphere can be expanded in terms of spherical harmonics
\begin{equation}
    \Delta T(\theta, \phi)=\sum_{\ell, m}a_{\ell m}Y_{\ell m}(\theta, \phi)\,,
\end{equation}
where $Y_{\ell m}$ are the spherical harmonics, and $a_{\ell m}$ are the complex harmonic-space coefficients.

In the standard inflationary paradigm, the curvature perturbation is described by a Gaussian random variable with zero mean. In a universe with metric \eqref{eq:flrw-metric} and trivial topology, i.e., simply-connected, the spectrum of the scalar Laplacian is continuous, allowing a complete and orthogonal Fourier decomposition of perturbations. Homogeneity and isotropy of the background then imply that perturbations with different wavenumbers are uncorrelated, and with a power spectrum depending only on the amplitude of the wavenumbers. This ultimately leads to vanishing off-diagonal correlations of temperature fluctuations in harmonic space, in which case the angular power spectrum, or \(C_{\ell}\), completely describes the statistics of temperature anisotropies:
\begin{equation}
    C_\ell^{\E{18}; TT}=4\pi \! \bigintsss \! \frac{\dd k}{k} \, \calPR(k) \, \Delta_\ell^{T}(k) \, \Delta_\ell^{T*}(k)\,.
    \label{eq:ang_pow_spec}
\end{equation}
Here, \(\calPR (k)\) denotes the primordial curvature power spectrum,  which determines the amplitude of the adiabatic curvature perturbation as a function of wavenumber \(k\), and $\Delta_\ell^{T}(k)$ is the angular transfer function, which relates the CMB temperature anisotropies to the adiabatic curvature perturbation.

This picture is drastically changed if the space is not simply-connected. In this case, non-trivial boundary conditions imply both discretization of the eigenspectrum of the Laplacian eigenmodes and correlations among different Fourier modes. When the resulting perturbations are projected onto the CMB sky, both of these effects imply the appearance of off-diagonal temperature correlations in harmonic space. Thus, a great deal of work consists in constructing the full covariance matrix, whether in Fourier or harmonic space. 

The focus of this work is on the six fully compact, orientable topologies \E{i} with $i \in \{1,...,6\}$. They can be constructed from a set of generators $g_{a_j}^{E_i}$ of the associated, discrete subgroup $\Gamma^{E_i}\subset E(3)$. For orientable manifolds, the generators $g_{a_j}^{E_i}$ are pure translations or corkscrew motions. Each generator acts on a point \(\xvec\in E^3\) as
\begin{equation}
    \xvec \mapsto \mathsf{M}_{a}^{\E{i}} (\xvec-\xvec_0) + \Tvec_{a_{j}}^{\E{i}}+\xvec_0\,,
\end{equation}
where the matrix \(\mathsf{M}_{a}^{\E{i}} \in \mathrm{SO(3)}\) (no reflections) encodes the rotational part of the generator and \(\Tvec_{a_{j}}^{\E{i}} \in \mathds{R}^{3}\) is its translational part. The index \(a\) labels the distinct type of isometries while \(j\) indexes repeated generators of the same type, and $\xvec_0$ indicates a point on the axis/intersection of the axes.\footnote{Consequently, the observer is located at $-\xvec_0$.}
These generators impose periodic boundary conditions on the scalar field, requiring the eigenmodes of the Laplacian, $\Upsilon_{\kvec}^{\E{i}}(\xvec)$, to be invariant under the group action. Consequently, the allowed wavevectors become discretized. To compute the CMB observables, we expand these spatial eigenmodes in the spherical harmonic basis,
\begin{equation}
    \Upsilon_{\kvec}^{\E{i}}(\xvec) = 4\pi\sum_{\ell, m}j_\ell (kr)\xi_{k\ell m}^{\E{i},\hatkvec} Y_{\ell m}(\theta,\phi)\,,
\end{equation}
where $j_\ell(kr)$ is the spherical Bessel function.
The coefficients $\xi_{k\ell m}^{\E{i},\hatkvec}$ from this expansion are then used to construct the full harmonic-space temperature covariance matrix, which (for \E{1}--\E{6}) is given by 
\begin{equation}
    C^{\E{i}; TT}_{\ell m \ell' m'} = \frac{(4\pi)^2}{V_{\E{i}}} \sum_{\nvec \in \mathcal{N}^{\E{i}}} \Delta^T_\ell(k_{\nvec}) \Delta^{T*}_{\ell'}(k_{\nvec}) \frac{2\pi^2 \calPR(k_{\nvec})}{k_{\nvec}^3} 
    \xi_{k_{\nvec} \ell m}^{\E{i}; \hatkvec_{\nvec}} \xi_{k_{\nvec} \ell' m'}^{\E{i}; \hatkvec_{\nvec} *}\,,
    \label{eq:harm_cov}
\end{equation}
where $V_{\E{i}}$ denotes the volume of the compact manifold and $\mathcal{N}^{\E{i}}$ labels the discrete set of $\kvec_{\nvec}$.

There is, however, one important aspect in this approach that was heretofore set aside for simplicity, even as it was understood to potentially be significant: when predicting the CMB covariance matrix in a non-trivial manifold, one usually adopts the standard power-law form for the primordial power spectrum, derived from canonical quantization rules in a simply-connected space. This approach is largely adopted, whether in simulating CMB maps in non-simply-connected manifolds~\cite{Aurich:2000br,Riazuelo:2002ct,Uzan:2003ea,Riazuelo2004:prd,Rocha:2002kk,Aurich:2004xa,Phillips:2004nc,Hipolito-Ricaldi:2005vst} or in inferences of detectability from such simulations~\cite{Cornish:1997ab,ShapiroKey:2006hm,Fabre:2013wia,Cornish:2003db,Aurich:2004fq,Niarchou:2007nn,Bielewicz:2011jz,Planck:2013okc,Planck:2015gmu,COMPACT:2022gbl}.

In fact, in previous work we provided general analytic constructions of scalar (spin-0) eigenmodes, together with their full harmonic-space covariance matrices, for both orientable \cite{COMPACT:2023rkp} and non-orientable \cite{COMPACT:2025adc} Euclidean three-manifolds, and assessed their detectability assuming the standard, isotropic primordial power spectrum. We also constructed tensor (spin-2) eigenmodes and their corresponding covariance matrices for orientable Euclidean manifolds \cite{Samandar:2025kuf}, again adopting a similar primordial power spectrum to study the statistical properties of CMB tensor (polarization) anisotropies and their implications for topology detectability \cite{Samandar:2025kuf} and parity violation \cite{COMPACT:2024cud}. The same assumption of a standard primordial power spectrum was likewise made in our earlier machine-learning--based exploration of the detectability of non-trivial cosmic topology signatures in harmonic space \cite{COMPACT:2024dqe}.

However, one may argue that in a fully developed topological description of the Universe, the derivation of the primordial power spectrum should be subject to the same boundary conditions imposed on cosmological observables. While quantization prescriptions for fields on non-trivial manifolds do exist \cite{Halliwell:1989myn}, they are known to be non-unique \cite{Fulling:1972md,wald1994quantum}. Moreover, non-trivial topologies can influence the classical dynamics of the early Universe by introducing inhomogeneities and/or anisotropies \cite{Negro2026}; however, the impact of these effects on the primordial power spectrum remains to be fully understood.

While such a fundamental description is still lacking, and given that the signatures of non-trivial topology manifest primarily on large angular scales, we adopt a pragmatic approach and investigate how functional modifications of the primordial power spectrum at these scales affect detectability. We model these deviations both as intrinsic consequences of non-trivial topology and as topology-independent uncertainties that may hinder its detection.

With this goal in mind, we systematically explore three phenomenological modifications of the primordial curvature power spectrum in the infrared (IR) regime: cutoffs (power suppression), enhancements (power amplification), and oscillations. These modifications redistribute power among different Fourier modes, thereby directly altering the amplitudes of the non-zero components of the CMB covariance matrix. We quantify the impact of such changes on topological detectability in harmonic space using both the Kullback--Leibler (KL) divergence \cite{kullback1951, kullback1959information} and a gradient-boosting machine-learning (ML) algorithm.
This dual approach enables us to assess the robustness of topology detection in the presence of uncertainties in early-universe physics.

In Section~\ref{secn:method}, we describe the considered modifications of the primordial power spectrum, the computation of the CMB covariance matrices, the theoretical background of the KL divergence, and the ML techniques employed. Section~\ref{subsecn:numerical_results} presents the numerical results together with their interpretation. Additional material is provided in the appendices: Appendix~\ref{app:eigenmodes} reviews the key properties of the scalar Laplacian eigenmodes and the associated correlation matrices; Appendix~\ref{app:kl} presents supplementary results for the KL divergence; and Appendix~\ref{app:ml} contains the ML feature-importance and interpretability analyses.

\section{Method}
\label{secn:method}
In the following, we describe the parametrizations and parameter ranges adopted for non-standard primordial power spectra. We then outline the numerical procedures used to compute CMB temperature covariance matrices and to evaluate the KL divergence for different topological models and power spectra. Finally, we discuss how simulated CMB realizations are classified according to their underlying topology using a gradient-boosting ML algorithm.

\subsection{Modifications to the primordial power spectrum}
\label{subsec:mods}
In standard \lcdm\ cosmology, the primordial power spectrum is typically assumed to follow a nearly-scale-invariant power-law form,
\begin{equation}
    \calPR(k) = A_\mathrm{s} \left(\frac{k}{k_*}\right)^{n_\mathrm{s} - 1}\,, \label{eq:powlaw}
\end{equation}
where \(A_\mathrm{s}\) is the amplitude at a pivot scale \(k_{*}\) and \(n_\mathrm{s}\) is the scalar spectral index. 
Deviations from this standard spectrum alter the distribution of power among different Fourier modes, and ultimately among different angular scales $\ell$ through \cref{eq:harm_cov}. 
Given a phenomenological modification of \(\calPR(k)\), a non-trivial topology, or both effects together, we compute the corresponding covariance matrices \(C^{\E{i}; TT}_{\ell m \ell' m'}\) and compare them with the baseline \lcdm\ prediction.
This analysis allows us to assess whether a given modification amplifies or suppresses topological signatures and to estimate the overall sensitivity of topology detection to uncertainties in the primordial power spectrum. 
Because signatures of non-trivial topology manifest primarily in the off-diagonal elements at large angular scales, modifications to \(\calPR (k)\) in the IR regime are of particular interest.

In this work, we classify such modifications as primordial power spectrum modifications that are a consequence of non-trivial topology itself and primordial power spectrum modifications that arise independently of the topology, e.g., due to features in the inflationary potential. In either case, a systematic phenomenological exploration of different modifications is necessary for assessing the robustness of topology detection methods. Therefore, we consider three general types of modifications:
\begin{itemize}
\item \textbf{Cutoffs:} suppression of power at large scales (and thus large angular scales);
\item \textbf{Enhancements:} amplification of power at large scales;
\item \textbf{Oscillations:} periodic modulations in the low-\(k\) region of \( \calPR(k)\).
\end{itemize}
Modifications that are a consequence of non-trivial topology are motivated by the characteristic scale of the non-trivial topology, which breaks the assumption of scale invariance. 
It would be reasonable to think that this may affect the power spectrum at scales near this characteristic scale. Since we do not know yet the exact effects, we model them with these types of modifications.

However, such large-scale features in \(\calPR(k)\) could also emerge from inflationary scenarios, independently of the underling topology.
An IR cutoff may result from models with a short period of inflation, with the detailed form of the cutoff depending on the assumed pre-inflationary state \cite{inf_IR_supp, inf_IR_supp2, inf_IR_supp3}.
Oscillatory features in \(\calPR(k)\)  can arise from phase transitions or step-like features in the inflaton potential \cite{inf_osci, inf_osci2} and are also a generic signature of anisotropies during inflation~\cite{Gumrukcuoglu:2007bx,Pitrou:2008gk}. 
Though oscillatory features usually arise at intermediate scales, here we focus on oscillations restricted to large scales. Enhancements in the power spectrum could result from periods of slowed inflaton rolling due again to the specifics of the inflaton potential or from extra dimensional effects as in \rcite{dark_dim}.

\begin{figure}[t]
  \centering
  \begin{subfigure}[b]{0.495\textwidth}
    \centering
    \includegraphics[width=\textwidth]{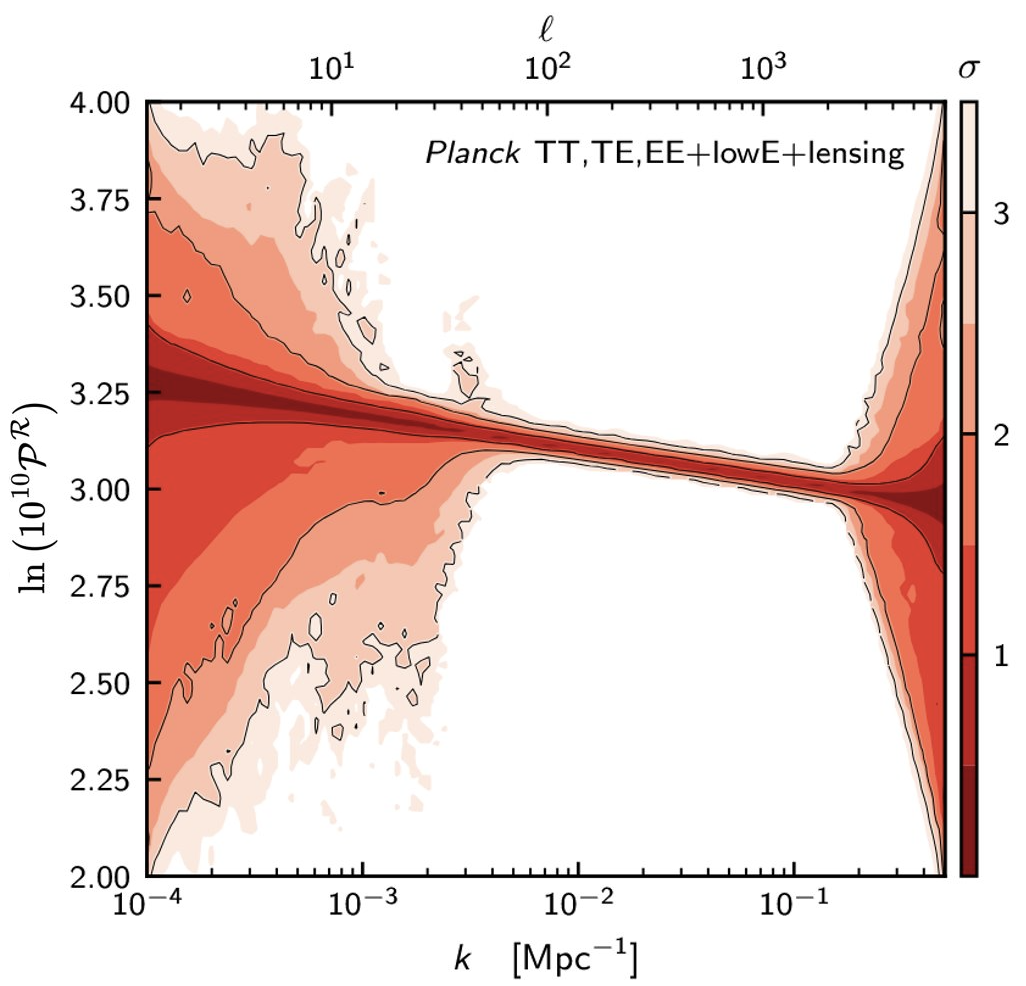}
  \end{subfigure}
  \hfill
  \begin{subfigure}[b]{0.465\textwidth}
    \centering
    \includegraphics[scale=1]{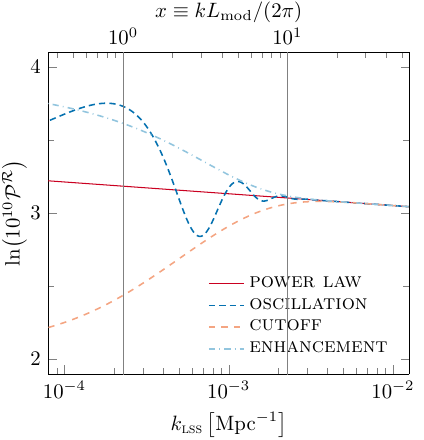}
  \end{subfigure}
  \caption{Left: Posterior distribution of \(\calPR(k)\) reconstructed from the {\it Planck} data using a Bayesian non-parametric reconstruction (Fig. 20 of  \rcite{planck18_inf}). Right: Modifications to the primordial power spectrum with parameters specified in \cref{eq:modifications} for \(\Lmod = \LLSS\).}
  \label{fig:planck_constraints}
\end{figure}

On the largest scales, {\t Planck} data do not tightly constrain \(\calPR(k)\). A Bayesian non-parametric reconstruction \cite{planck18_inf} shows a broad posterior distribution for \(\calPR(k)\) at these scales. However, it assumes the standard \lcdm\ model as a prior and is therefore not directly applicable to compact topologies; nonetheless, we use the \(3 \sigma\) bounds of this reconstruction as a reference for the magnitude of the modifications considered here (see \cref{fig:planck_constraints}).

We use the following parameterizations for the three types of modifications discussed above. While we explore a range of parameters (including values exceeding the reference bounds) to assess the dependency of the KL divergence, the specific parameters listed below represent the largest values that remain within the $3\sigma$ reference bounds. We adopt these as the standard maximal parameters for the results presented in this work.
\begin{align}
\textsc{cutoff: } \calPR_\text{mod} (k)&=\calPR(k)(1 - a e^{-x/x_c}), & a &= 0.7, \; x_c = 3.5, \nonumber \\ 
\textsc{enhancement: }\calPR_\text{mod} (k)&=\calPR(k)(1 + a e^{-x/x_c}), & a &= 0.8, \; x_c = 2.5, 
\label{eq:modifications} \\ 
\textsc{oscillation: }\calPR_\text{mod} (k)&=\calPR(k)(1 + a \sin(f x)e^{-x/x_c}), & a &= 1.2, \; f = 1.5, \; x_c = 2. \nonumber 
\end{align}
Here
\begin{equation}
x \equiv \frac{k \Lmod}{2 \pi}
\end{equation}
is a dimensionless, rescaled wavenumber. We define the global topological scale $\Lmod$ as the minimum distance between any point in the manifold and its nearest clone, 
\begin{equation} 
    \Lmod \equiv \min_{\vec{x}, g \in \Gamma^{E_i}} |g(\vec{x}) - \vec{x}|\,, 
\end{equation}
where the minimization is performed over all possible locations on the fundamental domain $\vec{x}$ and over all non-identity elements $g$ of the discrete group of isometries $\Gamma^{E_i}$ that define the topology. The actual physical scale at which the topology can be probed by a wave is determined by the topology-specific minimum wavenumber $k_\mathrm{min}$. For example, for the cubic \E{1} topology, \(\Lmod = L_1=L_2=L_3\) and \(x = 1\) coincides with the lowest allowed mode \(k_{\mathrm{min}} = 2\pi / \Lmod\), though, in other cases---non-cubic \E{1} or other topologies---$k_\mathrm{min}$ can be higher or lower.
Nevertheless, $\Lmod$ is a useful topology-agnostic phenomenological choice of scale.
Parametrizing spectral modifications in terms of $x\equiv \Lmod/\lambda$ ensures that the features in the primordial power spectrum scale naturally with the dimensions of the manifold. 
This means that modes with $x<1$ are not physically allowed in the cubic \E{1} and the excluded values of $x$ can extend to even higher values for other topologies. 
Nevertheless,  we treat the power spectrum as a continuous envelope within the Boltzmann code and define the modification for all $x$. 
The exponential decay term $e^{-x/x_c}$ ensures that the modification is suppressed for modes with $x\gg 1$, where the suppression starts at the characteristic wavelength $x_c$, to satisfy the {\it Planck} constraints for smaller scales.

The right panel of \cref{fig:planck_constraints} illustrates the resulting modifications for maximal parameter values and \(\Lmod = \LLSS\), the diameter of the last scattering surface. In this case, \(x = 1\) corresponds to \(k \approx \qty{2.3e-4}{\mega\parsec^{-1}}\). 

\subsection{Evaluation of CMB covariance matrices}
\label{subsecn:numerical_covariances}

The CMB temperature anisotropies are modeled as Gaussian random fields on the celestial sphere, with all statistical information fully encoded in the covariance matrices of their spherical harmonic coefficients. 

The covariance matrix differs significantly depending on the topology. For the simply-connected Euclidean space \E{18}, statistical isotropy enforces the diagonal covariance matrix
\begin{equation}
    C_{\ell m \ell' m'}^{\E{18}; TT}\equiv\biggl< \alm^{\E{18};T} a_{\ell' m'}^{\E{18};T*} \biggr> = C_{\ell}^{\E{18};TT}\delta^{(K)}_{\ell \ell'}\delta^{(K)}_{mm'}\,,
    \label{eq:stat_iso}
\end{equation}
with the Kronecker deltas $\delta^{(K)}_{\ell \ell'}$ and $\delta^{(K)}_{mm'}$, and the angular power spectrum $C_\ell$ as given by \cref{eq:ang_pow_spec}. The angular part of the three-dimensional integral over the wavevector \(\kvec\) can be computed trivially because of the unbroken isotropy, leaving the one-dimensional integral over the magnitude of \(\kvec\).

In contrast, non-trivial Euclidean topology breaks statistical isotropy, resulting in non-zero off-diagonal elements in the covariance matrix. 
For the fully compact, orientable Euclidean topologies, \cref{eq:harm_cov} describes the  CMB temperature-temperature  correlations.  Only a discrete set of wavevectors is allowed, so the integral over all \(\kvec\) is replaced by a sum over the allowed modes. This infinite sum is computationally expensive, and therefore, has to be truncated in practice.
This is accomplished with a multipole-dependent cutoff \(k_{\max} (\ell)\), which is defined using the ratio 
\begin{equation}
    R_\ell(k) \equiv \frac{C^{k}_\ell}{C^{\Lambda\text{CDM}}_\ell}\,,
    \label{eq:ratio_kmax}
\end{equation}
where
\begin{equation}
    C^{k}_\ell = 4\pi \int_0^{k} \dd k' \, \frac{\calPR(k')}{k'} \, \Delta_{\ell} (k')^{2}\,,
\end{equation}
and \(C^{\Lambda \text{CDM}}_\ell\) is the standard \lcdm\ angular power spectrum computed with \texttt{CAMB} \cite{CAMB} for the simply-connected topology \E{18}. 
We use the \textit{Planck} 2018 \lcdm\ parameters \cite{Planck:2018vyg} to determine the standard primordial power spectrum \(\calPR(k)\), and, as inputs to \texttt{CAMB}, to compute the transfer function \(\Delta_{\ell} (k)\). 
For the unmodified power spectrum, following the approach of \rcite{COMPACT:2023rkp}, the cutoff \(k_{\max} (\ell)\) is chosen such that at least \(99\%\) of the \lcdm\ power is recovered, i.e., \(R_\ell(k_\text{max}) \geq 0.99\). 
For off-diagonal elements \(\ell \neq \ell'\), the summation uses the largest of the two cutoffs, \(\max(k_\text{max}(\ell),k_\text{max}(\ell'))\), which, in practice, is \(k_\text{max}(\max(\ell,\ell'))\).
Further precision beyond 99\% minimally affects the KL divergence results. 
Increasing \(L/\LLSS\) increases the values of $\vert\vec{n}\vert$ that contribute to any specific \(\ell\).  
Hence, the computational cost of calculating a correlation matrix scales as \(\mathcal{O} (L^{3})\).
When evaluating the covariance matrices for modified primordial spectra, the same cutoffs \(k_{\max}(\max(\ell, \ell'))\) are maintained for consistency. 
Since the considered modifications are in the low-\(k\) regime, the  modes with  \(|\kvec|\gg|\kvec_{\max}(\ell)|\) contribute negligible power, just as in the standard power-law scenario. 
All computations are performed using a dedicated Python code developed by the COMPACT collaboration.\footnote{The code is available at \url{https://github.com/CompactCollaboration/CMBtopology}.}

It is worth noting that the pattern of non-zero elements in the covariance matrix depends on the orientation of the chosen coordinate system with respect to the fundamental domain.
Though, since rotations mix only \(\alm\) of different \(m\) within a given \(\ell\), many aspects of the correlation matrix will stay the same, such as the vanishing of certain blocks of fixed \(\ell\) and \(\ell'\). 
Since our focus is on comparing correlations in non-trivial topology  with the rotationally invariant covering space \E{18}, the KL divergence remains independent of coordinate orientation in the idealized case of noise-free observations over the full sky.

\subsection{Kullback-Leibler divergence}
\label{subsec:KL_div}
The KL divergence can be used to compare the predictions of two models for any observable, e.g., the CMB temperature anisotropies.
We compute the KL divergence between the standard, statistically isotropic model with temperature covariance as given by \cref{eq:stat_iso} and a non-trivial topology model with
\begin{equation}
    \left\langle \alm^{\E{i}; T} a_{\ell' m'}^{\E{i};T*} \right\rangle = C_{\ell m\ell'm'}^{\E{i};TT}\,.
\end{equation}
The KL divergence between the two resulting probability distributions, \(p(\{\alm\})\) and \(q(\{\alm\})\), is defined as
\begin{equation}
    \DKL (p||q) \equiv \int \dd \{\alm\} \, p(\{\alm\}) \ln\frac{p(\{\alm\})}{q(\{\alm\})}=\mathrm{E}_p\left[ \ln\frac{p(\{\alm\})}{q(\{\alm\})}\right]\,,
    \label{eq:KL_div}
\end{equation}
where \(\mathrm{E}_{p}\) is the expected value with respect to the distribution \(p\). The KL divergence can therefore be understood as the expected value of the log-Bayes ratio between the models $p$ and $q$ when the true model is the first one.
In information-theory terms, \(\DKL (p||q)\) quantifies the amount of information that is lost when the real data \(p\) is represented by the model \(q\). The greater the KL divergence, the more different (and distinguishable) the two probability distributions. 

In this work, we consider the probability distributions predicted for a non-trivial topology \E{i} (with or without a modified power spectrum) compared with the prediction of the trivial topology \E{18} (with or without a modified power spectrum). We can then write the KL divergence as
\begin{equation}
    \boxed{
        \DKL (\E{i}||\E{18}) = \frac{1}{2}\sum_{j=1}^{n} (-\ln\lambda_{j} + \lambda_{j} - 1)\,,
    }
    \label{eq:KL_eigenvalues}
\end{equation}
where \(\lambda_{j}\) are the eigenvalues of the matrix $(C_{\ell m\ell'm'}^{\E{18}})^{-1}C_{\ell m\ell'm'}^{\E{i}}$. We assume an ideal, noise-free experiment with full-sky coverage, so only cosmological and topological effects are present in the covariance matrices. The dimensionality \(n\) of the data vector \(\alm\), for multipoles up to \(\ell_\text{max}\), is given by
\begin{equation}
    n = \sum_{\ell=2}^{\ell_\text{max}}(2\ell+1).
    \label{eq:lmax}
\end{equation} 
The \emph{forward} KL divergence \(\DKL (\E{i}||\E{18})\) quantifies the information lost when CMB data from a universe with a non-trivial topology is modeled using the standard, simply-connected \lcdm\ model. 
In other words, it shows how distinguishable the two models are in principle, assuming \E{i} is the true topology of the Universe. 
For instance, a large value of \(\DKL (\E{i}||\E{18})\) implies that the trivial model would be a poor approximation to the true distribution, and thus, the models would be highly distinguishable. Alternatively, one may compute the \emph{backward} KL divergence \(\DKL (\E{18}||\E{i})\). As the eigenvalues of $(C_{\ell m\ell'm'}^{\E{i}})^{-1}C_{\ell m\ell'm'}^{\E{18}}$ are simply \(1 / \lambda_{j}\), this yields
\begin{equation}
    \DKL(\E{18}||\E{i}) = \frac{1}{2} \sum_{j=1}^{n} (\ln\lambda_{j} + \lambda_{j}^{-1} - 1)\,,
    \label{eq:KL_backward}
\end{equation}
which quantifies the mismatch when using a non-trivial topology \E{i} to describe the data from the trivial topology \E{18}. A large value implies that the non-trivial topological model imposes artificial features not supported by the data from the trivial topology, offering insight into the risk of false positives. In both directions \(\DKL = 0\), if and only if the probability distributions are identical almost everywhere (i.e., everywhere except in a zero-measure set). We use $\DKL=1$ as a threshold for detectability, i.e., two topologies are distinguishable for $\DKL\geq 1$.\footnote{
In Bayesian model comparison, one can compute which of two models \(M_{1}\) and \(M_{2}\) is more likely given some observed data \(d\) via the ratio of the probability of these models
\begin{equation}
    \frac{p(M_1|d)}{p(M_2|d)} = \frac{p(d|M_1)}{p(d|M_2)}\frac{p(M_1)}{p(M_2)} = B_{12} \, \frac{p(M_1)}{p(M_2)}\,,
\end{equation}
where \(B_{12}\) is the Bayes factor. 
It quantifies how much the relative odds between the two models have changed after taking into account the data. 
From the definition in \cref{eq:KL_div}, the KL divergence can be understood as the expected log-Bayes factor \cite{Fabre:2013wia}
\begin{equation}
    \DKL (M_1||M_2)=\left\langle \ln (B_{12})\right\rangle_{M_1}\,,
\end{equation}
where the Bayes factor is calculated under the assumption that the data follows the distribution \(p\) given by model \(M_{1}\). 
The Bayes factor can be interpreted using the (empirical) Jeffrey scale: for \(\ln B_{12} < 1\) the evidence is weak, while \(\ln B_{12} \gtrsim 2.5\) and  \(\ln B_{12} \gtrsim 5\) indicate moderate and strong evidence, respectively \cite{Trotta:2008qt}. 
\(\DKL \approx 1\) therefore corresponds to weak evidence for distinguishability, indicating where topological effects begin to imprint measurably on the CMB. 
For \(\DKL \ll 1\), the topologies are statistically indistinguishable, while for \(\DKL \gg 1\), they are clearly distinguishable. 
Actual detectability depends on several complicating factors such as cosmic variance, masking, foreground modeling, and instrumental noise.}

\subsection{Classification with machine learning}
\label{subsec:ML_method}
To classify CMB realizations from different topologies, we use the ML algorithm \catboost\footnote{The code is available at \url{https://github.com/catboost}.} \cite{catboost1, catboost2}, trained on the spherical harmonic coefficients \(\alm\) of the temperature anisotropies. 
\catboost\ is a gradient-boosting algorithm developed by \texttt{Yandex} that belongs to the family of gradient-boosted decision trees. 
Previously, in \rcite{COMPACT:2024dqe}, we found that gradient-boosting classifiers, e.g., \texttt{XGBoost}, outperformed random forests and neural networks on a similar task. 
The core idea of gradient boosting is to construct a strong classifier by iteratively adding weak learners, i.e., shallow decision trees.
At each iteration, a new tree is fitted to the negative gradient of the logarithmic loss function with respect to the current classifier's prediction, thereby minimizing the overall loss. This process is stage-wise, meaning that each tree is added sequentially to fix the remaining errors of the previous ensemble, and greedy, as the algorithm seeks the best possible improvement at the current step without revisiting earlier trees. The final classifier is a weighted sum of all trees, each contributing to correct the errors of its predecessors. 

Compared to other ML algorithms, such as multilayer-perceptrons (MLPs) and convolutional neural networks (CNNs), gradient-boosting algorithms tend to be less sensitive to training instabilities, overfitting and the order of the input features. 
\catboost\ requires minimal hyperparameter tuning (see Appendix \ref{app:hyperparams}), which makes it computationally efficient.
This is ideal since our goal is to assess whether informative features are present in the data rather than to optimize performance through highly specialized architectures. 
In addition, gradient-boosting models offer a high degree of interpretability through feature-importance analyses, which can help identify the spherical harmonic coefficients \(\alm\) that are most relevant for distinguishing different cosmic topologies.
\catboost\ includes \shap\ as a built-in tool for feature importance analysis (see Appendix \ref{app:shap}). Therefore, gradient boosting serves as an interpretable benchmark against which more complex deep-learning approaches can be compared.

Among available frameworks, we employ \catboost\ as it consistently matches or exceeds the accuracy of other gradient-boosting decision-tree methods while offering better generalization \cite{grad_boost}. 
In standard gradient boosting, each classifier is trained on the full dataset, including the true labels of all samples. 
This can lead to target leakage, where a classifier indirectly learns from the true label of a sample that it is supposed to predict. 
Gradually, the classifier can overfit, i.e.,  memorize the training data without learning features that generalize well to unseen data. 
\catboost\ avoids this using ``ordered boosting''---carefully controlling the information each classifier sees during training.
Instead of using all samples at once, \catboost\ shuffles the data and trains in a way that ensures that the prediction for any given sample is based only on samples that precede it in the shuffled order. 
This prevents the classifier from using the correct label of a data point to help predict its label. 
Ordered boosting  significantly reduces overfitting and improves the classifier's generalization to new, unseen data. 

Another key feature of \catboost\ is its use of symmetric (or balanced) trees as base predictors. 
In contrast to traditional gradient-boosting methods that construct trees either leaf-wise (e.g., \texttt{LightGBM}) or level-wise (e.g., \texttt{XGBoost}), \catboost\ grows trees symmetrically: all splits at a given depth use the same feature and threshold. Symmetric trees help regularize the learning process by limiting model complexity, thus promotes better generalization.

A central challenge in topology classification is the unknown orientation of the observer with respect to the topology's coordinate frame. 
While random rotations of the coordinate system leave the KL divergence unchanged (the KL divergence depends only on the eigenvalues of $(C_{\ell m\ell'm'}^{\E{18}})^{-1}C_{\ell m\ell'm'}^{\E{i}}$, which are rotationally invariant), they do alter the structure of the CMB covariance matrix. 
This results in the ML features being redistributed in harmonic space. It is computationally challenging to detect these redistributed features with ML, and performance can be affected significantly, especially for large manifolds \(L \gtrsim \LLSS\); see \rcite{COMPACT:2024dqe}. We will explore the sensitivity of ML algorithms to this effect in the future \cite{Tamosiunas2026pivb}. For now, we focus on a simulated dataset, where the observer's coordinate system is aligned with the symmetry axes of the fundamental domain.
In this unrotated case, the closest clone images are in the \(\pm\hat{x}\), \(\pm\hat{y}\), and \(\pm\hat{z}\) directions.

The goal is to train ML models capable of reliably classifying CMB realizations according to their underlying topology. The aim is for the classification to remain robust even in the presence of a non-standard primordial power spectrum. By testing the ML classifiers against simulated realizations with non-standard power spectra, we can determine whether the algorithm is genuinely learning topological features or if it is detecting deviations from the standard power-law. 
We can thereby ensure that ML-based topological probes are resilient to uncertainties in the primordial power spectrum.
A more ambitious goal is to use ML for parameter inference using the {\it Planck} data or other upcoming surveys by employing techniques such as likelihood-free and simulation-based inference.

\begin{figure}[t]
  \includegraphics[scale=1]{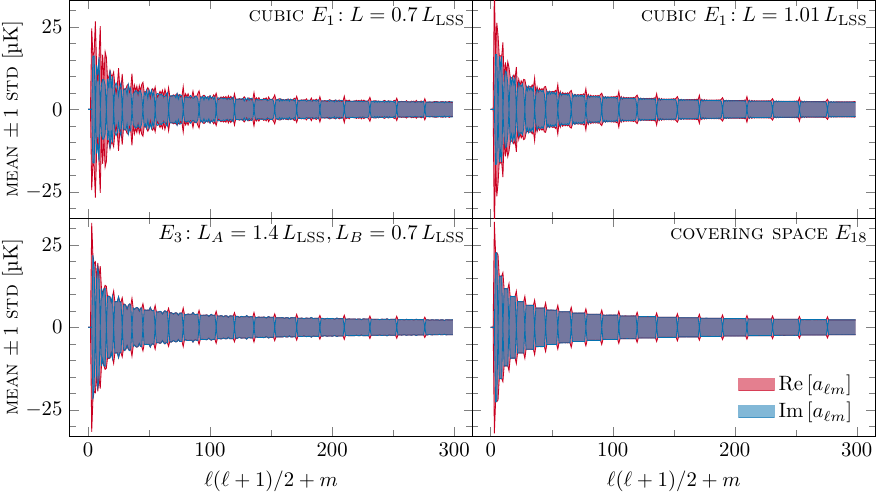}
  \caption{Mean and standard deviation of the real and imaginary parts of the first 300 \(\alm\) values in \((\ell, m)\) ordering from the full dataset of 200,000 realizations for selected classes. All shown realizations use the standard power-law primordial power spectrum.}
  \label{fig:CB_data}
\end{figure}

\para{Dataset and training.}
\catboost\ is trained on \(\alm\) realizations derived from the CMB temperature covariance matrix \eqref{eq:harm_cov} via Cholesky decomposition for \(\ell \in [2, 30]\). 
Cholesky decomposition factors the correlation matrix into a lower triangular matrix along with its conjugate \cite{COMPACT:2024dqe}.
Realizations of the temperature harmonic coefficients in the covering space are generated using the \texttt{synfast} function from the \texttt{healpy}\footnote{Available from \url{https://healpix.sourceforge.io/}.} package \cite{healpy, healpy2}. Each complex-valued realization is flattened into a real-valued input vector by separating the real and imaginary parts of the \(\alm\).\footnote{Following the \texttt{healpix} convention, we do not store negative $m$ values.}
For \(\ell_\text{max} = 30\), this results in $n=957$ independent features per sample; see \cref{eq:lmax}. 
While the ordering of these coefficients defines the structure of the feature vector, random forest implementations are generally invariant to feature permutations. Consistent with that, we find that using \((\ell, m)\) or \((m, \ell)\) ordering has no significant impact on training or test accuracy.
In the following, we use the \((\ell, m)\) ordering: \(\alm = [a_{00}, a_{10}, a_{11}, a_{20}, a_{21}, \dots, a_{\ell_\text{max}\ell_\text{max}}]\). 
 A specific entry in a particular data vector is denoted by \(\ell (\ell + 1) / 2 + m\). \cref{fig:CB_data} illustrates how the statistical properties of these coefficients vary across different classes. There are visible differences between the \(\alm\) of the small cubic \E{1} topology with \(L = 0.7 \, \LLSS\) and those of the covering space \E{18}. At low \(\ell\), for the non-trivial topology, the coefficients are overall lower and especially the imaginary components are smaller than the real ones. With increasing the size of the fundamental domain (top right), the \(\alm\) increasingly resemble those of the covering space. For a non-cubic fundamental domain (bottom left), the \(\alm\) are very similar to those of the covering space already at small side length \(L_{B} = 0.7 \, \LLSS\), where \(\LA = 1.4 \, \LLSS\) is fixed.

Each simulated CMB realization is labeled by the topology used to generate it (e.g., \E{1}, \E{2}, etc.) and the classifier is trained to predict this label. During training, \catboost\ minimizes the cross-entropy loss between the predicted class probabilities and the true labels. Once trained, the model outputs a posterior probability distribution over topologies for each new realization, allowing for both hard classification and probabilistic inference. Training and evaluation follow a supervised setup: for each class, 160,000 realizations are used for training and 40,000 for testing. The test data is completely unseen during training and thus provides a reliable measure of generalization performance. Training accuracy is useful during optimization to detect overfitting, while test accuracy reflects the actual effectiveness of the classifier on unseen data, i.e., how well the algorithm has learned features of the different topology classes. 

We start with binary classification, where the goal is to distinguish a specific manifold of a non-trivial topology, e.g., the cubic 3-torus \E{1} with size \(L = 1.01 \, \LLSS\), from the trivial Euclidean space \E{18}. In this case, \catboost\ minimizes the standard binary cross-entropy loss (see Appendix \ref{app:ml_Kldiv} for details)
\begin{equation}
    \mathcal{L}_\text{binary} = -\sum_{i = 1}^{N} [y_{i}\log p_{i} + (1 - y_{i}) \log(1 - p_{i})]\,,
    \label{eq:binary_loss}
\end{equation}
where \(N\) is the total number of samples in the training dataset, \(y_{i} \in \{0, 1\}\) is the true label of a sample \(i\), and \(p_{i}\) is the classifier's predicted probability for class 1. By training separate binary classifiers for various configurations, we evaluate how classification performance varies with topology, fundamental domain size, observer position, and different modifications to the primordial power spectrum.

In a second stage, we extend the analysis to multi-class classification, where the goal is to identify the topology of a given realization from among several candidates. This reflects a more realistic scenario in which the true topology is unknown and must be inferred from the data. In one case, we train the algorithm to also distinguish between realizations that have the standard or a modified primordial power spectrum. For multi-class classification, \catboost\ minimizes the softmax cross-entropy loss
\begin{equation}
    \mathcal{L}_{\text{multiclass}} = -\sum_{i = 1}^{N} \sum_{c = 1}^{C} y_{ic} \log\!\left(p_{ic}\right)\,,
\end{equation}
where \(y_{ic} \in \{0, 1\}\) denotes whether sample \(i\) belongs to class \(c\), and \(p_{ic}\) is the classifier's predicted probability for that class.

\section{Results and discussion}
\label{subsecn:numerical_results}

We quantify differences between the CMB temperature covariance matrices of the covering space \E{18} and those of representative manifolds \E{1}--\E{6} using the KL divergence. This analysis is carried out for both standard and modified primordial power spectra, enabling a systematic assessment of how such modifications influence the detectability of topological signatures. We then classify simulated realizations of the representative manifolds---again for both standard and modified power spectra---using the \catboost\ ML algorithm.

\subsection{KL divergence: on- and off-axis observer positions}
We now present our computed KL divergence for the fully compact, orientable topologies \E{1}--\E{6} as a function of the fundamental domain size. To ensure comparability with our previous results \cite{COMPACT:2023rkp}, we analyze the manifolds of \E{1}--\E{5} by fixing the dimension \(\LA = 1.4 \, \LLSS\) and varying \(L_{B}\);\footnote{$L_A$ and $L_B$ denote the translation lengths of the group generators: $L_B$ is the displacement of the corkscrew motion along the rotation axis (or of the simple translation for \E{1}), while $L_A$ is the length of the two translations spanning the orthogonal plane.} for \E{1}, this translates to \(L_{1} = L_{2} = 1.4 \, \LLSS\) and varying \(L_{3}\), and for \E{2}, we have \(\LA = L_{A_{1}} = L_{A_{2}}\). For the \E{6} topology, \(L = L_{A_{x}} = L_{B_{y}} = L_{C_{z}}\) is varied while \(r_{x} = r_{y} = r_{z} = 1/2\) are kept fixed.\footnote{For \E{6}, $L$ represents the translation lengths of the generators along each principal axis, while $r_i=1/2$ for the standard form as defined in \rcite{COMPACT:2023rkp}.}. We also include the standard cubic 3-torus \E{1} with \(L_{1} = L_{2} = L_{3}\)

The \E{1} topology is homogeneous, so the observer position does not affect observables. Here, we consider an untilted configuration for \E{1}. In contrast, the other topologies \E{2}--\E{6} are inhomogeneous. For each of these, our results are shown for two observer positions: an on-axis observer, where the rotation axis passes through the observer, at \(\xobs = (0, 0, 0)\), and an off-axis observer, where the axis does not pass through the observer but through \(\xobs \neq (0, 0, 0)\).
For these cases, the tilt parameters are set to zero as they are degenerate with the off-axis position of the observer. Off-axis positions where the nearest clone is closer than \(\LLSS\) would result in the observation of pairs of circles on the CMB sky around which the pattern of temperature fluctuations would ``match.''
Such circle pairs have been searched for in the Wilkinson Microwave Anisotropy Probe (WMAP) and {\it Planck} data and no statistically significant matched pairs have been identified; see \rcite{Cornish:1997ab, Cornish:2003db, ShapiroKey:2006hm, Vaudrevange:2012da, Planck:2013okc, Planck:2015gmu,COMPACT:2022nsu,COMPACT:2024qni}. 
These cases are therefore excluded.
Instead, the analyzed off-axis positions are situated midway between two excluded regions for \(L_{B} < \LLSS\), maximizing the number of nearest clones at equal distance from the observer. 
Such equidistant clones lead to coherent signals in the CMB sky due to the same physical fluctuations appearing in multiple directions, enhancing correlations across large angular separations. 
This contributes to the off-diagonal terms in the covariance matrix, increasing the KL divergence. 
For example, for the \E{2} topology with \(\LA = L_{A_{1}} = L_{A_{2}} = 1.4 \, \LLSS\) and an off-axis observer at \(-\xobs = (-0.35, 0, 0)\), there are four nearest clones at equal distance from the observer, located at 
\begin{align}
    \xvec &= \left((1.4 - 0.35)\LLSS,\, 0,\, \LB\right)^T, \quad & \xvec &= \left(-0.35 \, \LLSS,\, 0,\, \LB\right)^T, \\ \nonumber
    \xvec &= \left((1.4 - 0.35)\LLSS,\, 0,\,-\LB\right)^T, \quad & \xvec &= \left(-0.35 \, \LLSS,\, 0,\,-\LB\right)^T.
\end{align}
Each of these is at distance \(\left((0.7 \, \LLSS)^{2} + L_{B}^{2}\right)^{1/2}\) from the observer. Setting this equal to \(\LLSS\) defines the critical size, below which matched circles appear,
\begin{equation}
    L_{B} = \Lcircle = \sqrt{1 - 0.7^{2}} \, \LLSS \approx 0.71 \, \LLSS\,.
\end{equation}
For comparison, an on-axis observer at \(\xobs = (0, 0, 0)\) has only two nearest clones at \(\xvec = (0, 0, \pm \LB)^{T}\), both at distance \(L_{B}\), with critical size at \(L_{B} = \Lcircle = \LLSS\). Fewer equidistant clones lead to smaller topological correlations in the covariance matrix and therefore a lower KL divergence \cite{COMPACT:2023rkp}.

\begin{figure}[tb]
    \centering
    \includegraphics[scale=1]{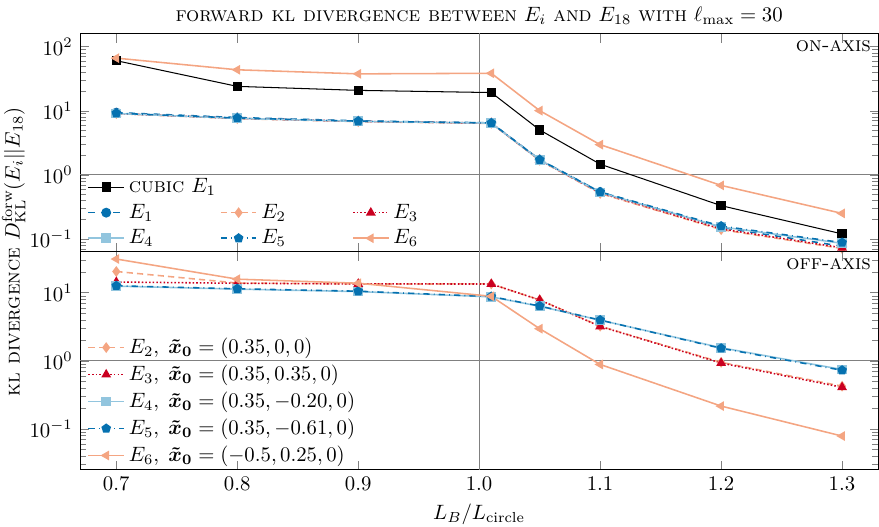}
    \caption{KL divergence for topologies \E{1}--\E{6}
    with an on-axis observer \(\xobs = (0, 0, 0)\) (upper panel) or an off-axis observer \(\xobs \neq (0, 0, 0)\) (lower panel). For the cubic \E{1} manifold, the \(x\)-axis indicates the length of all three sides. For the non-cubic manifolds, $L_B$ varies, while the length of the other sides is fixed at \(\LA = 1.4 \, \LLSS\) (for \E{1}, \(L_{1} = L_{2} = 1.4 \, \LLSS\) remain fixed while the varying \(L_{3}\) is shown on the \(x\)-axis). The \E{6} manifold has a single length scale \(L = L_{A_{x}} = L_{B_{y}} = L_{C_{z}}\), which is indicated by the $x$-axis, while \(r_{x} = r_{y} = r_{z} = 1/2\) are kept fixed.}
    \label{fig:KL_all_on_off_ax}
\end{figure}

Figure \ref{fig:KL_all_on_off_ax} displays the forward KL divergence \(\DKL (\E{i}||\E{18})\), comparing each non-trivial topology \E{i} to the trivial topology \E{18} as a function of the fundamental domain size. The horizontal dashed line \(\DKL = 1\) marks an approximate threshold for detectability, while the vertical dashed line denotes the critical size \(L = \Lcircle\), i.e., the maximum size for which pairs of matched circles in the sky can be detected. For an on-axis observer, this corresponds to the diameter of the last scattering surface \(L = \Lcircle = \LLSS\). 
As the size of the fundamental domain exceeds \(\Lcircle\), the topological imprint on the CMB becomes increasingly faint, and the KL divergence eventually drops below detectability \(\DKL < 1\). For fixed \(\LA = 1.4 \, \LLSS\), the KL divergence behaves very similarly across topologies \E{1}--\E{5}, both qualitatively and quantitatively.\footnote{Minor differences exist and increase with increasing \(\LB\), which might be related to the multipole-dependent cutoff \(|\kvec_{\text{max}} (\ell)|\) and its interplay with the distinct mode structures of each topology.} These nearly identical KL divergences indicate that for large \(\LA\), the distinguishability of non-trivial topologies from the trivial one is determined by the smaller dimension \(\LB\), regardless of the specific topology. For off-axis observers, \E{2} and \E{3} behave the same, as do \E{4} and \E{5}. This behavior of the KL divergence persists when using a non-standard primordial power spectrum, either in the non-trivial topologies \E{1}--\E{6} alone or in both the non-trivial and trivial ones. The complete set of KL divergence plots are included in Appendix \ref{app:other_conf}; here, we focus on representative cases.

\subsection{KL divergence: impact of modified power spectrum}
\begin{figure}[tb]
\centering
    \includegraphics[scale=1]{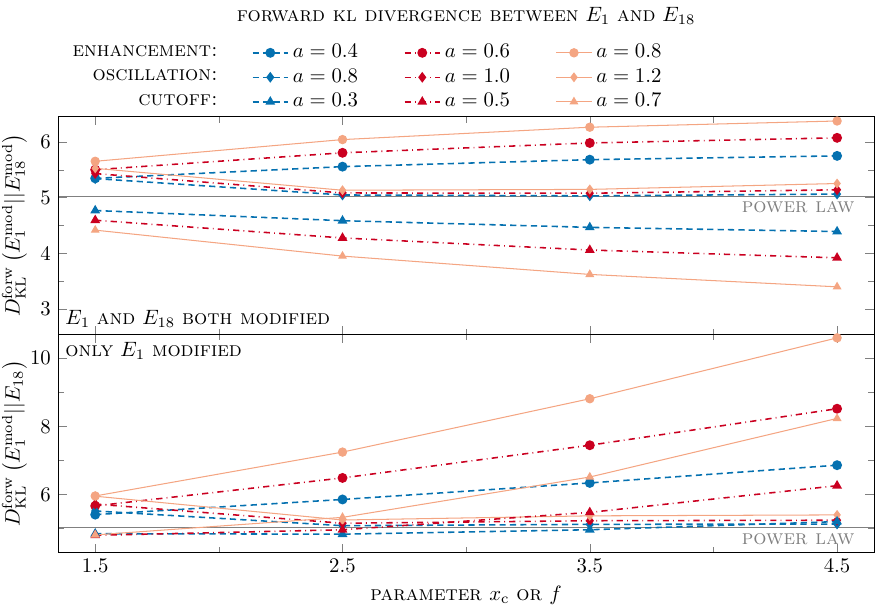}
    \caption{Influence of different parameters for the power spectrum modifications on the KL divergence for a cubic \E{1} manifold of size \(L = 1.05 \, \LLSS\). The parameters are those described in \cref{eq:modifications}. For the oscillation modification \(x_c = 2\) and the \(x\)-axis shows different frequencies.}
    \label{fig:KL_mod_params}
\end{figure}

As noted in \cref{subsec:mods}, we distinguish two possibilities for the origin of a power spectrum modification: it is either related or unrelated to the topology of space. 
Figure \ref{fig:KL_mod_params} shows the effect of the three types of modifications (enhancement, cutoff, and oscillation) on the KL divergence for different values of the parameters in \cref{eq:modifications}. 
The upper panel displays the KL divergence for modifications that are unrelated to the compactness of space, i.e., modifications that are present in \E{1} and \E{18}. 
In contrast, the lower panel shows the KL divergence for modifications that are related to the compactness of space, i.e., only the non-trivial topology has a modified primordial power spectrum. 
The KL divergence has been calculated between the cubic \E{1} topology with \(L = 1.05 \, \LLSS\) and the trivial topology \E{18}. 
Such a cubic \E{1} manifold with \(L > \LLSS\) is compatible with current observational constraints. 
For this \E{1} manifold, the forward KL divergence is \(\DKL (\E{1}||\E{18}) = 5.0\) for the standard power-law primordial power spectrum.

A power spectrum modification that is related to the non-trivial topology, i.e., a modification only in \E{1}, results in a relatively higher KL divergence compared to the standard power-law case. The higher the amplitude \(a\) and the wavenumber cutoff \(x_c =  k_c L /  (2 \pi)\), the higher the resulting KL divergence. Here, the low-\(k\) enhancement leads to the highest KL divergence. In the case of the oscillatory modification, \(x_c = 2\) is fixed, and the \(x\)-axis shows different frequencies; for \(f = 1.5\) the increase of the KL divergence is the strongest.

On the other hand, modifications unrelated to the topology of space have a different effect on the KL divergence: in this case, a low-\(k\) cutoff (in the power spectrum of both the \E{1} and \E{18} topologies being compared) causes a decrease in the KL divergence compared to the standard power-law case. An enhancement, i.e., an increase in power at low \(k\) has the contrary effect---it results in a higher KL divergence. Both effects are stronger: the amplitude \(a\) and the wavenumber cutoff \(x_c\) are both larger. The increase in the KL divergence for the oscillation modification is most significant for \(f = 1.5\).

In the following, we apply modifications with the maximal parameters that are within the \(3\sigma\) bound of the \textit{Planck} posterior (see \cref{eq:modifications}) to different-size manifolds of \E{1}--\E{6}.
We thereby analyze how the effect of modifications on the KL divergence varies depending on the topology and the fundamental domain size. 

\begin{figure}[t]
    \includegraphics[scale=1]{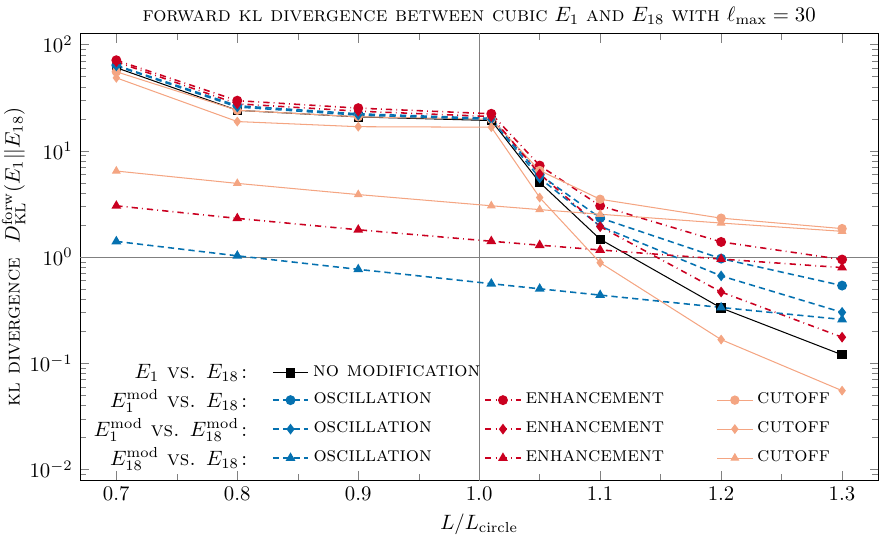}
    \caption{KL divergence for a cubic \E{1} topology as a function of \(L = L_{1} = L_{2} = L_{3}\) in units of the diameter of the last scattering surface \(\Lcircle = \LLSS\). Primordial power spectrum modifications are parametrized in terms of the dimensionless wavenumber \(x = k \Lmod / (2 \pi)\); therefore, the KL divergence depends on \(\Lmod\) even for the trivial topology \E{18}, where $\Lmod=L$ for the cubic \E{1} topology.}
    \label{fig:KL_E1}
\end{figure}

Figure \ref{fig:KL_E1} shows the KL divergence \(\DKL (\E{1}||\E{18})\) between the cubic \E{1} topology and the trivial topology \E{18} as a function of the size of the fundamental domain \(L = L_{1} = L_{2} = L_{3}\). The black solid line corresponds to the standard power-law primordial power spectrum, previously analyzed in Ref. \cite{COMPACT:2023rkp}, where the KL divergence drops below the detectability threshold (\(\DKL = 1\)) at \(L \approx 1.1 \, \LLSS\). For the modifications to the primordial power spectrum of the non-trivial topology only, the resulting KL divergences are shown in colored lines with circular markers for the three types of modifications: a low-\(k\) cutoff, oscillations, and a large-scale enhancement, with parametrization and parameters as stated in \cref{eq:modifications}. Because these modifications are parametrized in terms of the dimensionless wavenumber \(x = k \Lmod / (2 \pi)\), even for the trivial topology \E{18}, the KL divergence depends on \(\Lmod\).\footnote{For the non-cubic cases, \(\Lmod\) is always the smallest side length \(\LB\).} To isolate the effect of the modified primordial power spectrum alone, maintaining the trivial topology, we also compute \(\DKL (\E{18}^\text{mod}||\E{18})\), shown as colored lines with triangular markers. As \(L \to 1.2 \, \LLSS\), the KL divergence \(\DKL  (\E{1}^\text{mod}||\E{18})\) for the cutoff and enhancement cases converges toward \(\DKL (\E{18}^\text{mod}||\E{18})\), indicating that, for large topologies, most of the observed KL divergence arises from differences in the primordial power spectrum rather than from topological differences. 
For the oscillatory modification, \(\DKL (\E{18}^\text{mod}||\E{18})\) is very low at \(L \gtrsim 1.1 \, \LLSS\) and \(\DKL (\E{1}^\text{mod}||\E{18})\) does not approach it as closely as for the other two modifications. In this case, the difference between \(\DKL (\E{1}^\text{mod}||\E{18})\) and the standard, unmodified \(\DKL (\E{1}||\E{18})\) for large fundamental domains cannot be clearly attributed to differences in the primordial power spectrum.

\begin{figure}[t]
    \includegraphics[scale=1]{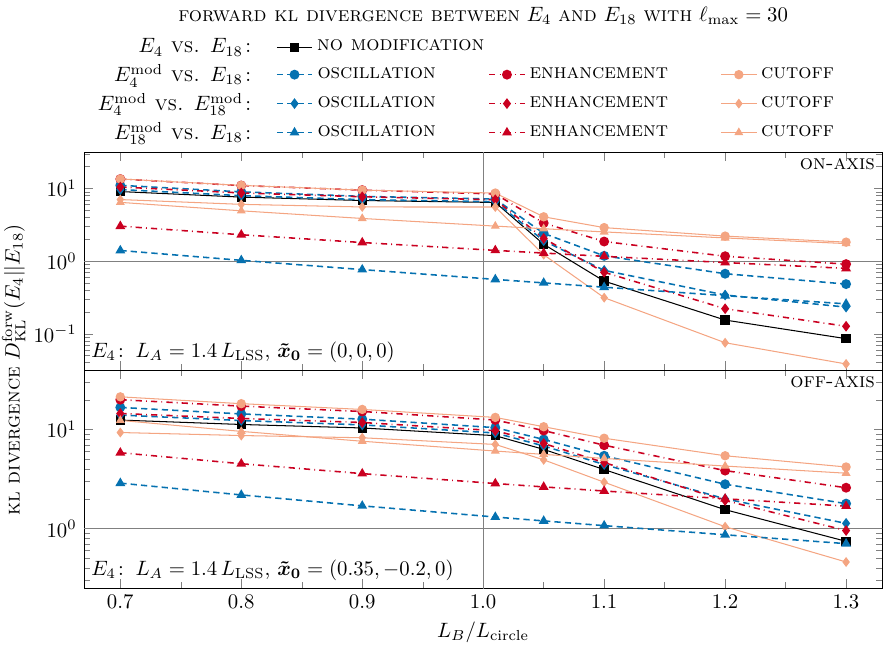}
    \caption{Upper panel: KL divergence for the \E{4} topology with \(\LA = 1.4 \, \LLSS\) for an on-axis observer with circle limit \(\LB = L=\Lcircle = \LLSS\). Lower panel: KL divergence for the \E{4} topology with \(\LA = 1.4 \, \LLSS\). The circle limit for an off-axis observer is \(\LB = L= \Lcircle = \sqrt{1-0.7^2} \, \LLSS \approx 0.71 \, \LLSS\). Primordial power spectrum modifications are parametrized in terms of the dimensionless wavenumber \(x = k \Lmod / (2 \pi)\); therefore, the KL divergence depends on \(\Lmod\) even for the trivial topology \E{18}.}
    \label{fig:KL_E4}
\end{figure}

In the second scenario, the non-trivial and trivial topologies are assigned the same modified primordial power spectrum. The corresponding KL divergences \(\DKL  (\E{1}^\text{mod}||\E{18}^\text{mod})\) are shown as colored lines with diamond-shaped markers. Notably, the relative ordering of these KL divergences across modifications is inverted compared to the previously discussed \(\DKL (\E{1}^\text{mod}||\E{18})\). For the low-\(k\) cutoff, the KL divergence \(\DKL (\E{1}^\text{mod}||\E{18}^\text{mod})\) is significantly lower than in the standard power-law case \(\DKL (\E{1}||\E{18})\) across all values of \(L\). This highlights that the low-\(k\) modes contain crucial topological information, and therefore, suppressing them diminishes the topological signature in the correlation matrix. This is further supported as, for the large-scale enhancement, the KL divergence \(\DKL (\E{1}^\text{mod}||\E{18}^\text{mod})\) is increased compared to the standard \(\DKL (\E{1}||\E{18})\) across all values of \(L\). Thus, amplifying the large-scale power reinforces the topological signal. The oscillatory case behaves more subtly: \(\DKL (\E{1}^\text{mod}||\E{18}^\text{mod})\) closely tracks the standard KL divergence \(\DKL (\E{1}||\E{18})\) up to \(L = 1.05 \, \LLSS\). Then, it decays more slowly than in the standard or enhancement case. The oscillation corresponds to an enhancement around \(x = 1\), followed by an attenuation around \(x = 3\), and so on, until it follows the standard power spectrum for \(x > 10\). This more complex shape makes it difficult to trace a clear correspondence with KL divergence behavior.

\begin{figure}[t]
    \includegraphics[scale=1]{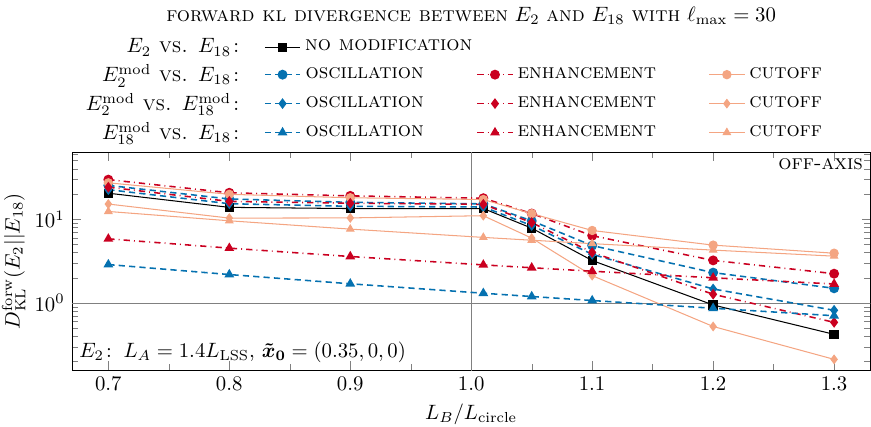}
    \caption{KL divergence for the \E{2} topology with \(\LA = 1.4 \LLSS\). The circle limit for an off-axis observer is \(L_{B} = \Lcircle = \sqrt{1-0.7^{2}} \LLSS \approx 0.71 \LLSS\). Primordial power spectrum modifications are parametrized in terms of the dimensionless wavenumber \(x = k \Lmod / (2 \pi)\); therefore, the KL divergence depends on \(\Lmod\) even for the trivial topology \E{18}.}
    \label{fig:KL_E2_oa}
\end{figure}

The top panel of \cref{fig:KL_E4} displays the KL divergence between the \E{4} topology with fixed side length \(\LA = 1.4 \, \LLSS\) and varying \(\LB\) and the trivial \E{18} topology. The qualitative impact of power spectrum modifications is similar to the cubic \E{1} case, with two notable differences: overall, the KL divergence is lower for the non-cubic \E{4} than for the cubic \E{1}. Moreover, for the low-\(k\) cutoff or enhancement applied only to the non-trivial topology, the KL divergence \(\DKL (\E{4}^\text{mod}||\E{18})\) is notably higher than the standard \(\DKL (\E{4}||\E{18})\) across all \(L\), not just for large values. As shown previously in \cref{fig:KL_all_on_off_ax}, the KL divergence is the same for all the manifolds of \E{1}--\E{5} topologies with the same side lengths and an on-axis observer.

The bottom panels of \cref{fig:KL_E4,fig:KL_E2_oa} show similar results for the manifolds of \E{4} and \E{2} with an off-axis observer. For an off-axis observer, the KL divergence behaves the same for the \E{2} and \E{3} topologies and for the \E{4} and \E{5} topologies, as shown in \cref{fig:KL_all_on_off_ax}. For an off-axis observer, the circle limit \(\Lcircle\), i.e., the smallest \(\LB\) for which no pairs of identical circles appear in the CMB, is reduced to \(\Lcircle \approx 0.71 \, \LLSS\). In the standard power-law case, the KL divergence falls below the detection threshold (\(\DKL = 1\)) at \(L \gtrsim 1.25 \, \Lcircle\), later than for the on-axis observer, in terms of \(\Lcircle\). Expressing the results in terms of \(L / \Lcircle\) facilitates comparison with observational circle search limits and provides a natural detectability scale. Inhomogeneous topologies exhibit a strong dependence on the observer location: the KL divergence varies significantly with observer position. However, the direction of change (increase or decrease of the KL divergence) induced by power spectrum modifications remains consistent across observer positions. For the \E{4} topology with an off-axis observer, the KL divergence \(\DKL (\E{4}^\text{mod}||\E{18})\) is higher than the standard \(\DKL (\E{4}||\E{18})\), and the difference is greater than for an on-axis observer. Very similar trends can be observed for the \E{2} topology with an off-axis observer.

\begin{figure}[t]
    \includegraphics[scale=1]{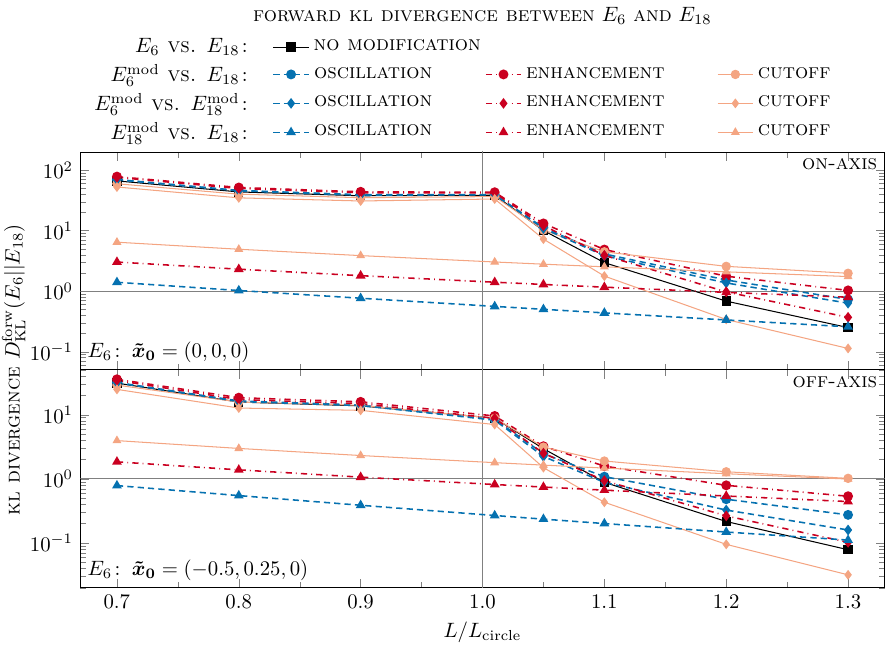}
    \caption{Upper panel: KL divergence for the \E{6} topology with \(L = L_{A_{x}} = L_{B_{y}} = L_{C_{z}}\) and \(r_{x} = r_{y} = r_{z} = 1 / 2\). The circle limit for an on-axis observer is \(\Lcircle = \LLSS / \sqrt{2} \approx 0.71 \, \LLSS\). Lower panel: KL divergence for the \E{6} topology with \(L = L_{A_{x}} = L_{B_{y}} = L_{C_{z}}\) and \(r_{x} = r_{y} = r_{z} = 1 / 2\). The circle limit for an off-axis observer is \(\Lcircle \approx 0.9 \, \LLSS\). Primordial power spectrum modifications are parametrized in terms of the dimensionless wavenumber \(x = k \Lmod / (2 \pi)\); therefore, the KL divergence depends on \(\Lmod\) even for the trivial topology \E{18}.}
  \label{fig:KL_E6}
\end{figure}

The \E{6} topology is structurally more complex. 
For equal side lengths \(L = L_{A_{x}} = L_{B_{y}} = L_{C_{z}}\) and parameters \(r_{x} = r_{y} = r_{z} = 1 / 2\), the on-axis observer at \(\xobs = (0, 0, 0)\) has their nearest clone at a distance \(\sqrt{2}L\), which sets the circle limit at \(L = \Lcircle = \LLSS / \sqrt{2} \approx 0.71 \, \LLSS\). 
For the off-axis position \(\xobs = (-0.5, 0.25, 0)\), which has been numerically found in \rcite{COMPACT:2023rkp}, the distance to the nearest clone is \(\sqrt{L^{2} + (0.5 \, \LLSS - L)^{2}}\), which yields the circle limit \(L = \Lcircle \approx 0.91 \, \LLSS\). 
Figure \ref{fig:KL_E6} shows that the KL divergence falls below unity later for the on-axis observer than for the off-axis one, in terms of \(L / \Lcircle\), suggesting that this off-axis position is suboptimal. Notably, for the on-axis observer, the KL divergence \(\DKL (\E{6}^\text{mod}||\E{18})\) for the cutoff modification is very similar to that of the standard power-law case for \(L \leq 1.05 \, \LLSS\)---this is different from the behavior observed for other topologies. All other observed increases and decreases in the KL divergence are similar to those seen for the previously considered topologies.

\subsection{Machine-learning classification with \catboost}
While the KL divergence provides a theoretical estimate of the information available in order to distinguish between different cases, it does not prescribe a functional classification mechanism. We employ the \catboost\ algorithm specified in \cref{subsec:ML_method} to classify simulated realizations of different topologies. This allows us to move beyond theoretical distinguishability and evaluate the practical robustness of our classifiers.

\begin{figure}[tb]
\centering
    \includegraphics[scale=1]{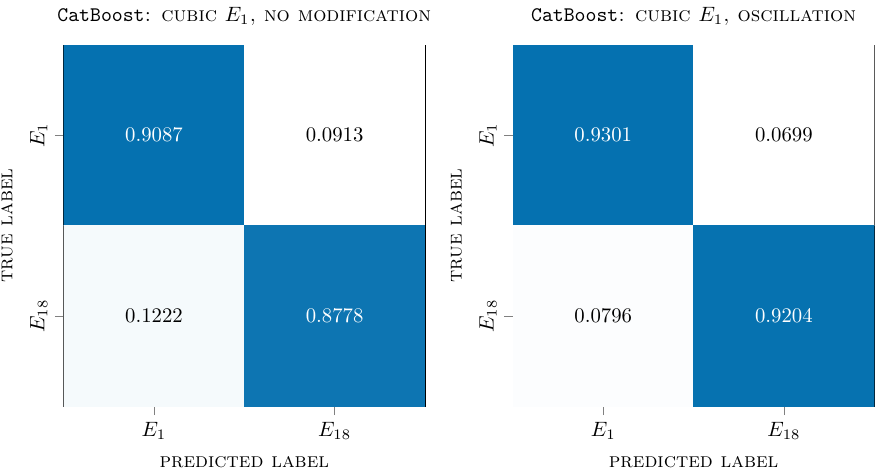}
    \caption{Normalized confusion matrices for a binary classifier distinguishing unrotated realizations of the cubic \E{1} of size \(L = L_{1} = L_{2} = L_{3} = 1.05 \, \LLSS\) from the trivial \E{18} topology. Left panel: unmodified power spectrum. Right panel: \E{1} realizations with an oscillation modification applied to the primordial power spectrum.}
    \label{fig:CM_CB_E1}
\end{figure}

Figure \ref{fig:CM_CB_E1} displays the normalized confusion matrix for the \catboost\ classifier trained to distinguish between the trivial topology \E{18} and a manifold of the non-trivial, cubic \E{1} topology of size \(L = 1.05 \, \LLSS\), i.e., larger than the limit for detection of matched circle pairs. 
The left panel corresponds to a classifier trained on realizations with the unmodified primordial power spectrum, while the right panel shows results from training on \E{1} realizations with an oscillation modification applied to their primordial power spectrum. The rows correspond to the true class labels and sum to one, while the columns denote predicted labels. 
Diagonal entries represent correct classifications, whereas off-diagonal terms indicate misclassifications. In both cases, the classifier achieves excellent test accuracy, with a slightly better performance when the modified primordial power spectrum is used.

Averaging over the class-specific test accuracies yields the overall test accuracy. For instance, for the left confusion matrix in \cref{fig:CM_CB_E1} (without modifications), the test accuracy is \(\hatp = 0.8933\), and with the oscillation modification (right confusion matrix), it is \(\hatp = 0.9253\). Since the confusion matrices are computed using the full test data set, the purely statistical uncertainty from the finite test size can be estimated under a binomial model as
\begin{equation}
    \sigma_{\hatp} = \sqrt{\frac{\hatp(1 - \hatp)}{N_\mathrm{test}}}\,, \quad \text{with} \quad N_{\mathrm{test}} = 80.000\,,
    \label{eq:ML_uncertainty}
\end{equation}
which is very small in practice---for this example, \(\hatp = 0.8933 \pm 0.0011\) for the unmodified power spectrum and \(\hatp = 0.9253 \pm 0.0009\) for the oscillation modification. However, this binomial error does not capture the variability between different realizations of the \(\alm\), i.e., the effect of cosmic variance. To obtain a more realistic error estimate, we split the test data set into \(K = 20\) disjoint subsets and compute the accuracy \(a_{i}\) separately on each subset. 
We then calculate the mean accuracy \(\bar{a}\) and the standard deviation across the subsets,
\begin{equation}
    s = \sqrt{\frac{1}{K-1} \sum_{i} (a_{i} - \bar{a})^{2}}\,,
\end{equation}
which reflects both finite-sample fluctuations and variations across realizations. 
This standard deviation, therefore, provides a more conservative and meaningful estimate of the uncertainty on the reported accuracies. For the example (confusion matrices in \cref{fig:CM_CB_E1}), this yields \(\bar{a} = 0.8933 \pm 0.0054\) for the unmodified power spectrum and \(\bar{a} = 0.9253 \pm 0.0052\) for the oscillation modification.

\begin{figure}[t]
    \centering
    \includegraphics[scale=1]{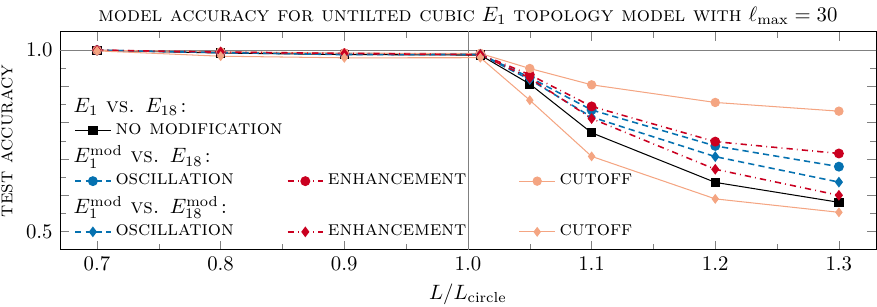}
    \caption{Test accuracy for the cubic \E{1} topology as a function of \(L = L_{1} = L_{2} = L_{3}\) in units of the diameter of the last scattering surface \(\LLSS = \Lcircle\).}
    \label{fig:Acc_E1}
\end{figure}

To assess the impact of topology size and power spectrum modifications on classification accuracy more broadly, many ML classifiers were trained across a range of fundamental domain sizes and primordial power spectrum alterations. Figure \ref{fig:Acc_E1} shows the test accuracy as a function of the fundamental domain size \(L\) for the cubic \E{1} topology. As expected, the test accuracy decreases with increasing \(L\) for all configurations. While the test accuracy remains above \(98\%\) for \(L \leq 1.01 \, \LLSS\), it drops sharply beyond \(L \geq 1.05 \, \LLSS\). This mirrors the behavior observed in the KL divergence in Figs. \ref{fig:KL_all_on_off_ax} to \ref{fig:KL_E6} and corresponds to the disappearance of matched circle pairs in the CMB once \(L > \Lcircle = \LLSS\), thereby reducing topological information in the CMB. The same set of primordial power spectrum modifications used for the KL divergence analysis (see \cref{eq:modifications}) is applied here. The colored lines with a circular marker correspond to modifications applied only to \E{1}, while the lines with a diamond-shaped marker indicate the same modification applied to both \E{1} and \E{18}. The ordering of the curves closely matches that of the KL divergences in \cref{fig:KL_E1}, confirming that modifications specific to the non-trivial topology enhance its detectability. If the modification to the primordial power spectrum is unrelated to the non-trivial topology, i.e., a modification in both \E{18} and \E{1}, then it affects detectability in a shape-dependent manner: suppression of power on large scales reduces both the KL divergence and classification accuracy (compared to the baseline of unmodified power spectra), thereby hindering detection. This suppression effect is more pronounced for \(L > \Lcircle\). Conversely, enhancements of large-scale power can improve detectability, though only significantly once \(L > \Lcircle\).

\begin{figure}[t]
    \centering
    \includegraphics[scale=1]{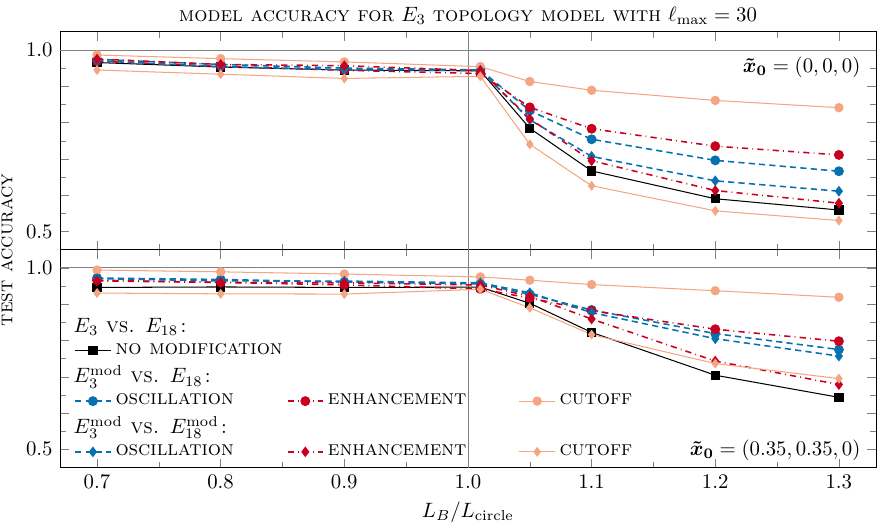}
    \caption{Test accuracy for the \E{3} topology as a function of \(\LB\) in units of \(\Lcircle\), the largest size for which pairs of matched circles can be detected in the sky. \(\LA = 1.4 \, \LLSS\) is fixed. Upper panel: on-axis observer, \(\Lcircle = \LLSS\). Lower panel: off-axis observer, \(\Lcircle = \sqrt{1 - 0.7^{2}} \, \LLSS \approx 0.71 \, \LLSS\).}
    \label{fig:Acc_E3}
\end{figure}

\begin{figure}[t]
    \centering
    \includegraphics[scale=1]{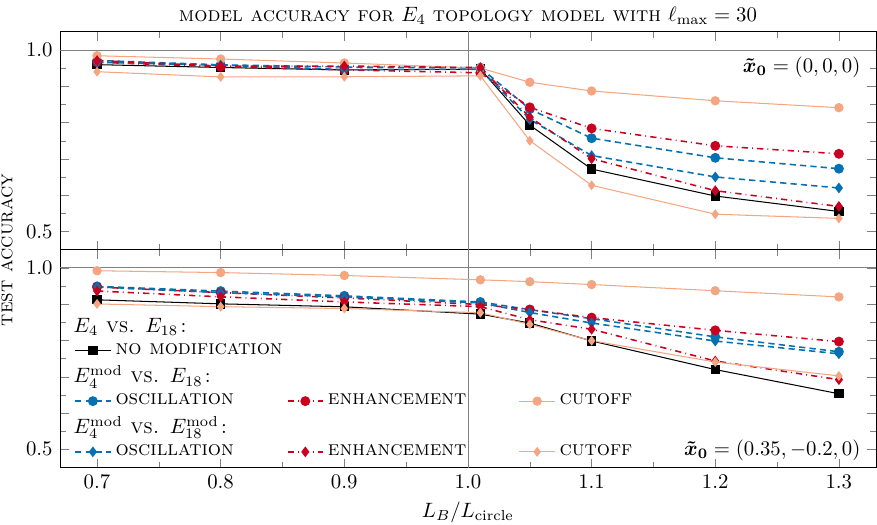}
    \caption{Test accuracy for the \E{4} topology as a function of \(\LB\) in units of \(\Lcircle\). \(\LA = 1.4 \, \LLSS\) is fixed. Upper panel: on-axis observer, \(\Lcircle = \LLSS\). Lower panel: off-axis observer, \(\Lcircle = \sqrt{1 - 0.7^{2}} \, \LLSS \approx 0.71 \, \LLSS\).}
    \label{fig:Acc_E4}
\end{figure}

\begin{figure}[h!]
    \centering
    \includegraphics[scale=1]{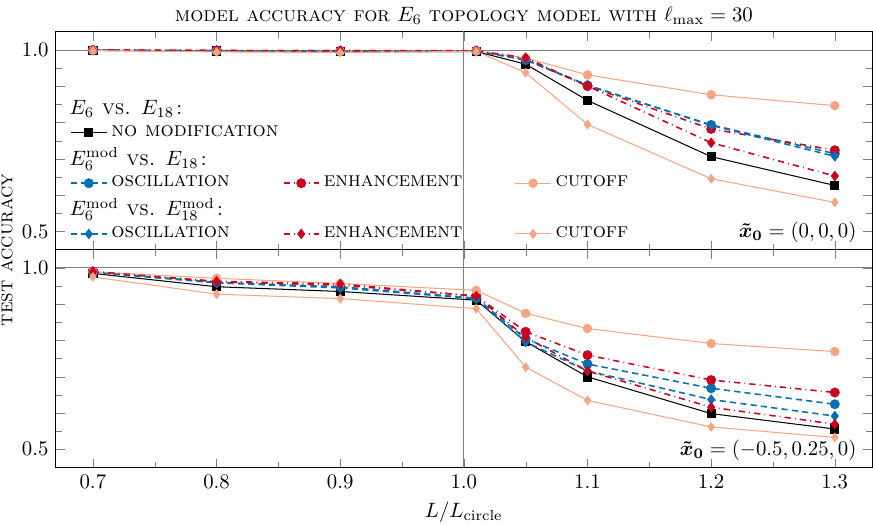}
    \caption{Test accuracy for the \E{6} topology as a function of \(L = L_{A_{x}} = L_{B_{y}} = L_{C_{z}}\) in units of \(\Lcircle\). Upper panel: on-axis observer, \(\Lcircle = \LLSS / \sqrt{2} \approx 0.71 \, \LLSS\); Lower panel: off-axis observer, \(\Lcircle \approx 0.9 \, \LLSS\).}
    \label{fig:Acc_E6}
\end{figure}

Figures \ref{fig:Acc_E3} to \ref{fig:Acc_E6} present the test accuracies for the \E{3}, \E{4}, and \E{6} topologies, for both on-axis and off-axis observers. These cases were selected due to their distinct KL divergence behavior: for on-axis observers, the KL divergence is identical for the non-cubic \E{1}--\E{5}, while for off-axis observers, \E{2} and \E{3} share the same KL divergence, as do \E{4} and \E{5} (see \cref{fig:KL_all_on_off_ax}). To further test the assumption that classification accuracy mirrors the KL divergence, we analyze both \E{3} and \E{4} for on-axis observers. As expected, their test accuracies are nearly identical. For an on-axis observer and \(L < \Lcircle\), the non-cubic \E{3} and \E{4} manifolds yield lower test accuracies than the cubic \E{1} manifold and the \E{6} manifold with a single length scale. 
This matches the trend in the KL divergence, which is lower for the non-cubic configurations with fixed side length \(\LA = 1.4 \, \LLSS\) than for the said \E{1} and \E{6} configurations. Another consistent feature is that for an on-axis observer in \E{6}, the test accuracy (and the KL divergence) associated with an enhancement modification to the primordial power spectrum of \E{6} only, surpasses that of the oscillation modification at large \(L\). Moreover, the classifier's performance for a low-\(k\) cutoff, applied to both \E{6} and \E{18}, exceeds the baseline power-law case at \(L \geq 1.2 \, \LLSS\), which has not been observed in the KL divergence.

Overall, the binary classification results reflect those of the KL divergence analysis closely. The empirical, simulation-based test accuracies and the theoretical KL divergence are in strong agreement. This supports the conclusion that topology-specific modifications to the primordial power spectrum significantly improve detectability, and that topology-unrelated modifications affect detectability in predictable ways, quantifiable via both KL divergence and ML techniques such as \catboost. For further analysis of the detected features and their importance to the classification, as assessed by SHAP analysis, see Appendix~\ref{app:shap}.

\begin{figure}[tb]
    \centering
    \includegraphics[scale=1]{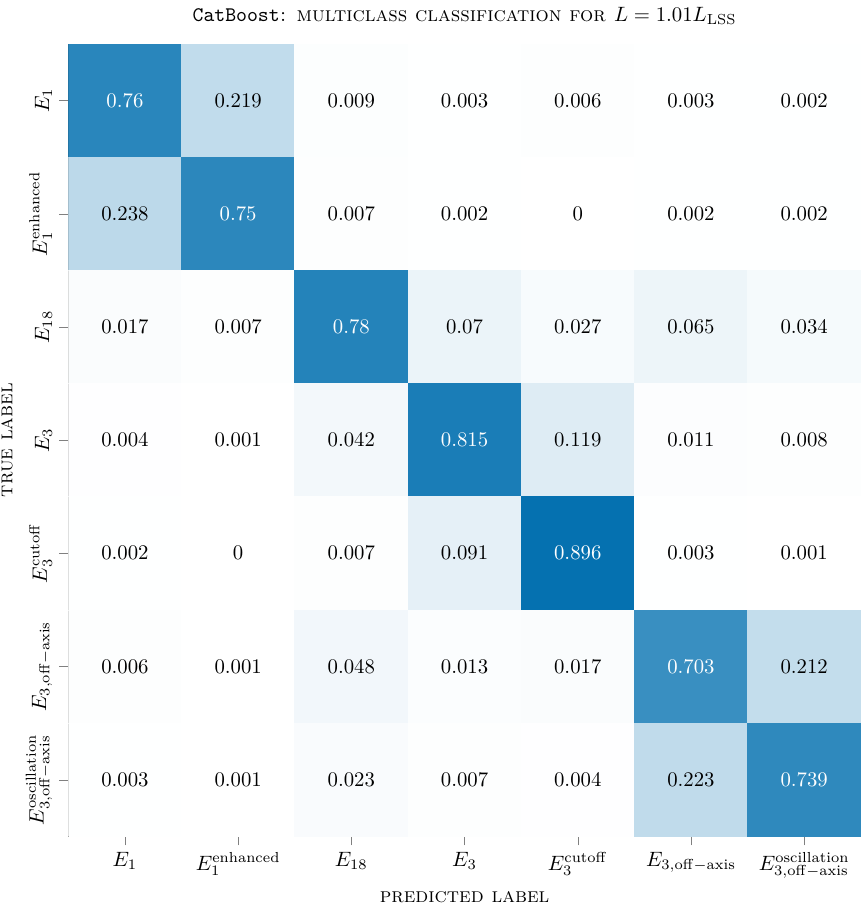}
    \caption{Confusion matrix for a seven-class classification task involving unrotated realizations of \E{18}, cubic \E{1} with \(L = L_{1} = L_{2} = L_{3} = 1.01 \LLSS\) with and without enhancement, and four \E{3} variants (on-/off-axis observer, with/without primordial power spectrum modification) with \(\LA = 1.4 \, \Lcircle\) and \(\LB = 1.01 \, \Lcircle\). \(\Lcircle = \LLSS\) for an on-axis observer, while for an off-axis observer at \(\xobs = (0.35, 0.35, 0)\), it is \(\Lcircle \approx 0.71 \, \LLSS\).}
    \label{fig:CM_CB_multi}
\end{figure}

While binary classification provides valuable insights and a direct comparison to the KL divergence, multiclass classification reflects a more realistic scenario where the goal is to infer the topology and the parameters like the fundamental domain size and the observer position from a given CMB realization. Naturally, the number of classes must be limited, but their careful selection allows us to test whether a ML classifier can accurately distinguish between \E{1} and \E{3} manifolds of the same size, or between two \E{3} configurations with different observer positions or power spectra. Figure \ref{fig:CM_CB_multi} displays the normalized confusion matrix for a classifier trained to distinguish between seven classes: the trivial topology \E{18} with the standard power-law primordial power spectrum, two versions of cubic \E{1} (with and without an enhancement modification), and four non-cubic \E{3} configurations (on-axis and off-axis observer, each with and without a modified primordial power spectrum). The classifier achieves high classification accuracy across all classes (diagonal elements in the confusion matrix). Considering that for seven classes the baseline accuracy is only about \(14\%\) (lower than for binary classification, where random guessing leads to \(50\%\) accuracy) accuracies above \(70\%\) in this multiclass case indicate strong performance. The confusion matrix shows that most classification errors occur between configurations that differ only in their power spectrum. In particular, there is notable confusion between the two \E{1} manifolds, and between the two \E{3} manifolds with off-axis observers. However, confusion between different topologies \E{1}, \E{18}, \E{3} or \E{3; \text{off-axis}} is low, indicating that the classifier has learned meaningful, topology-sensitive features.

\begin{figure}[t]
    \includegraphics[scale=1]{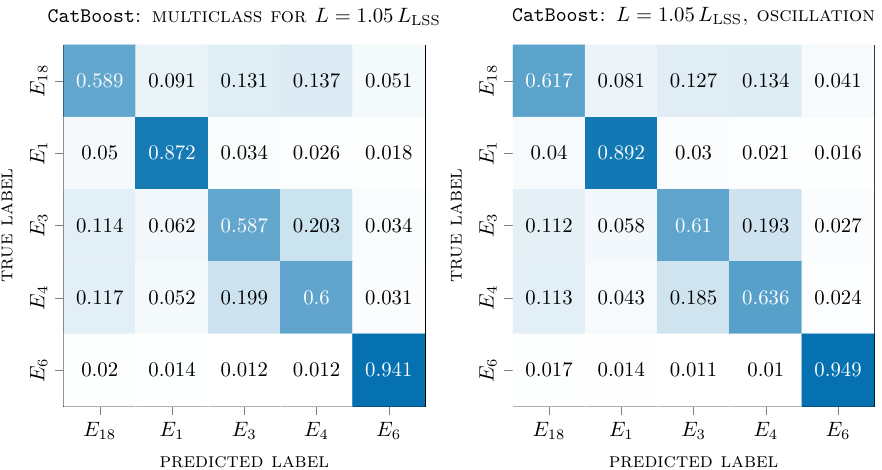}
    \caption{Confusion matrices for five-class classification of non-rotated realizations. Left panel: unmodified primordial power spectrum. Right panel: oscillation modification applied to all topologies. The \E{1} is cubic with \(L = L_{1} = L_{2} = L_{3} = 1.05 \, \LLSS\), while the \E{3} and \E{4} topologies are non-cubic with \(\LA = 1.4 \, \LLSS\) and \(L_{B} = 1.05 \, \LLSS\) and the \E{6} topology has \(L = L_{A_{x}} = L_{B_{y}} = L_{C_{z}} = 1.05 \, \Lcircle\) with \(\Lcircle = \LLSS / \sqrt{2} \approx 0.71 \, \LLSS\).}
    \label{fig:CM_CB_multi_osci}
\end{figure}

Figure \ref{fig:CM_CB_multi_osci} compares two five-class classifiers: the left classifier uses data with the standard power-law spectrum, while the right classifier uses data that includes an oscillation modification applied to all topologies. With a \(20\%\) baseline accuracy, both classifiers show good performance. In both cases, the classifier reliably identifies \E{1} and \E{6}, but shows greater confusion among \E{3}, \E{4}, and the trivial \E{18}. Apparently, the larger the fundamental domain sizes of the non-cubic \E{3} and \E{4} manifolds (with $L_A=1.4L_{\text{LSS}}$ and $L_B=1.05L_\text{LSS}$), the harder they are to be distinguished from infinite space than the \E{1} and \E{6} manifolds with a single length scale of $L=1.05L_{\text{LSS}}$. Applying the oscillation modification to all classes slightly improves the classification accuracy for each class by $1\%$-- $3\%$. Considering that we find the standard deviation across subsets to be $\sim 0.5\%$, this shows that even in multiclass scenarios, modifications to the primordial power spectrum can improve detectability.

\section{Conclusions}
\label{secn:conclusion}

In this work, we have investigated the detectability of the fully compact, orientable, Euclidean topologies \E{1}--\E{6} under deviations from the standard primordial power spectrum. 
While current observations are consistent with an infinite, simply-connected space and a nearly-scale-invariant $\calPR(k)$, we still do not know the global topology of the Universe nor the precise shape of the primordial power spectrum. 
Since large-scale anomalies in the CMB suggest that both assumptions deserve re-examination \cite{Planck:2019evm,Jones:2023ncn}, the focus of this work has been on quantifying how modifications of the primordial power spectrum affect the detectability of non-trivial topology. 
We have taken care to distinguish between modifications that are a direct consequence of the global shape of the Universe---and thus do not appear in the simply-connected covering-space power spectrum---and those produced by unrelated physical mechanisms---and therefore equivalently affect the power spectrum of non-trivial and trivial topologies.

The KL divergence between the predicted CMB statistics of a given non-trivial topology and those of the trivial, simply-connected case, yields a theoretic estimate of the detectability of cosmic topology. 
The KL divergence analysis has shown that a non-standard primordial power spectrum can amplify or suppress the KL divergence depending on how it affects the power at large angular scales. 
Crucially, a low-\(k\) cutoff, when applied to both topologies, suppresses topological signatures,  highlighting the importance of large-scale modes in topology detection.  
Enhancements at large scales or oscillatory features lead to an increase in the KL divergence, particularly in the regime of \(L > \Lcircle\). This implies that the detection of topological signatures becomes, in principle, more feasible, when there is such a modification. 
From a statistical perspective, applying modifications to the primordial power spectrum in both the trivial and non-trivial cases can be interpreted as exploring the response of the KL divergence to modifications of \(\calPR (k)\) within the allowed posterior range from the {\it Planck} data. 
Variations of the primordial power spectrum shift the relative weight of large-scale modes, thereby changing the strength of the topological signal. 
The spread in the KL divergence values across different modifications thus provides an estimate of how sensitive topology detection is to primordial power spectrum uncertainties. Observed variations have shown that these uncertainties can have a significant impact on detectability for large manifolds that yield KL divergences $\sim 1$. 

Finally, when modification of the power spectrum is a consequence of the underlying topology, and therefore only appears for the non-trivial topology, the KL divergence increases relative to the standard power-law case of the simply-connected covering space for all modifications considered. 
This indicates that the detectability of non-trivial cosmic topology can be enhanced if the topology itself induces a non-standard primordial power spectrum.

To complement the theoretical bounds, classification experiments with \catboost\ have been performed on simulated CMB temperature maps for various topologies, both with and without modifications to the primordial power spectrum. 
The results broadly mirror the KL divergence trends: realizations with enhanced large-scale power have been easier to classify, while suppressed power in realizations of both the trivial and non-trivial topology has reduced classification accuracy. 
This agreement between analytic information bounds and ML classifiers strengthens the conclusion that general uncertainties in the primordial power spectrum can improve or worsen the detectability of topology signatures in the CMB, while topology induced modifications at large scales enhance detectability.

There are several promising extensions to this work. First, realistic observational effects such as noise and sky masking should be included to evaluate detectability under practical conditions. Integrating polarization data and performing analysis directly in pixel space (i.e., working with maps) could facilitate detection. Second, future studies must account for the effect of an unknown observer orientation (with respect to the topology's coordinate frame) on ML classification. Furthermore, Bayesian likelihood inference frameworks that jointly fit topology parameters and the primordial power spectrum could reduce biases stemming from incorrect assumptions about either. Such approaches could be applied to the \textit{Planck} data and later extended to new observations. In particular, the \textit{LiteBIRD} satellite will provide full-sky measurements of large-scale CMB polarization, targeting the modes relevant for tests of CMB anomalies and cosmic topology \cite{LiteBird:2025}. Finally, while our current results demonstrate a robust proof of concept, even higher classification accuracies could be achieved with significantly larger datasets. However, the use of such datasets entails substantial memory and computational costs, and we therefore leave this for future work.

From a theoretical standpoint, exploring inflationary models in compact spaces could clarify whether certain power spectrum features are natural signatures of a non-trivial topology rather than independent phenomena.

In summary, this study has demonstrated that the theoretical detectability limits of cosmic topology depend on an interplay between the topology itself, the size of its fundamental domain, the observer location, and the shape of the primordial power spectrum. Modifications to the power spectrum that alter large-scale mode amplitudes can affect our ability to distinguish non-trivial topologies from the trivial case, either by amplifying their characteristic CMB imprints or by erasing them from observational reach. Therefore, accounting for uncertainties in the primordial power spectrum is essential for any realistic search for the global shape of our Universe.

\acknowledgments
A.T. is supported by the European Union's Horizon Europe research and innovation programme under the Marie Sk\l odowska-Curie grant agreement No. 101126636.
Y.A. acknowledges support by the Spanish Research Agency (Agencia Estatal de Investigaci\'on)'s grant RYC2020-030193-I/AEI/10.13039/501100011033, by the European Social Fund (Fondo Social Europeo) through the  Ram\'{o}n y Cajal program within the State Plan for Scientific and Technical Research and Innovation (Plan Estatal de Investigaci\'on Cient\'ifica y T\'ecnica y de Innovaci\'on) 2017-2020, by the Spanish Research Agency through the grant IFT Centro de Excelencia Severo Ochoa No CEX2020-001007-S funded by MCIN/AEI/10.13039/501100011033, by the Spanish National Research Council (CSIC) through the Talent Attraction grant 20225AT025, and by the Spanish Research Agency's Consolidaci\'on Investigadora
2024 grant CNS2024-154430.
G.A. is supported by the Spanish Research Agency's Consolidaci\'on Investigadora
2024 grant CNS2024-154430.
J.C.D. is supported by the Spanish Research Agency (Agencia Estatal de Investigaci\'on), the Ministerio de Ciencia, Inovaci\'on y Universidades, and the European Social Funds through grant JDC2023-052152-I, as part of the Juan de la Cierva program.
C.J.C., A.K., D.P.M. and G.D.S.\ acknowledge partial support from NASA ATP grant RES240737 and from NASA ADAP grant 24-ADAP24-0018;  A.H.J.\ from the Royal Society and STFC in the UK; and A.S. and G.D.S.\ from DOE grant DESC0009946.
D.P.M. acknowledges support by the Bulgarian National Science Fund program ``VIHREN--2024" project No. KP--06--DV/9/17.12.2024. 
M.M.B. acknowledges support by the Spanish Ministry of Science, Innovation and Universities under the FPU predoctoral grant FPU22/02306.
F.C.G\ was supported by the Presidential Society of STEM Postdoctoral Fellowship at Case Western Reserve University and by Ministerio de Ciencia, Innovaci\'on y Universidades, Spain, through a Beatriz Galindo Junior grant BG23/00061.
A.N.\ is supported by the Richard S.\ Morrison Fellowship.
T.S.P is supported by Funda\c{c}\~ao Arauc\'aria (NAPI Fen\^omenos Extremos do Universo, grant 347/2024 PD\&I). J.R.E. acknowledges support from the European Research Council under the Horizon 2020 Research and Innovation Programme (Grant Agreement No.~819478).
J.C.D. is and A.T. was supported by CSIC through grant No. 20225AT025. A.T.\ was also supported by the Richard S.\ Morrison Fellowship. 
We are thankful to the Istituto Nazionale di Fisica Nucleare in Italy.
This work is partially funded by the European Commission -- NextGenerationEU, through Momentum CSIC Programme: Develop Your Digital Talent. We acknowledge HPC support by Emilio Ambite, staff hired under the Generation D initiative, promoted by \url{Red.es}, an organization attached to the Spanish Ministry for Digital Transformation and the Civil Service, for the attraction and retention of talent through grants and training contracts, financed by the Recovery, Transformation and Resilience Plan through the EU's Next Generation funds.

\appendix

\section{Eigenmodes of the scalar Laplacian and correlation matrices}
\label{app:eigenmodes}

This appendix provides a brief summary of the derivation of the harmonic-space covariance matrices, which has been addressed in detail in previous work \cite{COMPACT:2023rkp}.
The 18 three-dimensional Euclidean topologies can be classified by the number of dimensions that are compact and by whether or not they are orientable, homogeneous, and isotropic. 
This work focuses on the fully compact, orientable topologies \E{1}--\E{6}, whose properties are summarized in Table \ref{tab:properties}. The trivial topology \E{18}, i.e., the Euclidean space \(E^{3}\) serves as the universal covering space for all flat topologies.

\begin{table}[t]
\newcommand{\highlightcolor}{white}
    \begin{tabular}{clcccc}
        \toprule
        \multirow{2}{*}{\textsc{symbol}} & \multirow{2}{*}{\textsc{name}} & \textsc{compact} & \multirow{2}{*}{\textsc{orientable}} & \textsc{homo-} & \multirow{2}{*}{\textsc{isotropic}} \\
        & & \textsc{dimensions} & & \textsc{geneous} & \\
        \midrule
        \addlinespace[0pt]
        \rowcolor{\highlightcolor}\E{1} & 3-torus & 3 & \textsc{yes} & \textsc{yes} & \textsc{no} \\
        \rowcolor{\highlightcolor}\E{2} & Half-turn& 3 & \textsc{yes} & \textsc{no} & \textsc{no} \\
        \rowcolor{\highlightcolor}\E{3} & Quarter-turn & 3 & \textsc{yes} & \textsc{no} & \textsc{no} \\
        \rowcolor{\highlightcolor}\E{4} & Third-turn & 3 & \textsc{yes} & \textsc{no} & \textsc{no} \\
        \rowcolor{\highlightcolor}\E{5} & Sixth-turn & 3 & \textsc{yes} & \textsc{no} & \textsc{no} \\
        \rowcolor{\highlightcolor}\E{6} & Hantzsche-Wendt & 3 & \textsc{yes} & \textsc{no} & \textsc{no} \\
        \E{18} & Covering space & 0 & \textsc{yes} & \textsc{yes} & \textsc{yes} \\
        \bottomrule
    \end{tabular}
    \caption{Properties of the 6 three-dimensional compact orientable Euclidean topologies that are the focus of this work, plus the covering space.}
    \label{tab:properties}
\end{table}

\subsection{The covering space, \E{18}}
\label{subsec:E18}
For the trivial topology \E{18}, the scalar eigenfunctions of the Laplacian are unrestricted plane waves
\begin{equation}
    \Upsilon_{\kvec}^{\E{18}}(\xvec) = e^{i \kvec\cdot(\xvec - \xobslss)},
\end{equation}
where \(\boldsymbol{x_{0}}\) indicates the position of an arbitrary origin relative to the observer's coordinate system\footnote{The inclusion of \(\boldsymbol{x_{0}}\) has no particular role for the covering space or for \E{1} but is crucial for \E{2}--\E{6}.}. These modes are normalized as
\begin{equation}
    \bigintsss_{\mathbb{R}^3} \frac{\dd^{3} x}{(2\pi)^{3}} \, \Upsilon_{\kvec}(\xvec) \Upsilon^{*}_{\kpvec}(\xvec) = \delta^{(D)}(\kvec - \kpvec),
\end{equation}
where \(\delta^{(D)}(\kvec - \kpvec)\) is the three-dimensional Dirac delta function. 
Alternatively, the eigenfunctions may be expanded in spherical waves as 
\begin{equation}
    \Upsilon_{\kvec}^{\E{18}}(\xvec) = \sum_{\ell, m}\xi_{k\ell m}^{\E{18},\hatkvec} \mathcal{Y}_{k\ell m}(\xvec) = e^{-i \kvec\cdot\xobslss}\sum_{\ell, m}i^\ell Y_{\ell m}^*(\hatkvec) \mathcal{Y}_{k\ell m}(\xvec),
\label{eq:planar_to_spherical2}
\end{equation}
with \(\hatkvec \equiv \kvec/|\kvec|\) and 
\begin{equation}
\mathcal{Y}_{k\ell m}(r, \theta, \phi) = 4\pi j_\ell(kr) \, Y_{\ell m}(\theta, \phi),
\end{equation}
where \(j_\ell (kr)\) is the spherical Bessel function, and \(Y_{\ell m}(\theta, \phi)\) is the spherical harmonic. The mode \( \mathcal{Y}_{k\ell m}\) is normalized according to
\begin{equation}
    \int_{\mathbb{R}^3} r^2\dd r \, \dd\!\left(\cos(\theta)\right) \dd\phi \, \frac{\mathcal{Y}_{k\ell m} \, \mathcal{Y}_{k'\ell'm'}^*}{(2\pi)^3} = \frac{1}{k^2} \, \delta^{(D)}(k - k') \, \delta^{(K)}_{\ell\ell'} \, \delta^{(K)}_{mm'},
\end{equation}
with the Kronecker deltas \(\delta^{(K)}_{\ell \ell'}\) and \(\delta^{(K)}_{m m'}\).

In standard inflationary cosmology, scalar perturbations are seeded by the adiabatic curvature perturbation \(\delta^{\calR}\), which can be expanded as
\begin{equation}
    \delta^\calR(\xvec) = \bigintsss \!\! \frac{\dd^{3} k}{(2\pi)^3} \, \delta^\calR(\kvec) \Upsilon_{\kvec}^{\E{18}}(\xvec),
    \label{eq:curv_pert}
\end{equation}
where the coefficients \(\delta^{\calR} (\kvec)\) are Gaussian random variables with zero mean and covariance fixed by the dimensionless power spectrum \(\calPR (k)\).  Other scalar fields, such as the CMB temperature anisotropies, are related to \(\delta^{\calR}(\kvec)\) via linear transfer functions \(\Delta^{X}\):
\begin{equation}
    \delta^{X} (\kvec) = \Delta^{X} \delta^{\calR} (\kvec).
\end{equation}
The two-point function of Fourier modes is then
\begin{equation}
    C^{\E{18}; XY}_{\kvec \kpvec} = \langle \delta^X(\kvec) \delta^{Y^*}(\kpvec)\rangle = (2\pi)^{3} \frac{2\pi^2}{k^3} \calPR(k)\Delta^X (k)\Delta^{Y*}(k)\delta^{(D)}(\kvec - \kpvec),
    \label{eq:exp_iso}
\end{equation}
which expresses both isotropy, since $\calPR$ is a function of $k=|\kvec|$, and homogeneity, since modes of different \(\kvec\) are uncorrelated due to the Dirac delta function \(\delta^{(D)}(\kvec - \kpvec)\). To connect with observed fields like CMB temperature or polarization anisotropies, one projects the perturbation field \(\delta^{X}\) onto the celestial sphere, integrating along the line of sight with the appropriate transfer function, yielding
\begin{equation}
    \delta^X(\theta, \varphi)=\sum_{\ell m}a_{\ell m}^{\E{18};X}Y_{\ell m}(\theta, \varphi),
    \label{eq:connection}
\end{equation}
with 
\begin{equation}
    \alm^{\E{18}; X}=\frac{4\pi}{(2\pi)^3} \int \dd^{3} k \, \delta^\calR(\kvec)\xi_{k\ell m}^{\E{18};\hat{\kvec}}\Delta_\ell^X(k),
    \label{eq:a_lm_18}
\end{equation}
where the angular transfer function \(\Delta_\ell^{X} (k)\) absorbs the \(j_{\ell} (kr)\) that contributed to the integrand of the radial integral. Because the \(\delta^{X} (\kvec)\) are zero-mean Gaussian random variables, the spherical harmonic coefficients \(\alm\) are also zero-mean Gaussian random variables, whose properties are entirely described by their covariance matrix. If we further impose statistical isotropy, it follows that this matrix is diagonal (i.e., the \(\alm\) are statistically independent), with the diagonal part depending only on $\ell$,
\begin{equation}
    C_{\ell m \ell' m'}^{\E{18}; XY}=\biggl< \alm^{\E{18};X} a_{\ell' m'}^{\E{18};* Y} \biggr> = C_{\ell}^{\E{18};XY}\delta^{(K)}_{\ell \ell'}\delta^{(K)}_{mm'},
\end{equation}
where the angular power spectrum is given by
\begin{equation}
\boxed{
    C_\ell^{\E{18}; XY}=4\pi \! \bigintsss \! \frac{\dd k}{k} \, \calPR(k) \, \Delta_\ell^{X}(k) \, \Delta_\ell^{Y*}(k).}
\end{equation}
This is the fundamental prediction of the standard model of cosmology for scalar perturbations in a simply-connected, spatially flat, homogeneous and isotropic universe.

\subsection{The fully compact, orientable topologies \E{1}--\E{6}}
\label{subsec:E1}
In \cite{COMPACT:2023rkp} we presented the generators, scalar-Laplacian eigenmodes, and various cosmological correlation matrices in orientable Euclidean three-manifolds. 
The 3-torus, also called \E{1}, is the simplest compact Euclidean topology and serves as a model for determining the eigenspectrum and eigenmodes of all the other compact Euclidean topologies. The \E{1} eigenmodes \(\Upsilon_{\kvec}^{\E{1}}(\xvec)\) are the subset of the \E{18} eigenmodes that respect the \E{1} symmetry conditions:
\begin{equation}
     \Upsilon_{\kvec}^{\E{1}} \left(g_{A_{j}}^{\E{1}}\xvec\right) = \Upsilon_{\kvec}^{\E{1}}(\xvec), \quad j = \{1, 2, 3\}.
     \label{eq:symmE1}
\end{equation}
Here \(g_{A_j}^{\E{1}}\) is a generator for the \E{1} topology (described in detail in \cite{COMPACT:2023rkp}), i.e., a translation. Invariance under a translation \(\Tvec_j^{\E{1}}\) implies
\begin{equation}
    e^{i \kvec \cdot[(\xvec - \xobslss) + \Tvec_j^{\E{1}}]} = e^{i \kvec\cdot(\xvec - \xobslss)},
\end{equation}
which leads to the quantization condition
\begin{equation}
    \kvec \cdot\Tvec_j^{\E{1}} = 2\pi n_j, \quad \text{for} \quad n_{j} \in \mathds{Z}.
    \label{eq:discrete_T}
\end{equation}

In order to simplify the discussion, we focus on the special case of the cubic \E{1}, where the three translations are in orthogonal directions that can be aligned along the three coordinate axes and are of equal length, \(L\). 
The fundamental domain is a cube of volume \(V_{\E{1}} = L^{3}\), and the allowed wavevectors are
\begin{equation}\label{eq:k_E1}
    \kvec_{\nvec} = \frac{2\pi}{L} \, \nvec = \frac{2 \pi}{L} \, (n_{1}, n_{2}, n_{3}).
\end{equation}
The corresponding eigenmodes are plane waves
\begin{equation}
    \Upsilon_{\kvec_{\nvec}}^{\E{1}}(\xvec) = e^{i \kvec_{\nvec}\cdot(\xvec - \xobslss)} \quad \text{for} \quad \nvec \in \mathcal{N}^{\E{1}}\equiv\{(n_1, n_2, n_3)|n_i \in \mathds{Z}\}\setminus(0, 0, 0).
\end{equation}
Analogous to \cref{eq:curv_pert}, the adiabatic curvature perturbation field can be expanded as
\begin{equation}
    \delta^{\calR}(\xvec)=\frac{1}{V_{\E{1}}}\sum_{\nvec \in \mathcal{N}^{\E{1}}}\delta^{\calR}_{\kvec_{\nvec}} \Upsilon_{\kvec_{\nvec}}^{\E{1}}(\xvec),
\end{equation}
where the primordial density fluctuations \(\delta^{\calR}_{\boldsymbol{k_{\nvec}}}\) are zero-mean, Gaussian random variables with variances determined by the primordial power spectrum \(\calPR(k_{\nvec})\), where \(k_{\nvec}=|\kvec_{\nvec}|\). 
Other scalar fields are related to \(\delta^{\calR}(k_{\nvec})\) via a transfer functions \(\Delta^{X}(k_{\nvec})\). 
Their two-point correlation in Fourier space is 
\begin{equation}
    \langle \delta^X_{\kvec_{\nvec}}\delta^{Y^*}_{\boldsymbol{k_n'}}\rangle=\frac{2\pi^2}{k_{\nvec}^3}V_{\E{1}}\calPR(k_{\nvec})\Delta^X(k_{\nvec})\Delta^{Y*}(k_{\nvec})\delta^{(K)}_{\kvec_{\nvec} \boldsymbol{k'_n}}.
\end{equation}
The spherical-harmonic coefficients can then be obtained via:
\begin{equation}
    \alm^{\E{1}; X} = \frac{4\pi}{V_{\E{1}}} \sum_{\nvec \in \mathcal{N}^{\E{1}}} \delta^{\calR}_{\boldsymbol{k_{\nvec}}} \xi_{k_{\nvec} \ell m}^{\E{1}; \hat{\kvec}_{\nvec}} \Delta^X_\ell(k_{\nvec}),
    \label{eq:alm_E1}
\end{equation}
where 
\begin{equation}
\boxed{
     \xi_{k_{\nvec} \ell m}^{\E{1}; \hatkvec_{\nvec}} = e^{-i \boldsymbol{k_{\nvec}} \cdot \xobslss} i^\ell Y_{\ell m}^*(\hatkvec_{\nvec}).}
\end{equation}
An analogous procedure applies for \E{2}-\E{6} and for the fully compact, orientable topologies \E{1}--\E{6}, the harmonic-space covariance matrix is given by 
\begin{equation}
\boxed{
    C^{\E{i}; XY}_{\ell m \ell' m'} = \frac{(4\pi)^2}{V_{\E{i}}} \sum_{\nvec \in \mathcal{N}^{\E{i}}} \Delta^X_\ell(k_{\nvec}) \Delta^{Y*}_{\ell'}(k_{\nvec}) \frac{2\pi^2 \calPR(k_{\nvec})}{k_{\nvec}^3} 
    \xi_{k_{\nvec} \ell m}^{\E{i}; \hatkvec_{\nvec}} \xi_{k_{\nvec} \ell' m'}^{\E{i}; \hatkvec_{\nvec} *}.}
\end{equation}
Each of the spaces \E{2}--\E{6} can be viewed as a quotient of \E{1} by additional discrete isometries, which introduce new features compared to \E{1}. 
These features are essential when searching for topological signatures in the CMB. 
The key difference lies in the nature and order \(N\) of the corkscrew motion(s) in each topology, which involve a discrete rotation by \(2 \pi / N\) about a fixed axis. 
The eigenmodes are then constructed as linear combinations of \(N\) plane waves, respecting the symmetry conditions of the topology. While \E{2} admits \(N = 2\) combinations of wavevectors due to it's half-turn symmetry, the eigenmodes in \E{3}, \E{4} and \E{5} involve \(N = 4\), \(N = 3\) and \(N = 6\) wavevectors, respectively. In contrast, \E{6} features three independent screw motions of order two and requires invariance under all three simultaneously. This leads to more restrictive conditions on the wavevectors, often resulting in more sparse mode sets and thus stronger statistical anisotropy.

\section{Additional KL divergence results}
\label{app:kl}
In this appendix, we present the backward KL divergences for all configurations considered in \cref{subsecn:numerical_results}, and also the forward and backward KL divergences for additional configurations.

\subsection{Backward KL divergence}
\label{app:back_kl}

\begin{figure}[t]
\centering
    \includegraphics[scale=1]{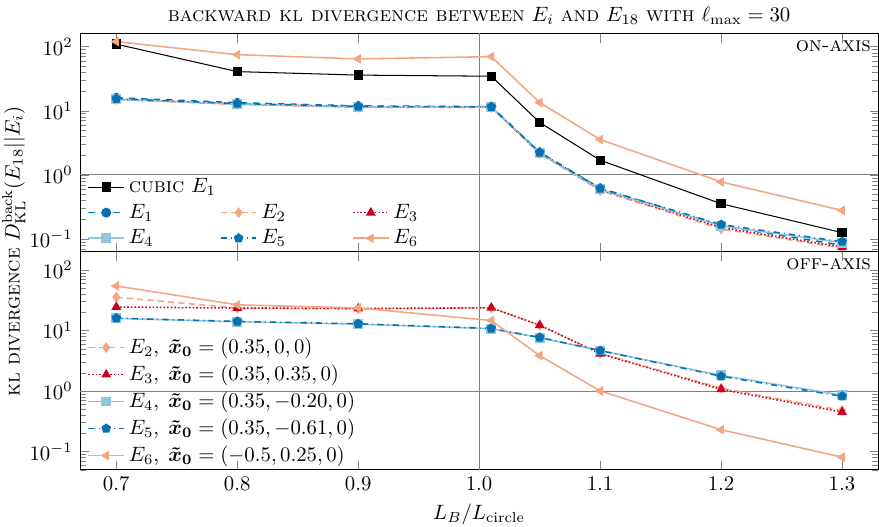}
    \caption{Backward KL divergence for topologies \E{1}--\E{6}
    with an on-axis observer \(\xobs = (0, 0, 0)\) (upper panel) or an off-axis observer \(\xobs \neq (0, 0, 0)\) (lower panel). For cubic \E{1}, the \(x\)-axis indicates the length of all three sides, while for non-cubic \E{1} the length of the other sides is \(\LA = 1.4 \, \Lcircle\) (for \E{1}, \(L_{1} = L_{2} = 1.4 \, \Lcircle\) remain fixed, while the varying \(L_{3}\) is shown on the \(x\)-axis). The \E{6} manifold has a single length scale \(L = L_{A_{x}} = L_{B_{y}} = L_{C_{z}}\), which is indicated by the $x$-axis, while \(r_{x} = r_{y} = r_{z} = 1/2\) are kept fixed.}
    \label{fig:bKL}
\end{figure}

\begin{figure}[t]
\centering
  \includegraphics[scale=1]{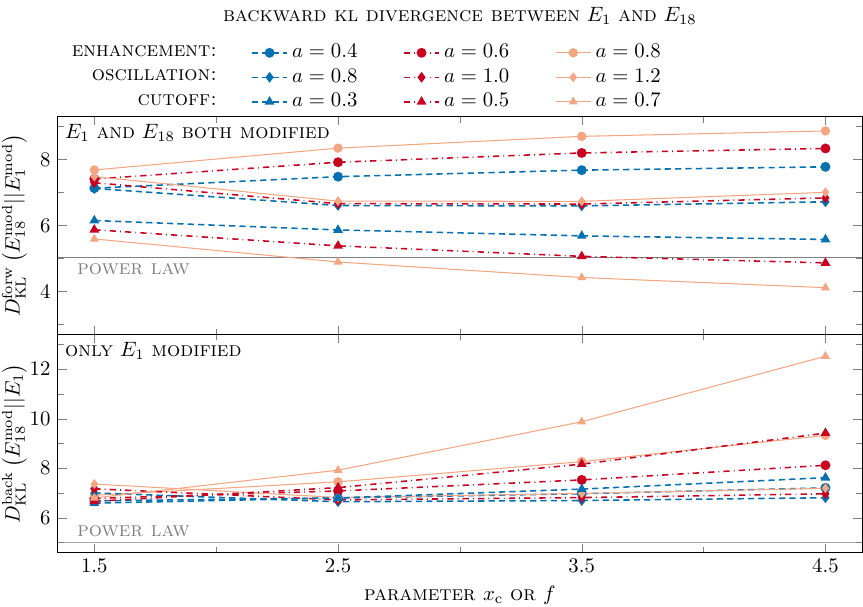}
  \caption{Influence of different parameters for the power spectrum modifications on the backward KL divergence for a cubic \E{1} topology of size \(L = 1.05 \, \LLSS\). The parameters are those described in \cref{eq:modifications}, for the oscillation modification \(x_c = 2\), and the \(x\)-axis shows different frequencies.}
  \label{fig:bKL2}
\end{figure}

\begin{figure}[h]
    \includegraphics[scale=1]{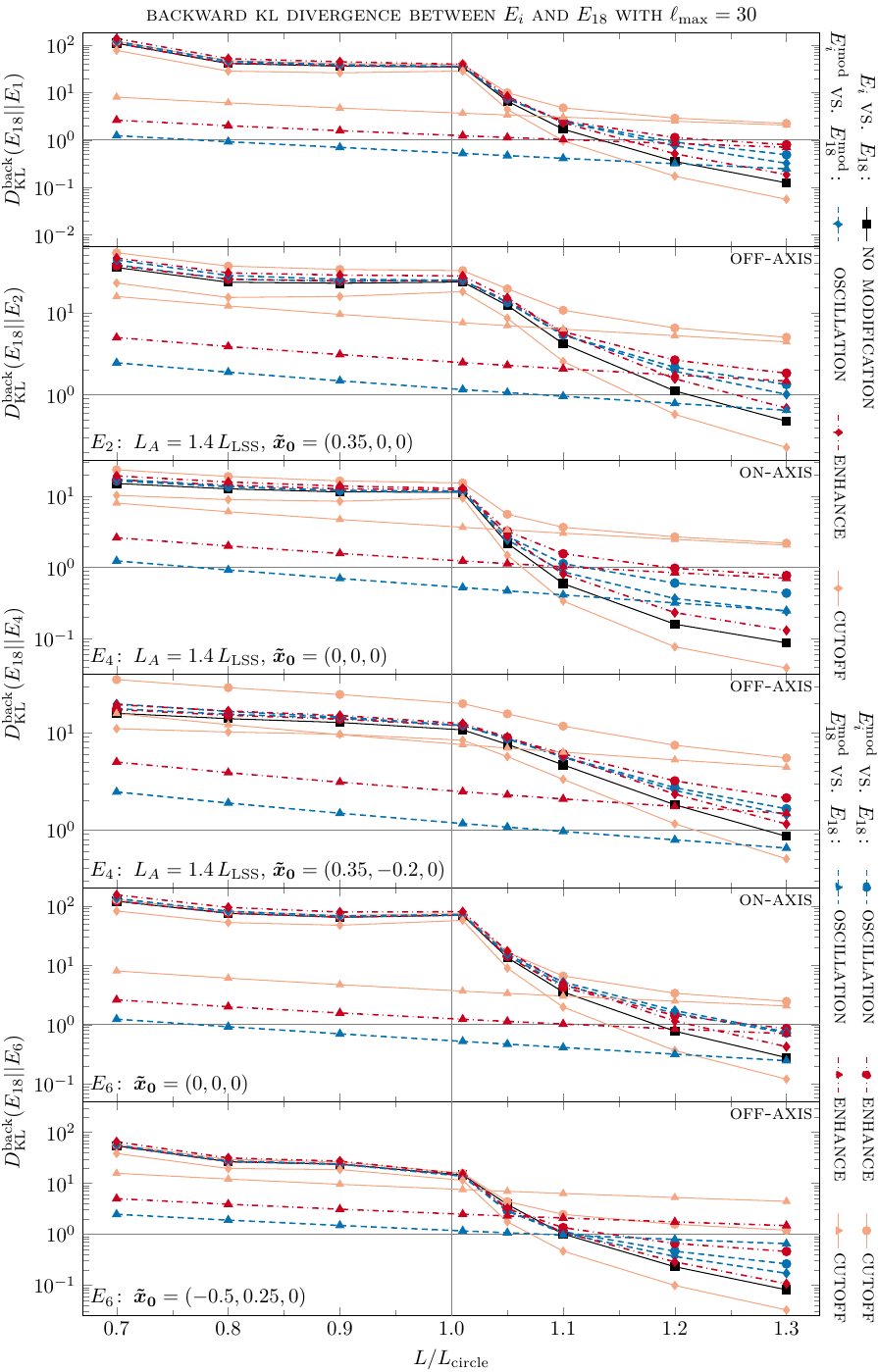}
    \caption{Backward KL divergence for the configurations of \cref{fig:KL_E1,fig:KL_E4,fig:KL_E2_oa,fig:KL_E6}.}
    \label{fig:bKL3}
\end{figure}

In \cref{fig:bKL,fig:bKL2,fig:bKL3}, we include the backward KL divergences corresponding to \cref{fig:KL_all_on_off_ax} to \ref{fig:KL_E6}. It behaves qualitatively like the forward KL divergence, but with overall higher values. Although both quantities are mathematically related and often display similar trends, their physical interpretations are distinct. In particular, the forward KL divergence is more relevant in the context of cosmological observations.
If the true spatial topology of the Universe were non-trivial, current observational analyses, which almost universally assume statistical isotropy and the trivial topology \E{18}, would incur the information loss quantified by the forward KL divergence. 
The backward KL divergence on the other hand, can be connected to the notion of false positives. Specifically, if the topology of the Universe is truly trivial (\E{18}), the backward KL divergence quantifies the mismatch incurred when attempting to fit the data with a specific non-trivial topology \E{i}. It provides a measure of how likely it is that trivial data could be incorrectly interpreted as evidence for a non-trivial topology. A small backward KL divergence would indicate that the non-trivial model can mimic the trivial case fairly well, increasing the risk of false positives. Conversely, larger values suggest that such a misinterpretation is unlikely.

\begin{figure}[h]
    \includegraphics[scale=1]{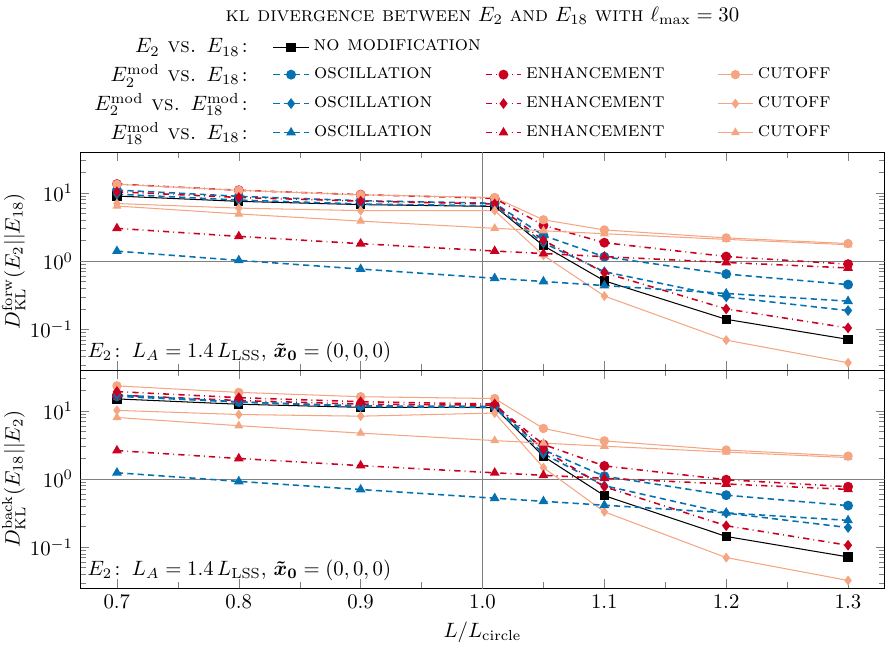}
    \caption{Forward (top panel) and backward (bottom panel) KL divergence for the \E{2} topology with \(\LA = 1.4 \, \LLSS\). The circle limit for an on-axis observer is \(L_{B} = \Lcircle = \LLSS\).}
    \label{fig:aKL_E2}
\end{figure}

\begin{figure}[h]
    \includegraphics[scale=1]{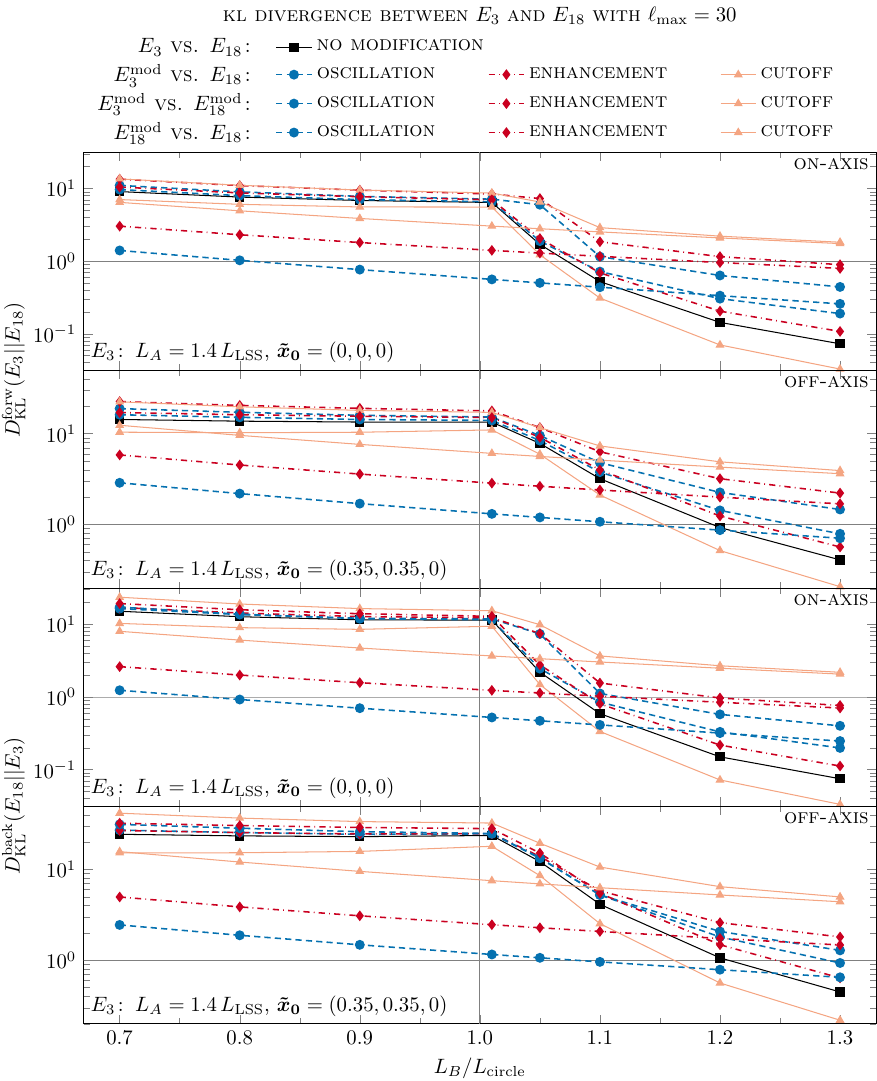}
    \caption{Forward and backward KL divergence for the \E{3} topology with \(\LA = 1.4 \, \LLSS\) for on- and off-axis observers. The circle limit for an on-axis observer is \(\LB = \Lcircle = \LLSS\).}
    \label{fig:aKL_E3}
\end{figure}

\subsection{KL divergence for other \E{1}--\E{5} configurations}
\label{app:other_conf}
\begin{figure}[t]
    \includegraphics[scale=1]{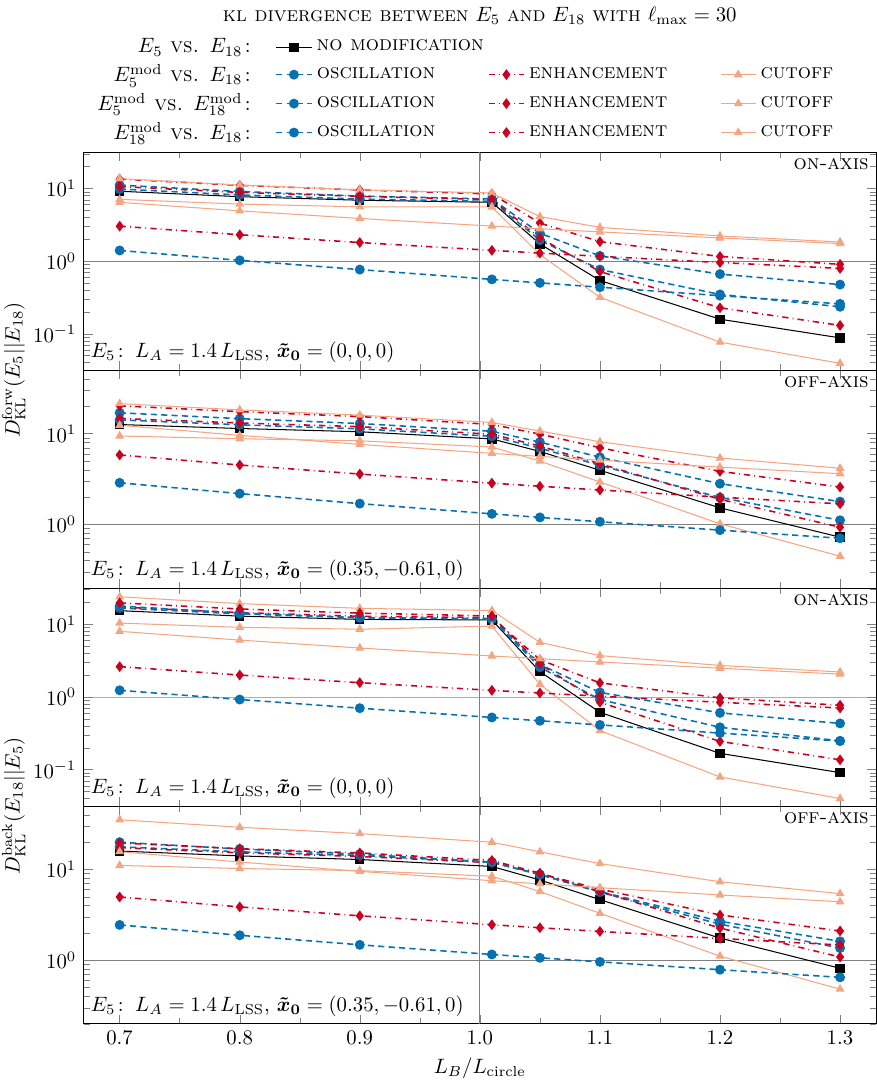}
  \caption{Forward and backward KL divergence for the \E{5} topology with \(\LA = 1.4 \, \LLSS\) for on- and off-axis observers. The circle limit for an on-axis observer is \(\LB = \Lcircle = \LLSS\).}
  \label{fig:aKL_E5}
\end{figure}

As we focused on representative cases in the main text, we will include the other results here for completeness. In \cref{fig:KL_all_on_off_ax}, we showed that for an on-axis observer, manifolds of \E{1}--\E{5} with the same side lengths, all have the same KL divergence. We also showed that for an off-axis observer, manifolds of \E{2} and \E{3} with the same side lengths, have the same KL divergence and so do those of \E{4} and \E{5}. While this was explicitly demonstrated for the standard power-law spectrum, we did not show it for the modified primordial power spectra. Therefore, we include in \cref{fig:aKL_E2,fig:aKL_E3,fig:aKL_E5} the full KL divergence results for \E{3} and \E{5} as well as those for \E{2} with an on-axis observer. These results confirm that the observed equivalences hold not only for the standard power-law case but also for modified primordial power spectra.

\clearpage
\section{Machine learning}
\label{app:ml}
In the following, we briefly describe how the cross-entropy loss in ML classification is related to the KL divergence and the likelihood. Then, we present the results from using the interpretability framework \texttt{SHAP} to analyze the decision-making process of the ML classifier used in \cref{subsecn:numerical_results}. Finally, we give some details on optimizing the hyperparameters for the \catboost\ algorithm, to achieve optimal classification accuracy.

\subsection{Cross-entropy, KL divergence, and likelihood:} 
\label{app:ml_Kldiv}
When doing ML classification with \catboost, the input data are simulated realizations \(x\) (spherical harmonic coefficients \(\alm\)) with labels \(y\) (the corresponding topology class). In this case, the true conditional distribution \(p (y|x)\) is a one-hot vector: if the true label is for example \E{1}, then \(p (y = \E{1}|x) = 1\) and all others entries are zero. The model distribution \(q_{\theta} (y|x)\) is the classifier's softmax output, where \(\theta\) denotes the learned model parameters (the tree splits and leaf values). The cross-entropy between the true label distribution \(p\) and the model's predicted distribution \(q_{\theta}\) is
\begin{equation}
    H(p,q_\theta)=-\mathrm{E}_p[\ln q_\theta(y|x)].
\end{equation}
In practice, the expectation is approximated by the empirical average over a dataset \(\{x^{(i)}, y^{(i)}\}\), leading to the categorical cross-entropy loss
\begin{equation}
    L=-\frac{1}{N}\sum_{i=1}^N\ln q_\theta(y^{(i)}|x^{(i)}),
\end{equation}
which is minimized during training. Recall that the forward KL divergence is defined as
\begin{equation}
      \DKL(p||q_\theta)=\mathrm{E}_p\left[ \ln\frac{p(x)}{q_\theta(x)}\right]=\underbrace{\mathrm{E}_p\left[ -\ln q_\theta(x)\right]}_{=H(p,q_\theta)}-\underbrace{\mathrm{E}_p\left[ -\ln p(x)\right]}_{=H(p)}.
\end{equation}
The second term is the true entropy of the label distribution, \(H (p)\), which is independent of the model parameters \(\theta\). Therefore, minimizing cross-entropy loss is equivalent to minimizing the forward KL divergence \(\DKL (p||q_\theta)\) between the true label distribution (one-hot) and the model's predicted distribution. Finally, note that the empirical cross-entropy loss is just the negative log-likelihood of the training data under the model:
\begin{equation}
    L(\theta)=-\frac{1}{N}\sum_{i=1}^{N} \ln q_\theta(y^{(i)}|x^{(i)})=-\frac{1}{N}\ln \prod_{i=1}^N q_\theta(y^{(i)}|x^{(i)})=-\frac{1}{N}\ln \mathcal{L(\theta)}.
\end{equation}
Thus, minimizing cross-entropy loss is the same as maximizing the likelihood of the data under the model.

\subsection{Interpretability with Shap}
\label{app:shap}

Interpretability has become an important aspect of modern ML, especially in scientific contexts where it is crucial to understand how classifiers arrive at their decisions to trust their predictions. To gain insight into the decision-making process of the \catboost\ classifier, we employed the interpretability framework \shap\ (SHapley Additive exPlanations).\footnote{Available from \url{https://github.com/shap/shap}.} 
\shap\ is a model-agnostic approach grounded in cooperative game theory that is particularly well-suited for tree-based models \cite{shap}. 
\shap\ assigns each input feature an importance value (Shapley value), which quantifies that feature's contribution to a given prediction. 
Interpretability can be evaluated at two complementary levels: local and global. Local explanations attribute individual predictions to specific input features, while global explanations summarize the overall importance of features across the entire dataset. For local explanations, \shap\ estimates the contribution of each feature, i.e., the real and imaginary part of each coefficient \(\alm\), by computing the expected change in the classifier output when that feature is added to different subsets of input features. This process is repeated across all possible feature orderings, ensuring that the final attributions reflect an unbiased average contribution for every feature. The Shapley value \(\phi_{i} (f, x)\) represents the average marginal contribution\footnote{The term inside the brackets is the marginal contribution, where \(f_{x} \left(P_{i}^{m} \cup \{i\}\right)\) is the classifier's prediction when it ``knows'' feature \(i\) and all the features that came before it in that specific permutation. \(f_{x} \left(P_{i}^{m}\right)\) is the classifier's prediction when it ``knows'' only the features that came before \(i\). The difference indicates how much feature \(i\) changes the prediction at that specific step.} of feature \(i\) across all permutations of feature orderings:
\begin{equation}
    \phi_{i} (f, x) = \sum_{m \in M} \frac{1}{N!}\Bigl[f_{x}\left(P_i^m \cup \{i\}\right) - f_{x} \left(P_{i}^{m}\right)\Bigr],
\end{equation}
where \(M\) is the set of all permutations of feature orderings, \(P_{i}^{m}\) denotes the set of features preceding feature \(i\) in permutation \(m\), \(N\) is the total number of features and \(f_{x}\) denotes the expected value of the classifier's prediction for instance \(x\), i.e., a specific realization. Exact computation of Shapley values become computationally too expensive for large \(N\), so \shap\ employs model-specific approximations, such as \texttt{TreeSHAP}\footnote{Available from \url{https://github.com/ModelOriented/treeshap}.} for tree ensembles \cite{treeshap}. For global interpretability, the mean absolute Shapley values are computed for each feature across the entire dataset. This yields an importance ranking of features, highlighting the \(\alm\) coefficients that consistently drive classification. 

\begin{figure}[htbp]
    \centering
    \includegraphics[width=\textwidth]{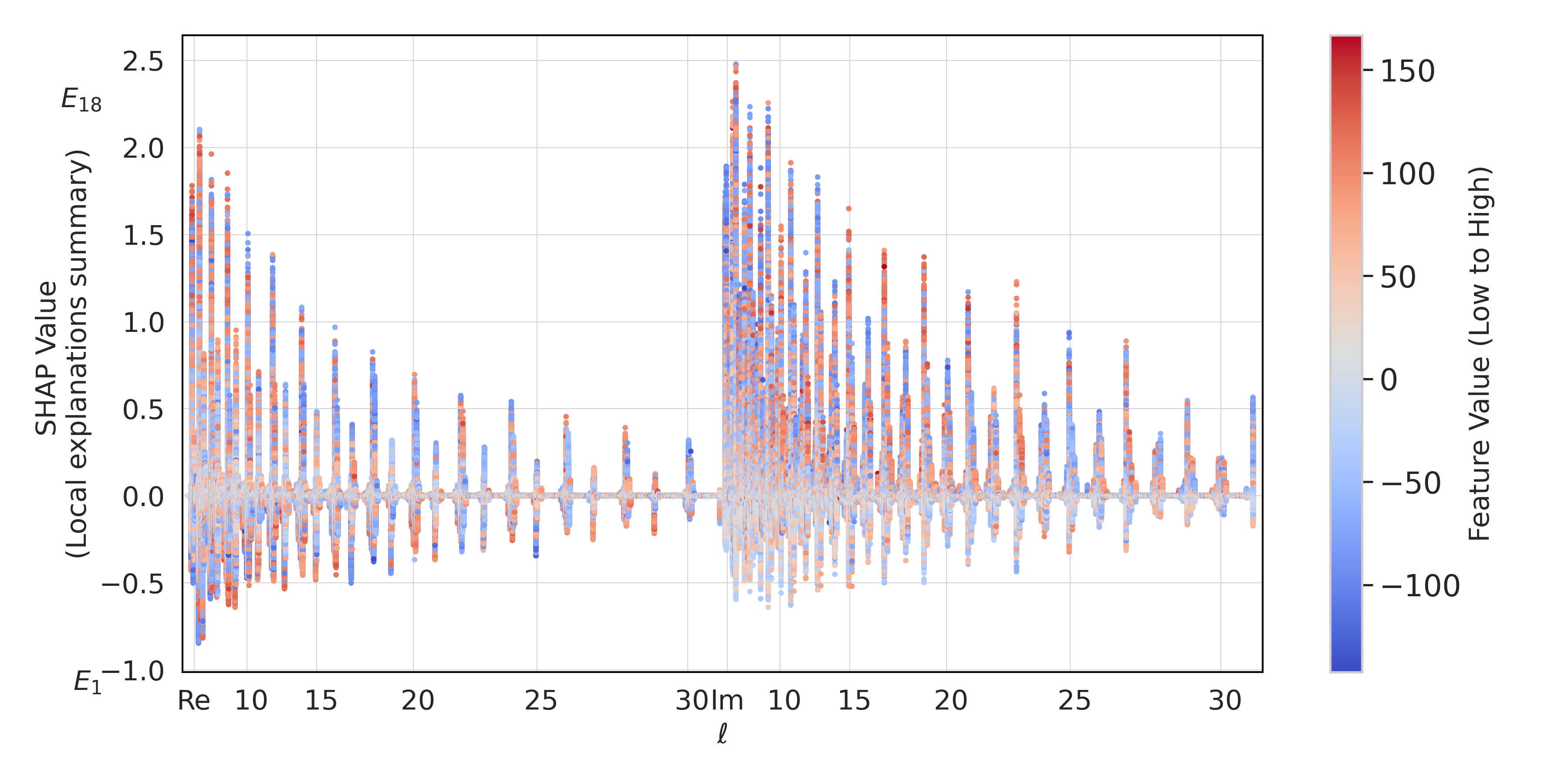}
    \caption{Local interpretability with \shap\ values for binary classification between \E{18} and the cubic \E{1} of size \(L = 1.01 \, \LLSS\). The features, i.e., the \(\alm\) coefficients are shown on the \(x\)-axis, ordered first in \(\ell\) and then in \(m\), and the feature values are indicated by color. The \(y\)-axis shows the \shap\ values for each feature of each realization.}
    \label{fig:Shap_local_E1_L1p01}
\end{figure}

\begin{figure}[htbp]
    \centering
    \includegraphics[width=\textwidth]{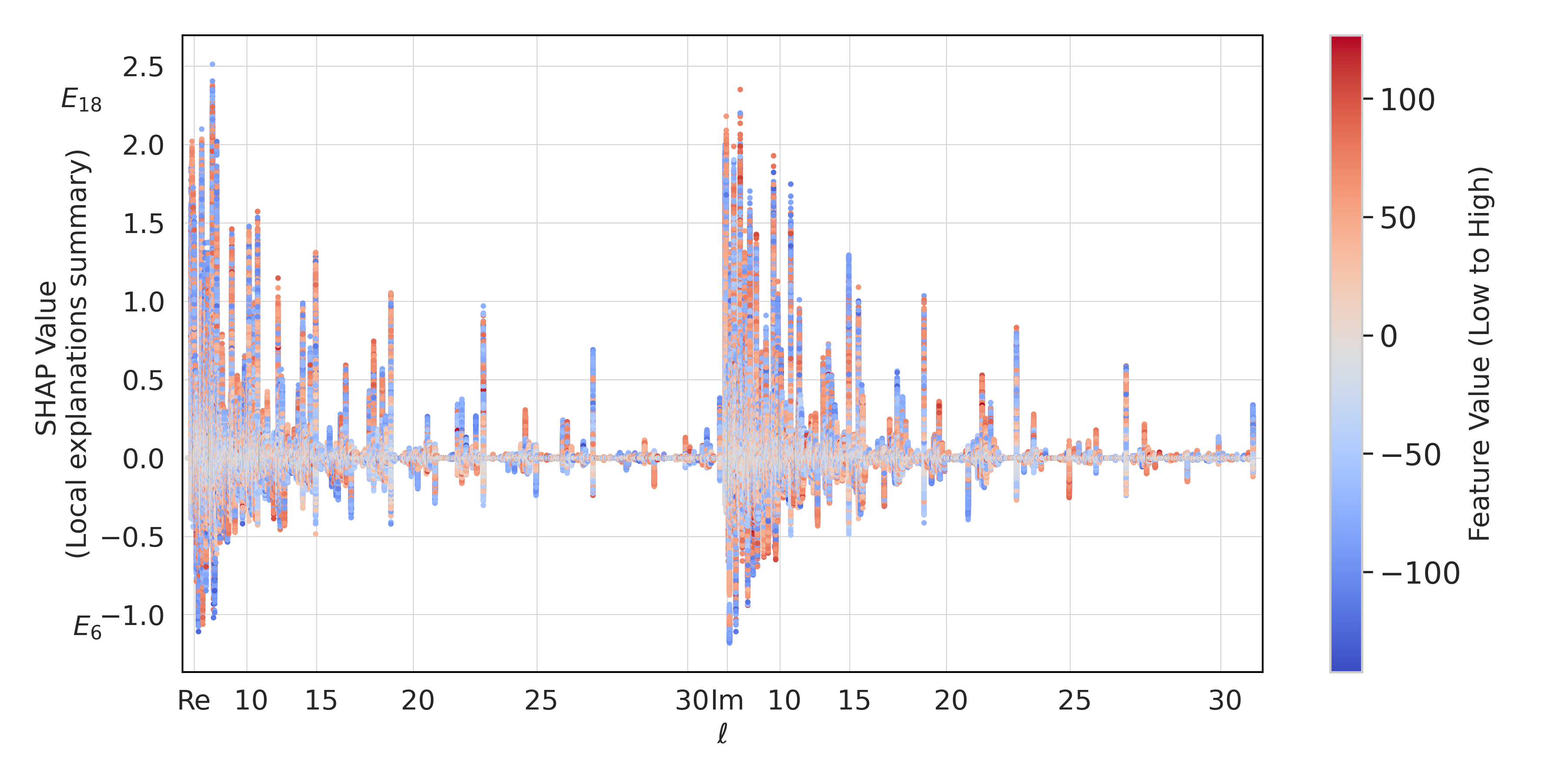}
    \caption{Local interpretability with \shap\ values for binary classification between \E{18} and an \E{6} manifold of size \(L = L_{A_{x}} = L_{B_{y}} = L_{C_{z}} = 1.01 \, \Lcircle\) with \(r_{x} = r_{y} = r_{z} = 1 / 2\). The features, i.e., the \(\alm\) coefficients are shown on the \(x\)-axis, ordered first in \(\ell\) and then in \(m\), and the feature values are indicated by color. The \(y\)-axis shows the \shap\ values for each feature of each realization.}
    \label{fig:Shap_local_E6_L1p01}
\end{figure}

\begin{figure}[htbp]
    \centering
    \includegraphics[width=\textwidth]{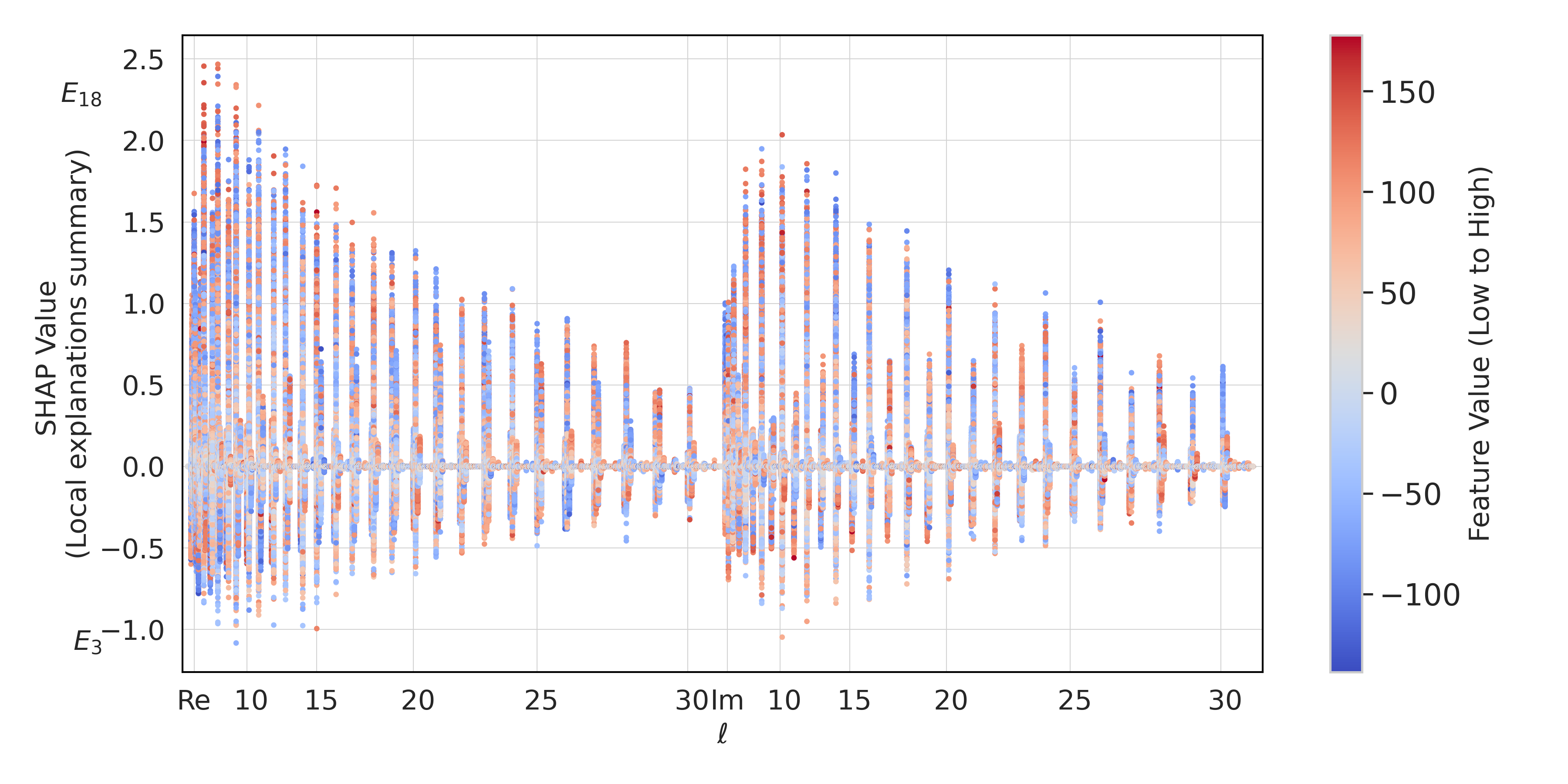}
    \caption{Local interpretability with \shap\ values for binary classification between \E{18} and \E{3} of size \(\LA = 1.4 \, \LLSS\) and \(\LB = 1.01 \, \LLSS\). The features, i.e., the \(\alm\) coefficients are shown on the \(x\)-axis, ordered first in \(\ell\) and then in \(m\), and the feature values are indicated by color. The \(y\)-axis shows the \shap\ values for each feature of each realization.}
    \label{fig:Shap_local_E3_L1p01}
\end{figure}

\begin{figure}[htbp]
    \centering
    \includegraphics[width=\textwidth]{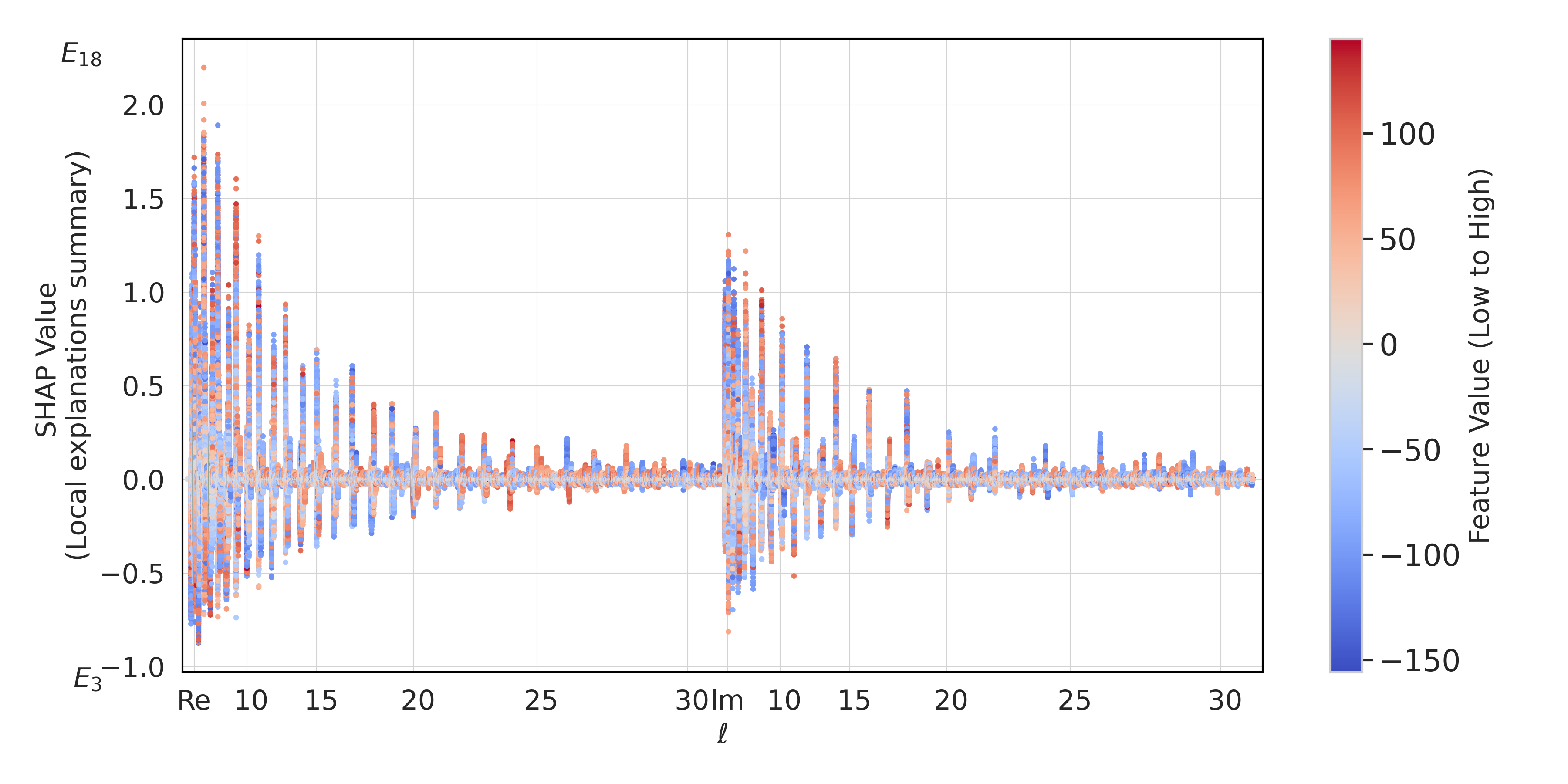}
    \caption{Local interpretability with \shap\ values for binary classification between \E{18} and \E{3} of size \(\LA = 1.4 \, \LLSS\) and \(\LB = 1.05 \, \LLSS\). The features, i.e., the \(\alm\) coefficients are shown on the \(x\)-axis, ordered first in \(\ell\) and then in \(m\), and the feature values are indicated by color. The \(y\)-axis shows the \shap\ values for each feature of each realization.}
    \label{fig:Shap_local_E3_L1p05}
\end{figure}

\cref{fig:Shap_local_E1_L1p01} presents local explanations with \shap\ values for binary classification between the trivial topology \E{18} and the cubic \E{1} of size \(L = 1.01 \, \LLSS\) evaluated on 40,000 realizations of the test dataset. The $n=957$ independent features (real and imaginary parts of the \(\alm\) coefficients for up to \(\ell_{\text{max}} = 30\)) are shown on the \(x\)-axis, ordered first in \(\ell\) and then in \(m\). The values of these features are indicated by color, blue corresponds to low and red to high values. The \(y\)-axis shows the \shap\ values for each feature of each realization. Values near zero indicate that the corresponding feature has negligible influence. The higher a positive \shap\ value, the stronger did the corresponding feature push the classification result toward the trivial \E{18} while negative \shap\ values indicate how strongly a certain feature has pushed the prediction towards \E{1}. The figure reveals that low-\(\ell\) coefficients are especially influential, consistent with topology signatures being primarily contained in the large angle correlations. For the imaginary parts of the \(\alm\) coefficients, large positive \shap\ values (favoring \E{18}) tend to occur for features with extreme values, whereas classification as \E{1} appears to be related to feature values around zero. For the real part of the \(\alm\) coefficients, very high or low feature values can also indicate towards \E{1}. 
Importantly, classification is not based on any single \(\alm\) coefficient but results from complex combinations or interactions among multiple features. \cref{fig:Shap_local_E6_L1p01}, which shows \shap\ values for binary classification between the trivial topology \E{18} and an \E{6} manifold with a single length scale \(L = 1.01 \, \Lcircle\), confirms this but also shows clear differences. In this case, the coefficients for \(\ell < 10\) are especially important to classify a realization as \E{6}. Moreover, there are fewer \(\alm\) coefficients with high \shap\ values and their distribution among the \(\ell\)-values is different. The \shap\ values for binary classification between the trivial topology \E{18} and \E{3} of size \(\LA = 1.4 \, \LLSS\) and \(\LB = 1.01 \, \LLSS\) are shown in \cref{fig:Shap_local_E3_L1p01}. The distribution of \(\alm\) coefficients with high \shap\ values among the \(\ell\)-values is similar to that for \E{1} but both can be clearly distinguished. Thus, for manifolds slightly larger than the limit for circle-detection, the \shap\ values of the \(\alm\) coefficients have a characteristic distribution, which depends on the topology. \cref{fig:Shap_local_E3_L1p05} shows the \shap\ values for binary classification between the trivial topology \E{18} and an \E{3} manifold of size \(\LA = 1.4 \, \LLSS\) and \(\LB = 1.05 \, \LLSS\), i.e., a slightly larger fundamental domain than in the previous figure. While the characteristic \E{3} distribution of \shap\ values across the \(\alm\) coefficients is still visible for \(\ell < 20\), it is clearly attenuated. For manifolds significantly larger than the limit for circle detection, topological signatures are present almost only in large scale correlations.  

\begin{figure}[htbp]
  \centering
    \includegraphics[width=0.89\textwidth]{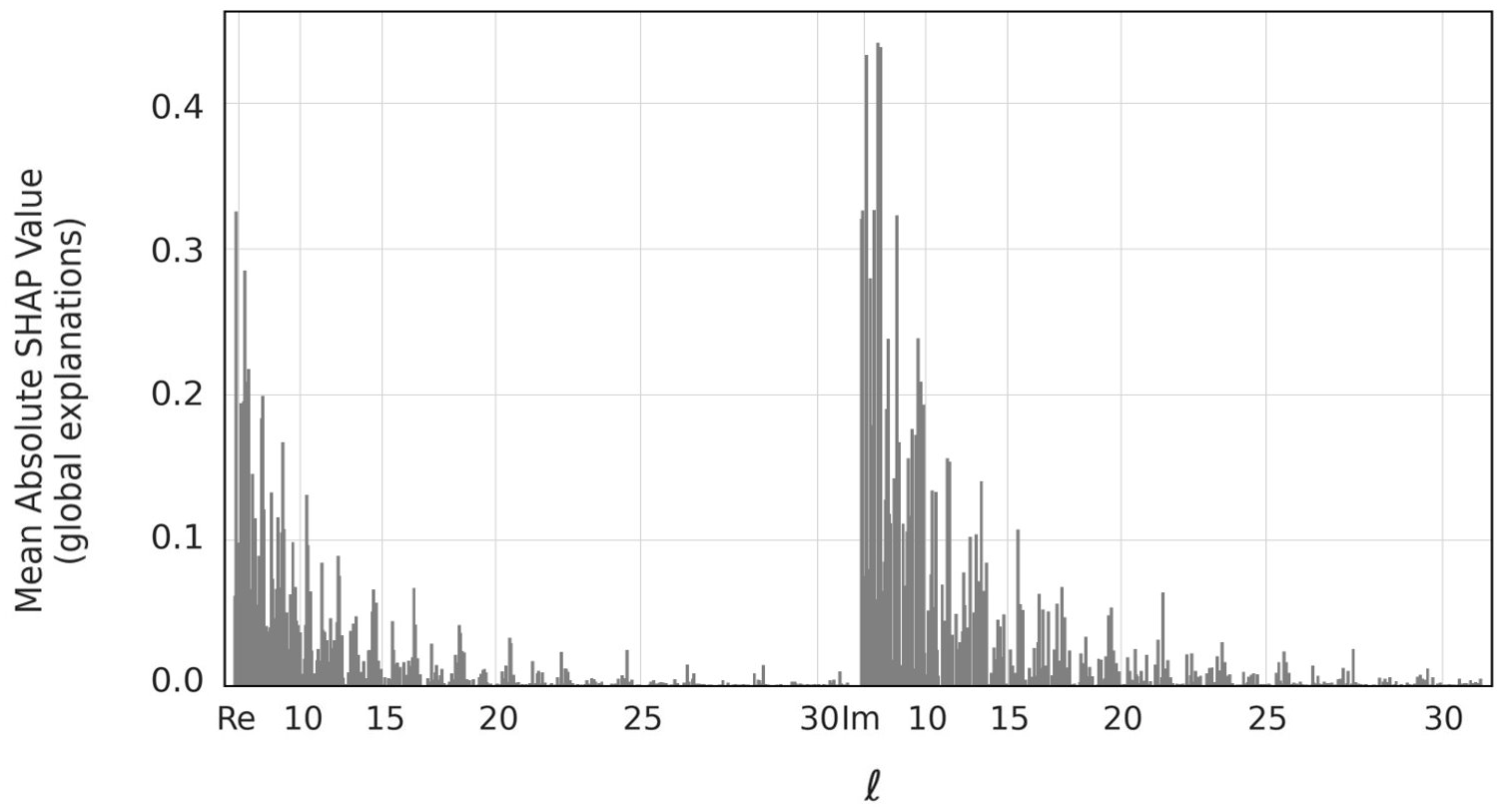}
    \includegraphics[width=0.9\textwidth]{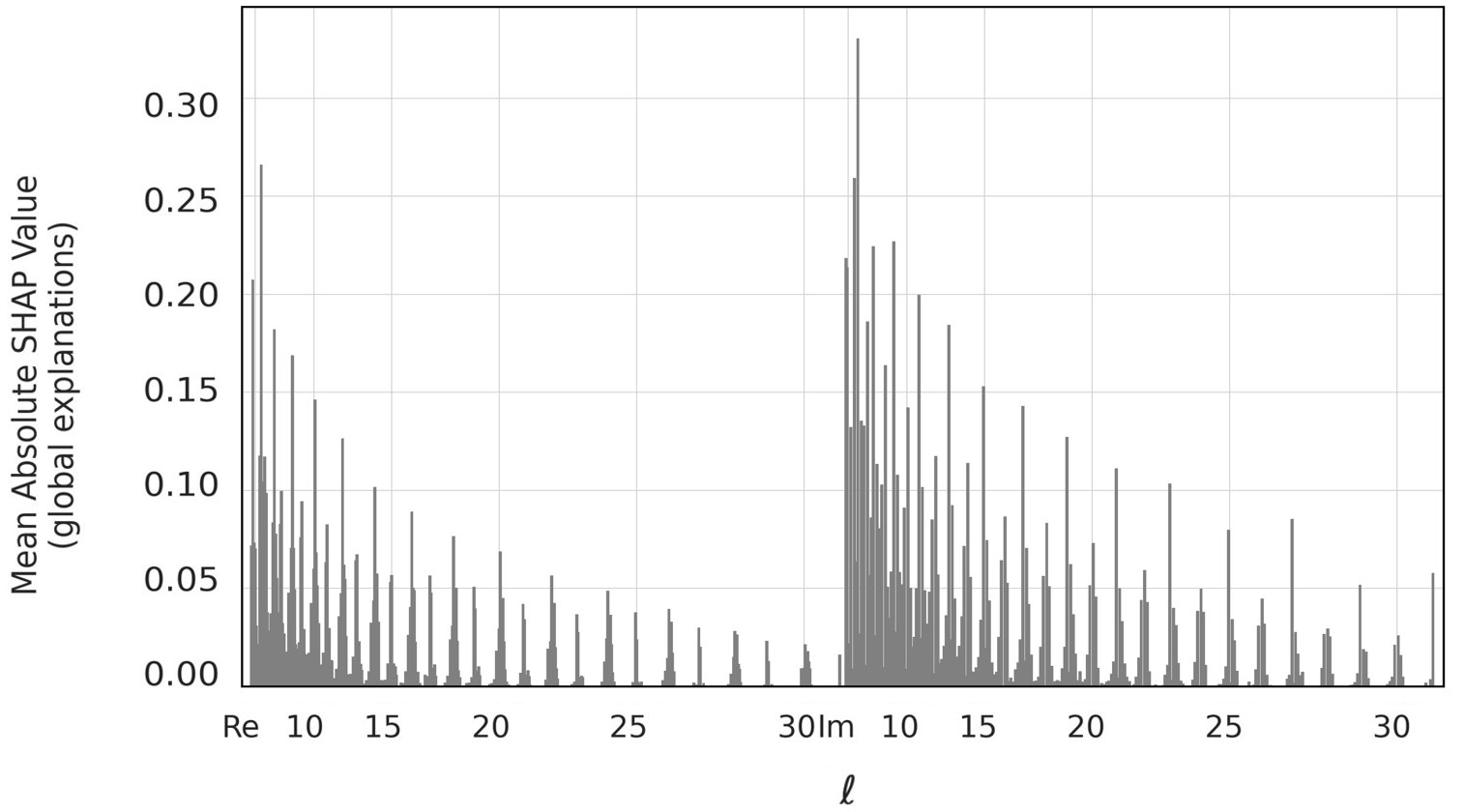}
\caption{Global interpretability from the mean absolute \shap\ values for binary classification between \E{18} and the cubic \E{1} of size $L=0.8\LLSS$ (top panel) and $L=1.01\LLSS$ (bottom panel).}
    \label{fig:Shap_global}
\end{figure}

\begin{figure}[htbp]
  \centering
    \includegraphics[width=0.9\textwidth]{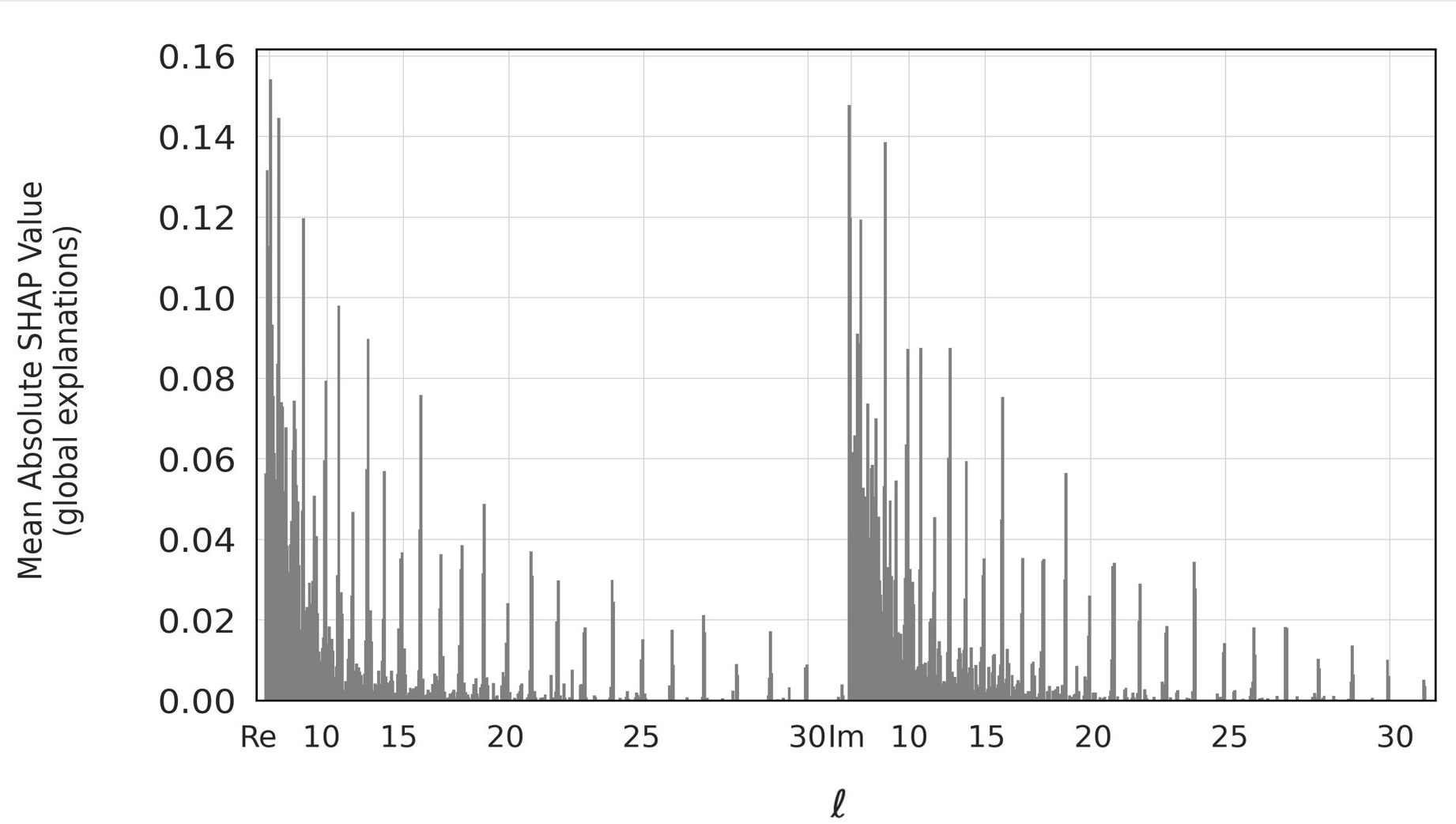}
    \includegraphics[width=0.9\textwidth]{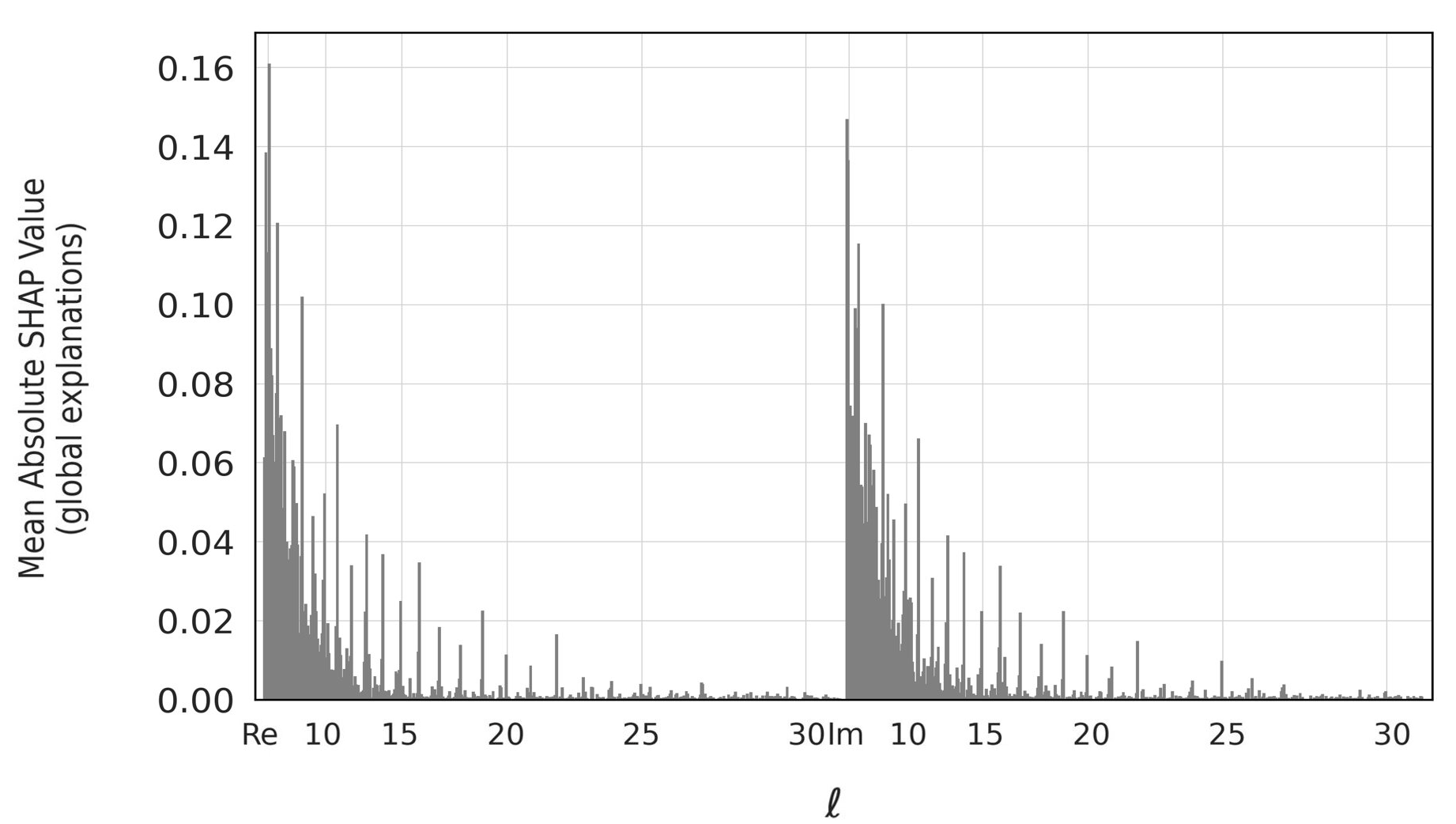}
\caption{Global interpretability from the mean absolute \shap\ values for binary classification between \E{18} and the \E{6} manifold of size $L=L_{A_x}=L_{B_y}=L_{C_z}=1.01\LLSS$ (top panel) and $L=1.05\LLSS$ (bottom panel) with $r_x=r_y=r_z=1/2$ and an off-axis observer with \(\xobs = (-0.5, 0.25, 0)\).}
    \label{fig:Shap_global_E6}
\end{figure}

The global explanations given by the mean absolute \shap\ values are shown in \cref{fig:Shap_global} for binary classification of the trivial \E{18} and the cubic \E{1} manifold of size \(L = 0.8 \, \LLSS\) (top) and size \(L = 1.01 \, \LLSS\) (bottom). For certain \(\alm\), sharp peaks in feature importance appear for manifolds slightly larger than the circle-detection limit \(\Lcircle\). These may be related to the residual correlations from the no longer visible matching circle pairs. The effect weakens for \(L = 1.05 \, \LLSS\) and becomes indistinct for larger fundamental domains. It is also not present for small manifolds with \(L \leq \Lcircle\) and no clear peaks are visible for off-axis observer in \E{3} and \E{4}. \cref{fig:Shap_global_E6} shows the mean absolute \shap\ values for binary classification of the trivial \E{18} and the \E{6} manifold with an off-axis observer, and with a single length scale of size \(L = 1.01 \, \LLSS\) (top) and size \(L = 1.05 \, \LLSS\) (bottom). For the slightly larger \E{6} manifold of size $L=1.05\LLSS$, no peaks are visible anymore and only the low-\(\ell\) coefficients have significant feature importance. While these characteristic peaks are clearly visible in the structure of the \shap\ values, others, like for example the topological differences for manifolds $L\geq 1.1\LLSS$, are subtler and cannot be linked to a distinct distribution of \shap\ values across the \(\alm\) coefficients. Similarly, modifications to the primordial power spectrum influence the \shap\ values in subtle ways, making it difficult to distinguish genuine physical effects from variability due to model stochasticity. 
\clearpage

\subsection{CatBoost hyperparameters}
\label{app:hyperparams}

In supervised ML, parameters are values that are learned during training (e.g. the weights of decision tree leaves) while hyperparameters are set before training and control the learning process. Setting appropriate hyperparameters requires balance: if the model is too complex (low regularization), it might overfit, i.e., fit the noise in the training data, leading to poor generalization. Conversely, an overly simple model (high regularization), might underfit and fail to capture important patterns in the data. There are a number of hyperparameters in \catboost, which can be tuned to achieve optimal performance for a given classification task. 
A key parameter is the number of boosting iterations, which specifies the maximum number of trees in an ensemble. 
We set a deliberately high upper limit of 15,000 iterations, in combination with early stopping after 100 rounds. Early stopping halts training if the \texttt{LogLoss} (for binary classification) or \texttt{MultiClass} (for multiclass classification) metric on the validation set does not improve for 100 consecutive iterations. i
It thereby ensures that the model does not overfit to the training data. Among the available hyperparameters, we focus on optimizing the following four:

\begin{itemize}
    \item \texttt{learning\_rate}: controls the contribution of each individual tree to the overall prediction. Smaller values generally improve generalization but require more boosting iterations.

    \item \texttt{depth}: refers to the maximum depth of trees in the \catboost\ ensemble. Limiting tree depth is a form of regularization because it regulates model complexity. Deeper trees can capture more intricate patterns but are more prone to overfitting.

    \item \texttt{l2\_leaf\_reg}: L2 regularization adds a quadratic penalty term to the loss function
    \begin{equation}
        L (\theta) = L_{0} (\theta) + \lambda ||\theta||_{2}^{2},
    \end{equation}
    where \(L_{0} (\theta)\) is the original loss function, \(\lambda\) is the regularization strength and \(||\theta||_{2}^{2}\) is the squared \(L2\) norm of the model parameters. This discourages the model from assigning too much importance to any single feature, thereby improving generalization, especially when the number of boosting iterations is large.

    \item \texttt{random\_strength}: During split selection, \catboost\ computes a score for each possible feature and threshold, \texttt{random\_strength} adds Gaussian noise to these scores before selecting the best split. For small values, split selection is nearly deterministic, while for larger values, more randomness is injected, which reduces sensitivity to small fluctuations in the data, thereby lowering the risk of overfitting.

\end{itemize}

We perform a coarse grid search over a wide range of \texttt{learning\_rate} and \texttt{depth} values to identify promising regions in hyperparameter space. This is followed by a refined grid search around the best-performing combinations. 
The resulting optimal region is then explored using Bayesian optimization,\footnote{Bayesian optimization is a sequential model-based approach to hyperparameter tuning. Instead of exhaustively or randomly searching the parameter space, it builds a probabilistic surrogate model (in \catboost, typically based on a tree-structured model) of the objective function. An acquisition function then determines which hyperparameters to try next, balancing exploration of uncertain regions with exploitation of areas likely to yield good results. 
This strategy is particularly sample-efficient, requiring fewer evaluations of the expensive training process compared to grid or random search.} which efficiently searches for the best parameters. 
Once \texttt{learning\_rate} and \texttt{depth} are fixed, we optimize \texttt{l2\_leaf\_reg} and \texttt{random\_strength} directly via Bayesian optimization.\par

\begin{figure}[t]
    \centering
    \includegraphics[scale=1]{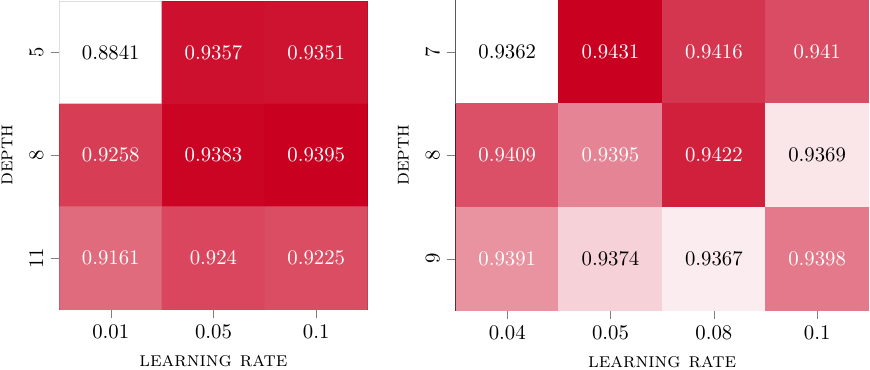}
    \caption{Grid search results for the \E{3} topology of size \(\LA = 1.4 \, \LLSS\) and \(L_{B} = 1.01 \, \LLSS\) with an off-axis observer with \(\xobs = (0.35, 0.35, 0)\), for a wide grid (left panel) and a consequent more refined one (right panel).}
    \label{fig:grid_search}
\end{figure}

\begin{figure}
    \centering
    \includegraphics[width=0.75\linewidth]{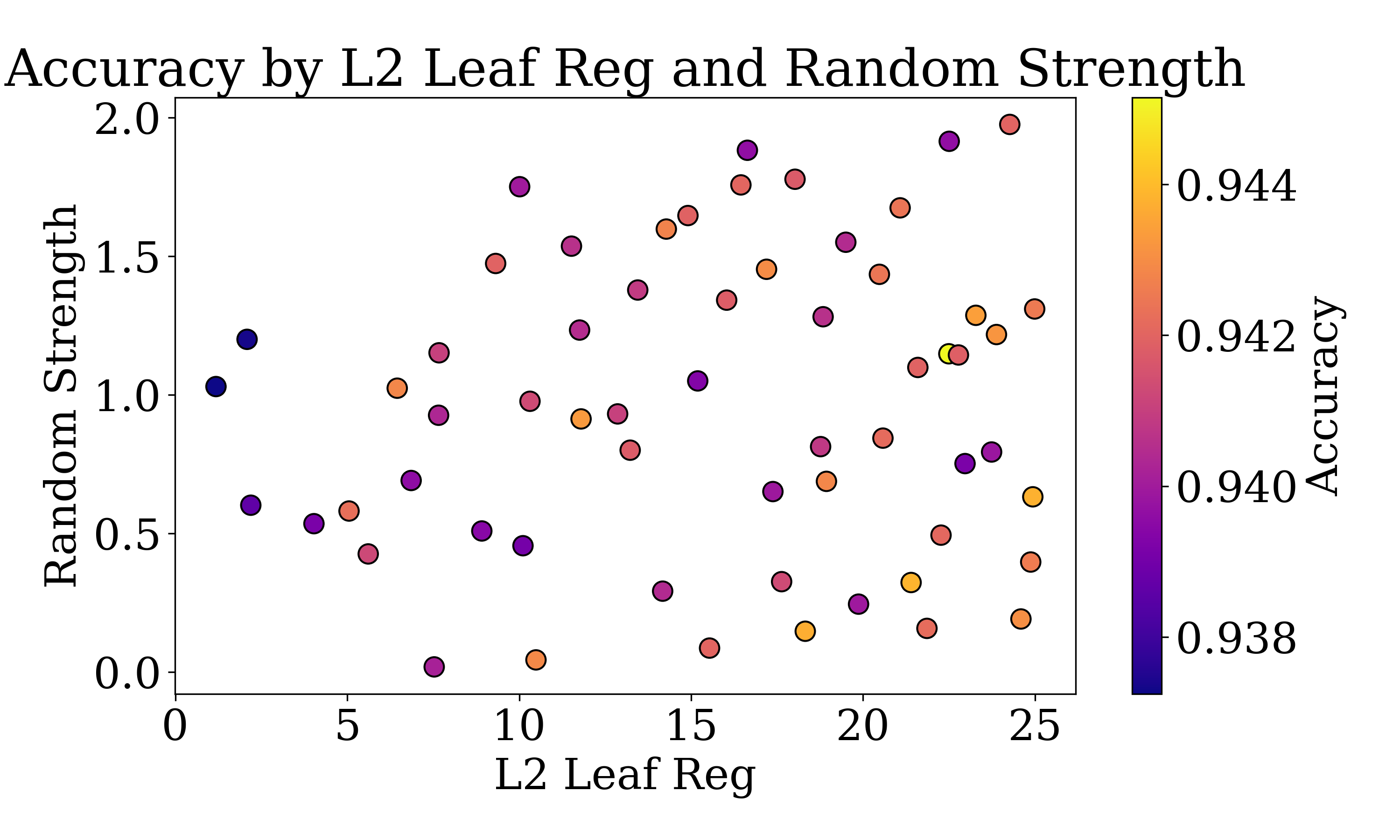}
    \caption{Bayesian optimization of the hyperparameters \texttt{l2\_leaf\_reg} and \texttt{random\_strength} for the \E{3} topology with an off-axis observer \(\xobs = (0.35, 0.35, 0)\) of size \(\LA = 1.4 \, \LLSS\) and \(L_{B} = 1.01 \, \LLSS\).}
    \label{fig:bayesian}
\end{figure}

For binary classification, hyperparameter tuning was performed on realizations of size \(L = 1.01 \, \Lcircle\) (or \(L_{B} = 1.01 \, \Lcircle\)) with an unmodified primordial power spectrum. Even though optimal hyperparameters may depend on the dataset, the procedure was not repeated for all fundamental domain sizes and power spectrum modifications, due to its high computational cost. \cref{fig:grid_search} shows an example grid search for the \E{3} topology with an off-axis observer, where optimal hyperparameters are first searched over a wide and then a refined grid. Note that \catboost\ is not fully deterministic\footnote{\catboost\ uses randomized elements in tree building through e.g. ordered boosting and Bayesian bootstrapping (\(\texttt{bagging\_temperature} = 1\) by default), and split selection (\(\texttt{random\_strength} \neq 0\)). For large datasets with \(> 100.000\) samples, run-to-run differences in accuracy are usually \(< 0.3\%\). We have performed optimization with fixed cross validation folds, so the folds are deterministic.} and thus when repeating with exactly the same hyperparameters, the accuracy can slightly differ between runs. After the grid search, \texttt{learning\_rate} and \texttt{depth} are finetuned simultaneously via Bayesian optimization over a narrowed range. \cref{fig:bayesian} shows the consequent Bayesian optimization of the hyperparameters \texttt{l2\_leaf\_reg} and \texttt{random\_strength}. The optimization process was repeated for the different topologies and observer positions considered. The final hyperparameters used in each scenario are listed in \cref{tab:cat_hyper}.

\begin{table}[b]
    \centering
    \begin{tabular}{l c c c c}
        \toprule
        \textsc{topology \&} & \multirow{2}{*}{\texttt{learning\_rate}} & \multirow{2}{*}{\texttt{depth}} & \multirow{2}{*}{\texttt{l2\_leaf\_reg}} & \texttt{random} \\
        \textsc{observer position} & & & & \texttt{strength} \\
        \midrule
        \E{1}: \(\xobs = (0, 0, 0)\) & 0.02 & 7 & 15 & 1.5 \\
        \E{3}: \(\xobs = (0, 0, 0)\) & 0.02 & 8 &  20 & 1.5 \\
        \E{3}: \(\xobs = (0.35, 0.35, 0)\) & 0.06 & 7 & 22 & 1.1 \\
        \E{4}: \(\xobs = (0, 0, 0)\) & 0.01 & 10 & 15 & 0.4 \\
        \E{4}: \(\xobs = (0.35, -0.20, 0)\) & 0.06 & 7 & 24 & 1.6 \\
        \E{6}: \(\xobs = (0, 0, 0)\) & 0.06 & 8 & 19 & 0.1 \\
        \E{6}: \(\xobs = (-0.5, 0.25, 0)\) & 0.05 & 8 & 20 & 1.9 \\
        \textsc{multiclass} & 0.015 -- 0.04 & 8 -- 10 & 6 -- 15 & 1.0 \\
        \bottomrule
    \end{tabular}
    \caption{Hyperparameters used for Binary and Multiclass classification with \texttt{CatBoost}. In all cases, the \texttt{bagging\_temperature} was kept at its default value of \(1.0\).}
    \label{tab:cat_hyper}
\end{table}

\begin{figure}[htbp]
  \centering
  \begin{subfigure}[b]{0.48\textwidth}
    \centering
    \includegraphics[width=\textwidth]{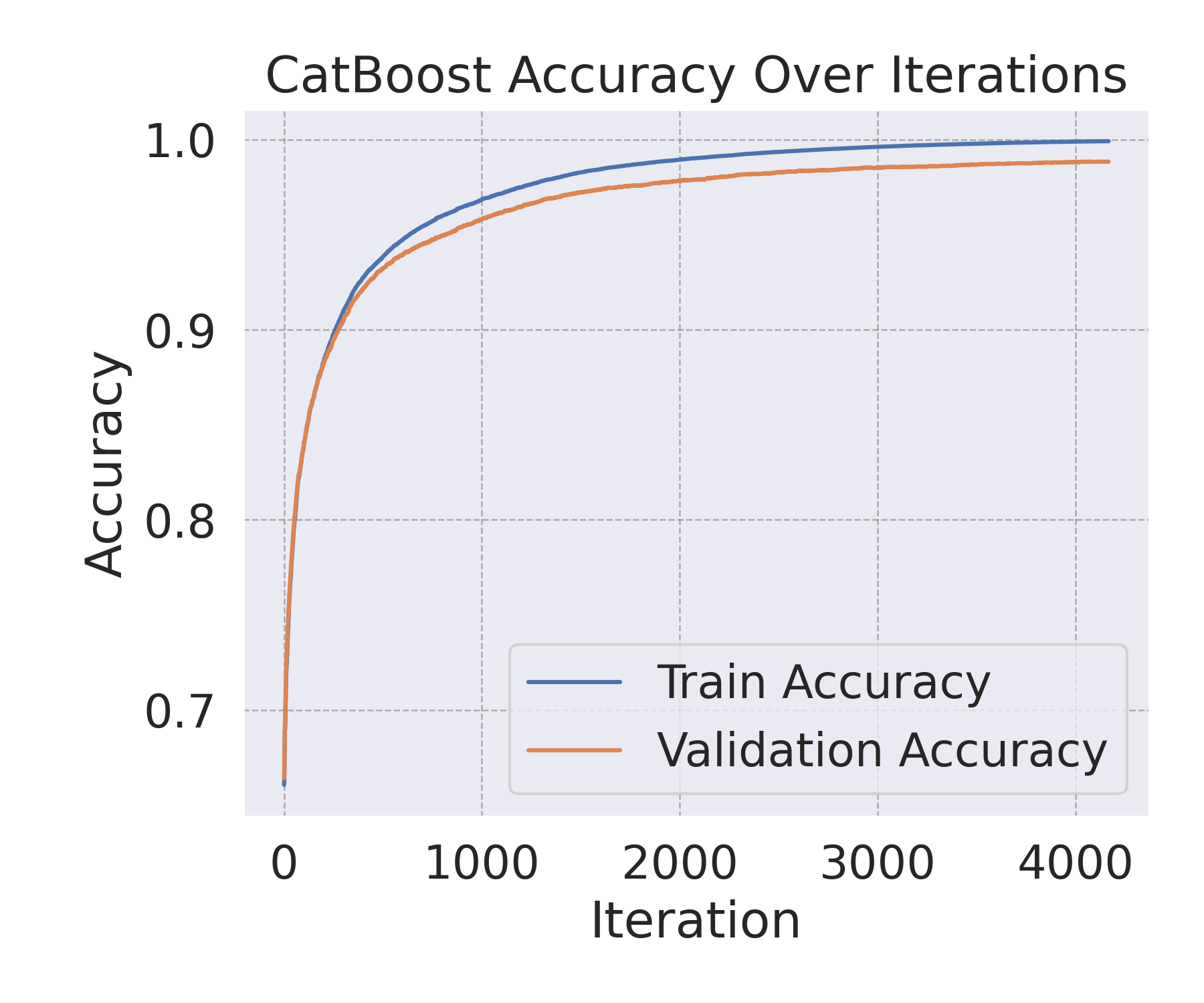}
  \end{subfigure}
  \hfill
  \begin{subfigure}[b]{0.48\textwidth}
    \centering
    \includegraphics[width=\textwidth]{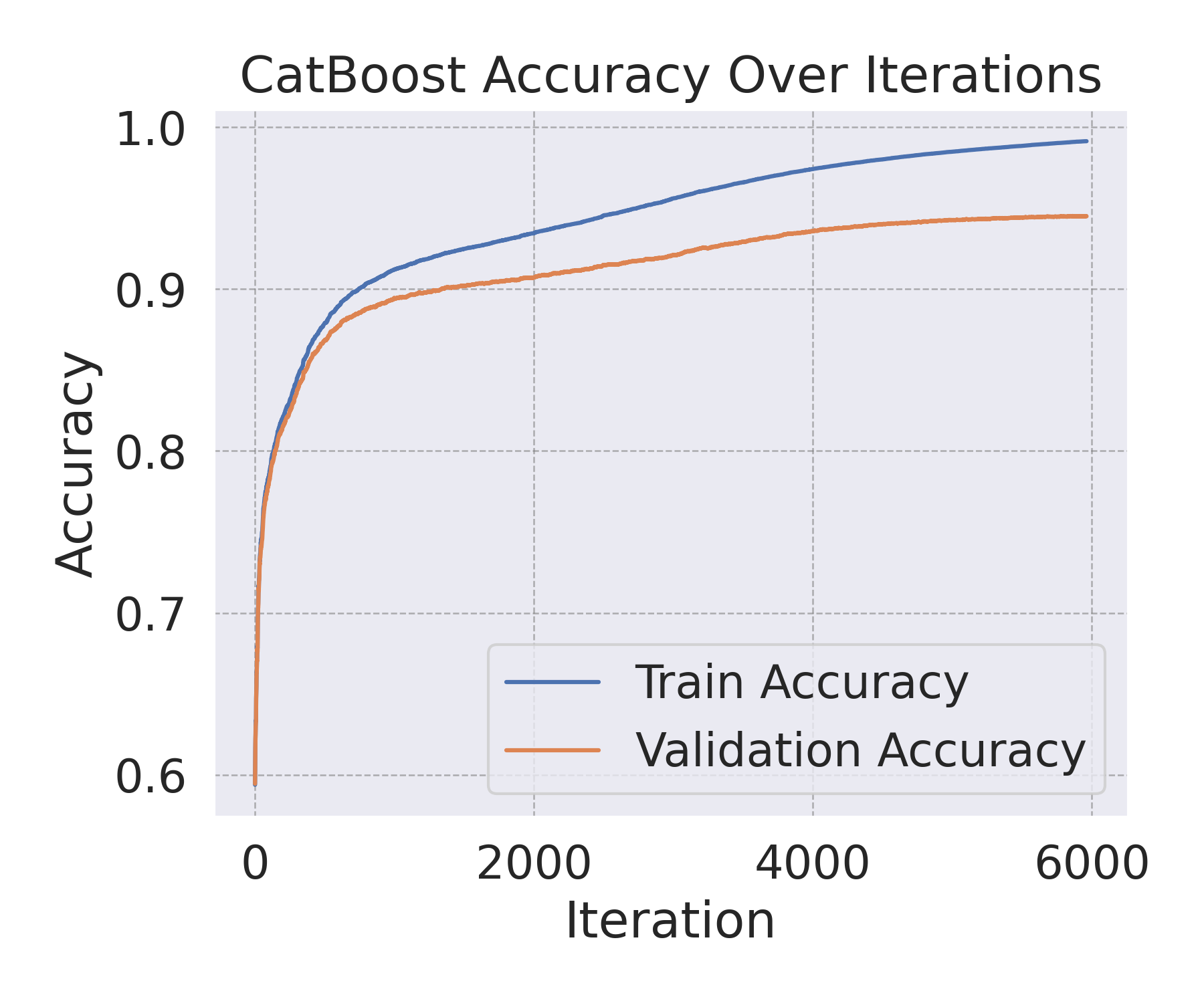}
  \end{subfigure}
\caption{Training and test accuracy for cubic \E{1} with \(L = L_{1} = L_{2} = L_{3} = 1.01 \, \LLSS\) (left panel) and for the \E{3} topology with \(\LA = 1.4 \, \LLSS\), \(\LB = 1.01 \, \LLSS\) and an on-axis observer (right panel). No modifications are applied to the primordial power spectrum.}
    \label{fig:Acc_train_test}
\end{figure}

\cref{fig:Acc_train_test} shows the evolution of training and validation accuracy during boosting. For the non-cubic \E{3} and \E{4} topologies, the validation accuracy converges to lower values than for \E{1} and \E{6}. This is consistent with the fact that the non-cubic cases also exhibit lower KL divergences (see \cref{fig:KL_all_on_off_ax}), indicating that those realizations are intrinsically harder to distinguish from the covering space \E{18}. The reduced accuracy is therefore a reflection of weaker topological information, rather than suboptimal hyperparameters. In line with this interpretation, the non-cubic cases also show moderate overfitting, with larger gaps between training and validation accuracy. Importantly, in all cases the accuracy curves converge smoothly and stably, indicating that the selected hyperparameters provide a good balance between model complexity, regularization, and learning rate.

\clearpage
\bibliographystyle{utphys}
\bibliography{topology, additional}
\end{document}